\documentclass[a4paper,11pt,openright,msc,final]{ufmgthesis}

\usepackage[brazil]{babel}      
\usepackage[latin1]{inputenc}
\usepackage[T1]{fontenc}
\usepackage{type1ec}
\usepackage{graphicx}

\usepackage{amsthm,amsfonts,amssymb,amsmath}
\usepackage{wasysym}
\usepackage{etex}
\usepackage[all,xdvi,knot]{xy}
\usepackage{proof}
\usepackage{stmaryrd}
\usepackage{palatino, url, multicol}
\usepackage{mathrsfs}
\def\grammareq {\mathrel{\raise.4pt\hbox{::}{=}}}%

\def\rulem {{\textsf{m}}}%
\def\rules {\textsf{ s }}%
\def\ruleaidown {{\textsf{a}\mkern-1.5mu\textsf{i}{\downarrow  }}}%
\def\ruleaiup {{\textsf{a}\mkern-1.5mu\textsf{i}{\uparrow  }}}%
\newcommand{\dual}[1]{\bar{#1}}

\newcommand{\at}[1]{\textsf{at} \ #1}
\def\ruleis{\textsf{is}}
\def\rulelis{\textsf{lis}}
\def\ruleqidown{\textsf{q$_{1} \downarrow$}}
\def\ruleqiidown{\textsf{q$_{2} \downarrow$}}
\def\rulelqiiidown{\textsf{lq$_{3} \downarrow$}}

\def\ruleiqidown{\textsf{iq$_{1} \downarrow$}}
\def\ruleliqiiidown{\textsf{liq$_{3} \downarrow$}}
\def\ruleliqivdown{\textsf{liq$_{4} \downarrow$}}
\def\ruleps{\textsf{ps}}
\def\rulepqdown{\textsf{pq}\downarrow}
\def\BVs{\textsf{BVs}}
\def\BVsl{\textsf{BVsl}}
\def\QVI{\textsf{QVI}}
\def\BVi{\textsf{BVi}}
\def\BVp{\textsf{BVp}}

\def\cosprover{\texttt{CoSProver}}

\def\mix{\textsf{mix}}
\def\MLL{\textsf{MLL}}

\newcommand{\conjunto}[2]{*+++[o][F-]{\txt{\{$#1$\} \\ \textbf{$#2$}}}}

\newcommand{\negsymbol}[1]{\~{}#1}
\def\unidade{*}
\newcommand{\pr}[1]{[#1]}
\newcommand{\cpr}[1]{(#1)}

\def\ruleodown{\circ \downarrow}%

\newcommand{\relweb}[1]{\textsf{w} \ #1}

\newtheorem{Theorem}{Teorema}
\newtheorem{Corollary}[Theorem]{Corolário}
\newtheorem{Proposition}[Theorem]{Proposição}
\newtheorem{Lemma}[Theorem]{Lema}
\newtheorem{Conjecture}[Theorem]{Conjectura}

\theoremstyle{definition}
\newtheorem{Definition}[Theorem]{Definição}
\newtheorem{Remark}[Theorem]{Observação}
\newtheorem{Example}[Theorem]{Exemplo}

\newtheorem{Notation}[Theorem]{Notação}

\newcommand{\aincnumber}[3]{<#1>_{#2,#3}^{\#}}
\newcommand{\incnumber}[3]{\{#1\}_{#2,#3}^{\#}}
\newcommand{\occ}[1]{{\textsf{occ}} \ #1}
\newcommand{\FBV}{{\textsf{FBV}}}
\newcommand{\BV}{{\textsf{BV}}}
\newcommand{\V}{{\textsf{V}}}

\def\ruleqdown {{\textsf{q}{\downarrow  }}}%
\newcommand{\LK}{\textsf{LK}}
\newcommand{\LJ}{\textsf{LJ}}
\newcommand{\LL}{\textsf{LL}}
\newcommand{\Gtc}{\textsf{G3c}}

\long\def\hide#1\endhide{}
\newcommand\cimp{\Rightarrow}
\newcommand\Seq[2]{#1\vdash #2}

\newcommand\limp{\mathbin{-\hspace{-0.70mm}\circ}}
\newcommand\iimp{\supset}
\newcommand\lpar{\bindnasrepma}

\newcommand\zero{0}
\newcommand\one{1}
\newcommand\bottom{\perp}
\newcommand\bang{\mathop{!}}
\newcommand\quest{\mathop{?}}
\newcommand\tensor\otimes
\newcommand\with{\mathbin{\&}}

\newcommand\lforall{\forall_{l}}
\newcommand\lexists{\exists_{l}}

\newcommand\Five[6]{#1\colon#2;#3\mathrel{\buildrel #4\over\longrightarrow}#5;#6}
\newcommand\Four[5]{#1\colon #2;#3\longrightarrow#4;#5}

\newcommand\Ascr{\mathcal{A}}

\def\lrgldel {\mathchoice{(}{(}{\langle}{\langle}}%
\def\lrgrdel {\mathchoice{)}{)}{\rangle}{\rangle}}%
\def\aprldel {\mathchoice
   {\mathopen {\setbox0=\hbox{$\displaystyle     \lrgldel$}\hbox to\wd0
                        {\hfil$\displaystyle     (       $\hfil}}}%
   {\mathopen {\setbox0=\hbox{$\textstyle        \lrgldel$}\hbox to\wd0
                        {\hfil$\textstyle        (        $\hfil}}}%
   {\mathopen {\setbox0=\hbox{$\scriptstyle      \lrgldel$}\hbox to\wd0
                        {\hfil$\scriptstyle      (        $\hfil}}}%
   {\mathopen {\setbox0=\hbox{$\scriptscriptstyle\lrgldel$}\hbox to\wd0
                        {\hfil$\scriptscriptstyle(        $\hfil}}}}%
\def\aprrdel {\mathchoice
   {\mathclose{\setbox0=\hbox{$\displaystyle     \lrgrdel$}\hbox to\wd0
                        {\hfil$\displaystyle     )       $\hfil}}}%
   {\mathclose{\setbox0=\hbox{$\textstyle        \lrgrdel$}\hbox to\wd0
                        {\hfil$\textstyle        )        $\hfil}}}%
   {\mathclose{\setbox0=\hbox{$\scriptstyle      \lrgrdel$}\hbox to\wd0
                        {\hfil$\scriptstyle      )        $\hfil}}}%
   {\mathclose{\setbox0=\hbox{$\scriptscriptstyle\lrgrdel$}\hbox to\wd0
                        {\hfil$\scriptscriptstyle)        $\hfil}}}}%
\def\seqldel {\mathchoice
   {\mathopen {\setbox0=\hbox{$\displaystyle     \lrgldel$}\hbox to\wd0
                        {\hfil$\displaystyle     \langle  $\hfil}}}%
   {\mathopen {\setbox0=\hbox{$\textstyle        \lrgldel$}\hbox to\wd0
                        {\hfil$\textstyle        \langle  $\hfil}}}%
   {\mathopen {\setbox0=\hbox{$\scriptstyle      \lrgldel$}\hbox to\wd0
                        {\hfil$\scriptstyle      \langle  $\hfil}}}%
   {\mathopen {\setbox0=\hbox{$\scriptscriptstyle\lrgldel$}\hbox to\wd0
                        {\hfil$\scriptscriptstyle\langle  $\hfil}}}}%
\def\seqrdel {\mathchoice
   {\mathclose{\setbox0=\hbox{$\displaystyle     \lrgrdel$}\hbox to\wd0
                        {\hfil$\displaystyle     \rangle  $\hfil}}}%
   {\mathclose{\setbox0=\hbox{$\textstyle        \lrgrdel$}\hbox to\wd0
                        {\hfil$\textstyle        \rangle  $\hfil}}}%
   {\mathclose{\setbox0=\hbox{$\scriptstyle      \lrgrdel$}\hbox to\wd0
                        {\hfil$\scriptstyle      \rangle  $\hfil}}}%
   {\mathclose{\setbox0=\hbox{$\scriptscriptstyle\lrgrdel$}\hbox to\wd0
                        {\hfil$\scriptscriptstyle\rangle  $\hfil}}}}%
\def\parldel {\mathchoice
   {\mathopen {\setbox0=\hbox{$\displaystyle     \lrgldel$}\hbox to\wd0
                        {\hfil$\displaystyle     [       $\hfil}}}%
   {\mathopen {\setbox0=\hbox{$\textstyle        \lrgldel$}\hbox to\wd0
                        {\hfil$\textstyle        [        $\hfil}}}%
   {\mathopen {\setbox0=\hbox{$\scriptstyle      \lrgldel$}\hbox to\wd0
                        {\hfil$\scriptstyle      [        $\hfil}}}%
   {\mathopen {\setbox0=\hbox{$\scriptscriptstyle\lrgldel$}\hbox to\wd0
                        {\hfil$\scriptscriptstyle[        $\hfil}}}}%
\def\parrdel {\mathchoice
   {\mathclose{\setbox0=\hbox{$\displaystyle     \lrgrdel$}\hbox to\wd0
                        {\hfil$\displaystyle     ]       $\hfil}}}%
   {\mathclose{\setbox0=\hbox{$\textstyle        \lrgrdel$}\hbox to\wd0
                        {\hfil$\textstyle        ]        $\hfil}}}%
   {\mathclose{\setbox0=\hbox{$\scriptstyle      \lrgrdel$}\hbox to\wd0
                        {\hfil$\scriptstyle      ]        $\hfil}}}%
   {\mathclose{\setbox0=\hbox{$\scriptscriptstyle\lrgrdel$}\hbox to\wd0
                        {\hfil$\scriptscriptstyle]        $\hfil}}}}%
\def\aprs #1{\aprldel #1\aprrdel}%
\def\pars #1{\parldel #1\parrdel}%
\def\seqs #1{\seqldel #1\seqrdel}%
\newdimen\dercldim                                
\newdimen\derccdim                                
\newdimen\dercrdim                                
\newdimen\derldim                                 
\newdimen\dercdim                                 
\newdimen\derrdim                                 
\newdimen\derdim                                  
\newdimen\derdldim                                
\newdimen\derdrdim                                
\newbox\derboxone                                 
\newbox\derboxtwo                                 
\newbox\derboxthree                               
\newbox\derboxfour                                
\newdimen\derquad\derquad=\fontdimen6\textfont2
\newdimen\deropen\deropen=\fontdimen5\textfont2\divide\deropen by3
\def\leaf #1{\global\setbox\derboxone=\hbox{\strut$#1$}%
   \global\derldim=0pt                            
   \global\dercdim=\wd\derboxone                  
   \global\derrdim=0pt                            
   }%
\def\rootaux #1#2#3{\setbox\derboxtwo=\hbox{\unhbox\derboxone}%
   \setbox\derboxthree=\hbox 
      {$\smash{\lower\fontdimen22\textfont2\hbox{$#1$}}$}%
   \setbox\derboxfour=\hbox 
      {$\smash{\lower\fontdimen22\textfont2\hbox{$#2$}}$}%
   \leaf{#3}
   \derdim=\dercdim\advance\derdim by-\derccdim\divide\derdim by2 
   \global\derldim=\dercldim\global\advance\derldim by-\derdim
   \global\derrdim=\dercrdim\global\advance\derrdim by-\derdim
   \deropen=\fontdimen5\textfont2\divide\deropen by3
   \setbox\derboxone=\hbox{\vbox{\offinterlineskip
         \hbox{\ifdim\derldim<0pt\kern-\derldim\fi
               \box\derboxtwo
               \ifdim\derrdim<0pt\kern-\derrdim\fi}%
         \kern\deropen
         \hbox{\ifdim\dercldim>\derldim
                  \ifdim\derldim>0pt\kern\derldim\fi
                  \else\kern\dercldim\fi
               \hbox to0pt{\hss\copy\derboxthree}%
               \vbox{\ifdim\derccdim>\dercdim\hsize=\derccdim
                                        \else\hsize=\dercdim \fi
                    \hrule height.2pt depth.2pt width\hsize}%
               \hbox to0pt{\copy\derboxfour\hss}%
               \ifdim\dercrdim>\derrdim
                  \ifdim\derrdim>0pt\kern\derrdim\fi
                  \else\kern\dercrdim\fi}%
         \kern\deropen
         \hbox{\ifdim\derldim>0pt\kern\derldim\fi
               \box\derboxone
               \ifdim\derrdim>0pt\kern\derrdim\fi}}}%
   \ifdim\derldim<0pt\global\derldim=0pt\fi       
   \ifdim\derrdim<0pt\global\derrdim=0pt\fi       
   \derdldim=\wd\derboxthree\advance\derdldim by-\dercldim
   \derdrdim=\wd\derboxfour \advance\derdrdim by-\dercrdim
   \ifdim\derdim<0pt
      \ifdim\derdldim<0pt
         \derdldim=0pt                            
      \fi
      \ifdim\derdrdim<0pt
         \derdrdim=0pt                            
      \fi
   \else
      \ifdim\derldim>0pt
         \ifdim\derdldim>-\derdim
            \advance\derdldim by\derdim           
         \else                                            
            \derdldim=0pt                         
         \fi                                      
      \else
         \advance\derdldim by\dercldim            
      \fi
      \ifdim\derrdim>0pt
         \ifdim\derdrdim>-\derdim
            \advance\derdrdim by\derdim           
         \else                                            
            \derdrdim=0pt                         
         \fi                                      
      \else
         \advance\derdrdim by\dercrdim            
      \fi
   \fi
   \global\setbox\derboxone=\hbox
      {\kern\derdldim\unhbox\derboxone\kern\derdrdim}%
   \global\advance\derldim by\derdldim            
   \global\advance\derrdim by\derdrdim            
   }%
\def\rootr #1#2#3#4{{#4}%
   \dercldim=\derldim
   \derccdim=\dercdim
   \dercrdim=\derrdim
   \rootaux{#1}{#2}{#3}}%
\def\rrootr #1#2#3#4#5{\derquad=\fontdimen6\textfont2
   {#4}%
           \dercldim  =\derldim
   \setbox\derboxtwo=\hbox{\unhbox\derboxone\kern\derquad}%
           \derccdim  =\dercdim
   \advance\derccdim by\derrdim
   \advance\derccdim by\derquad
   {#5}%
   \setbox\derboxone=\hbox{\unhbox\derboxtwo\unhbox\derboxone}%
   \advance\derccdim by\derldim
   \advance\derccdim by\dercdim
           \dercrdim  =\derrdim
   \rootaux{#1}{#2}{#3}}%
\def\rrrootr #1#2#3#4#5#6{\derquad=\fontdimen6\textfont2
   {#4}%
           \dercldim  =\derldim
   \setbox\derboxtwo=\hbox{\unhbox\derboxone\kern\derquad}%
           \derccdim  =\dercdim
   \advance\derccdim by\derrdim
   \advance\derccdim by\derquad
   {#5}%
   \setbox\derboxtwo=\hbox{\unhbox\derboxtwo\unhbox\derboxone\kern\derquad}%
   \advance\derccdim by\derldim
   \advance\derccdim by\dercdim
   \advance\derccdim by\derrdim
   \advance\derccdim by\derquad
   {#6}%
   \setbox\derboxone=\hbox{\unhbox\derboxtwo\unhbox\derboxone}%
   \advance\derccdim by\derldim
   \advance\derccdim by\dercdim
           \dercrdim  =\derrdim
   \rootaux{#1}{#2}{#3}}%
\def\root       #1#2#3{\rootr  {#1\;}{\;}{#2}{#3}}%
\def\rootnote #1#2#3#4{\rootr  {#1\;}{\;#2}{#3}{#4}}%

\newbox\stembox
\def\stemaux #1#2#3{\setbox\derboxtwo=\hbox{\unhbox\derboxone}%
   \setbox\derboxthree=\hbox{$#1$}   
   \setbox\derboxfour =\hbox{$#2$}   
      {\global\setbox\derboxone=\hbox{$#3$}%
      \global\derldim=0pt                         
      \global\dercdim=\wd\derboxone               
      \global\derrdim=0pt                         
      }
   \derdim=\dercdim\advance\derdim by-\derccdim\divide\derdim by2 
   \global\derldim=\dercldim\global\advance\derldim by-\derdim
   \global\derrdim=\dercrdim\global\advance\derrdim by-\derdim
   \deropen=\fontdimen5\textfont2\divide\deropen by3
   \setbox\derboxone=\hbox{\vbox{\offinterlineskip
         \hbox{\ifdim\derldim<0pt\kern-\derldim\fi
               \box\derboxtwo
               \ifdim\derrdim<0pt\kern-\derrdim\fi}%
         \kern-\deropen\kern-\ht\strutbox\kern-\dp\strutbox
         \hbox{\ifdim\dercldim>\derldim
                  \ifdim\derldim>0pt\kern\derldim\fi
                  \else\kern\dercldim\fi
               \hbox to0pt{\hss\copy\derboxthree}%
               \vbox{\ifdim\derccdim>\dercdim\hsize=\derccdim
                                        \else\hsize=\dercdim \fi
                    \hbox{$\vcenter{
                    \vbox{\offinterlineskip
                       \hbox{$\copy\stembox$}\kern-1\dp\stembox
                       \hbox{$\copy\stembox$}}}$}}%
               \hbox to0pt{\copy\derboxfour\hss}%
               \ifdim\dercrdim>\derrdim
                  \ifdim\derrdim>0pt\kern\derrdim\fi
                  \else\kern\dercrdim\fi}%
         \kern-\deropen
         \hbox{\ifdim\derldim>0pt\kern\derldim\fi
               \box\derboxone
               \ifdim\derrdim>0pt\kern\derrdim\fi}}}%
   \ifdim\derldim<0pt\global\derldim=0pt\fi       
   \ifdim\derrdim<0pt\global\derrdim=0pt\fi       
   \derdldim=\wd\derboxthree\advance\derdldim by-\dercldim
   \derdrdim=\wd\derboxfour \advance\derdrdim by-\dercrdim
   \ifdim\derdim<0pt
      \ifdim\derdldim<0pt
         \derdldim=0pt                            
      \fi
      \ifdim\derdrdim<0pt
         \derdrdim=0pt                            
      \fi
   \else
      \ifdim\derldim>0pt
         \ifdim\derdldim>-\derdim
            \advance\derdldim by\derdim           
         \else                                    
            \derdldim=0pt                         
         \fi                                      
      \else
         \advance\derdldim by\dercldim            
      \fi
      \ifdim\derrdim>0pt
         \ifdim\derdrdim>-\derdim
            \advance\derdrdim by\derdim           
         \else                                    
            \derdrdim=0pt                         
         \fi                                      
      \else
         \advance\derdrdim by\dercrdim            
      \fi
   \fi
   \global\setbox\derboxone=\hbox
      {\kern\derdldim\unhbox\derboxone\kern\derdrdim}%
   \global\advance\derldim by\derdldim            
   \global\advance\derrdim by\derdrdim            
   }%
\def\stemauxx #1#2#3{\setbox\derboxtwo=\hbox{\unhbox\derboxone}%
   \setbox\derboxthree=\hbox{$#1$}   
   \setbox\derboxfour =\hbox{$#2$}   
   \leaf{#3}
   \derdim=\dercdim\advance\derdim by-\derccdim\divide\derdim by2 
   \global\derldim=\dercldim\global\advance\derldim by-\derdim
   \global\derrdim=\dercrdim\global\advance\derrdim by-\derdim
   \deropen=\fontdimen5\textfont2\divide\deropen by3
   \setbox\derboxone=\hbox{\vbox{\offinterlineskip
         \hbox{\ifdim\derldim<0pt\kern-\derldim\fi
               \box\derboxtwo
               \ifdim\derrdim<0pt\kern-\derrdim\fi}%
         \kern\deropen
         \hbox{\ifdim\dercldim>\derldim
                  \ifdim\derldim>0pt\kern\derldim\fi
                  \else\kern\dercldim\fi
               \hbox to0pt{\hss\copy\derboxthree}%
               \vbox{\ifdim\derccdim>\dercdim\hsize=\derccdim
                                        \else\hsize=\dercdim \fi
                    \hbox{\hfil}}%
               \hbox to0pt{\copy\derboxfour\hss}%
               \ifdim\dercrdim>\derrdim
                  \ifdim\derrdim>0pt\kern\derrdim\fi
                  \else\kern\dercrdim\fi}%
         \hbox{\ifdim\derldim>0pt\kern\derldim\fi
               \box\derboxone
               \ifdim\derrdim>0pt\kern\derrdim\fi}}}%
   \ifdim\derldim<0pt\global\derldim=0pt\fi       
   \ifdim\derrdim<0pt\global\derrdim=0pt\fi       
    \derdldim=\wd\derboxthree\advance\derdldim by-\dercldim
   \derdrdim=\wd\derboxfour \advance\derdrdim by-\dercrdim
   \ifdim\derdim<0pt
      \ifdim\derdldim<0pt
         \derdldim=0pt                            
      \fi
      \ifdim\derdrdim<0pt
         \derdrdim=0pt                            
      \fi
   \else
      \ifdim\derldim>0pt
         \ifdim\derdldim>-\derdim
            \advance\derdldim by\derdim           
         \else                                    
            \derdldim=0pt                         
         \fi                                      
      \else
         \advance\derdldim by\dercldim            
      \fi
      \ifdim\derrdim>0pt
         \ifdim\derdrdim>-\derdim
            \advance\derdrdim by\derdim           
         \else                                    
            \derdrdim=0pt                         
         \fi                                      
      \else
         \advance\derdrdim by\dercrdim            
      \fi
   \fi
   \global\setbox\derboxone=\hbox
      {\kern\derdldim\unhbox\derboxone\kern\derdrdim}%
   \global\advance\derldim by\derdldim            
   \global\advance\derrdim by\derdrdim            
   }%
\def\stemr #1#2#3#4{{#4}%
   \dercldim=\derldim
   \derccdim=\dercdim
   \dercrdim=\derrdim
   \stemaux{#1}{#2}{#3}}%
\def\stemrr #1#2#3#4{{#4}%
   \dercldim=\derldim
   \derccdim=\dercdim
   \dercrdim=\derrdim
   \stemauxx{#1}{#2}{#3}}%
\def\stem #1#2#3#4{\setbox\stembox=\hbox{$\|$}%
   \stemrr{  }{  }{#3              }  {
   \stemr {{\scriptstyle #1}\;}{\;{\scriptstyle #2}}{\kern\wd\stembox} {
   \stemrr{  }{  }{\kern\wd\stembox}{
   #4                             }}}}%
\def\stempr #1#2#3{\setbox\stembox=\hbox{$\|$}%
   \stemrr{  }{  }{#3              }  {
   \stemr {{\scriptstyle #1}\;}{\;{\scriptstyle #2}}{\kern\wd\stembox} {
   \stemrr{  }{  }{\kern\wd\stembox}{
      {\global\setbox\derboxone=\hbox{%
         \vbox to0pt{\vss\hbox{$-$}\vss\kern-2\deropen}}%
      \global\derldim=0pt                            
      \global\dercdim=\wd\derboxone                  
      \global\derrdim=0pt                            
      }                             }}}}%

\def\deraux {\derldim=0pt\dercdim=0pt\derrdim=0pt}%
\def\der       #1#2#3{\deraux\root  {#1}{#2}{#3}        \box\derboxone}%
\def\dernote       #1#2#3#4{\deraux\rootr  {#1\;}{\;#2}{#3}{#4}\box\derboxone}%
\def\ddernote    #1#2#3#4#5{\deraux\rrootr {#1\;}{\;#2}{#3}{#4}{#5}\box
                                                                   \derboxone}%
\def\inf       #1#2#3{\der  {#1}{#2}{\leaf{#3}}}%

\def\strpr  #1#2#3{\stempr{#1}{#2}{#3}\box\derboxone}%
\def\strder #1#2#3#4{\stem{#1}{#2}{#3}{#4}\box\derboxone}%

\newbox\derskelboxone \newbox\derskelboxtwo \newbox\derskelboxthree
\newbox\derskelboxfour \newdimen\derskeldimenone
\newdimen\derskeldimentwo \newdimen\derskeldimenthree
\newdimen\derskeldimenfour \newdimen\derskeldimenfive
\newdimen\derskeldimensix \newdimen\derskeldimenseven
\newdimen\derskeldimeneight
\def\derskel #1#2#3#4{%
   \setbox\derskelboxone=\hbox{$#1$\strut}%
   \derskeldimenone=\ht\derskelboxone
   \advance\derskeldimenone by\dp\derskelboxone
   \derskeldimentwo=\wd\derskelboxone
   \divide\derskeldimentwo by2
   \setbox\derskelboxone=\hbox to0pt{%
      \hss\raise\dp\derskelboxone\box\derskelboxone\hss}%
   \ht\derskelboxone=0pt
   \dp\derskelboxone=0pt
   \setbox\derskelboxtwo=\hbox{$#3$\strut}%
   \derskeldimenthree=\ht\derskelboxtwo
   \advance\derskeldimenthree by\dp\derskelboxtwo
   \derskeldimenfour=\wd\derskelboxtwo
   \divide\derskeldimenfour by2
   \setbox\derskelboxtwo=\hbox to0pt{%
      \hss\raise\dp\derskelboxtwo\box\derskelboxtwo\hss}%
   \ht\derskelboxtwo=0pt
   \dp\derskelboxtwo=0pt
   \ifdim\derskeldimenone>\derskeldimenthree
      \else\derskeldimenone=\derskeldimenthree\fi
   \setbox\derskelboxthree=\hbox{$#4$\strut}%
   \derskeldimenfive=\ht\derskelboxthree
   \advance\derskeldimenfive by\dp\derskelboxthree
   \derskeldimensix=\wd\derskelboxthree
   \divide\derskeldimensix by2
   \setbox\derskelboxthree=\hbox to0pt{%
      \hss\lower\ht\derskelboxthree\box\derskelboxthree\hss}%
   \ht\derskelboxthree=0pt
   \dp\derskelboxthree=0pt
   \setbox\derskelboxfour=\hbox{$#2$\strut}%
   \derskeldimenseven=\ht\derskelboxfour
   \advance\derskeldimenseven by\dp\derskelboxfour
   \derskeldimeneight=\wd\derskelboxfour
   \divide\derskeldimeneight by2
   \setbox\derskelboxfour=\hbox to0pt{%
      \hss\raise\dp\derskelboxfour\box\derskelboxfour\hss}%
   \ht\derskelboxfour=0pt
   \dp\derskelboxfour=0pt
   \ifdim\derskeldimenone>\derskeldimenseven
      \else\derskeldimenone=\derskeldimenseven\fi
   \derskeldimenthree=\derskeldimentwo
   \advance\derskeldimenthree by2\derskeldimeneight
   \advance\derskeldimenthree by\derskeldimenfour
   \advance\derskeldimenthree by2em
   \divide\derskeldimenthree by2
   \advance\derskeldimensix by-\derskeldimenthree
   \derskeldimenseven=\derskeldimensix
   \advance\derskeldimensix by-\derskeldimentwo
   \advance\derskeldimenseven by-\derskeldimenfour
   \ifdim\derskeldimensix>0pt
      \else\derskeldimensix=0pt\fi
   \ifdim\derskeldimenseven>0pt
      \else\derskeldimenseven=0pt\fi
   \vbox{\kern\derskeldimenone\hbox{\kern\derskeldimensix
         \kern\derskeldimentwo
         \xy
         <-\derskeldimenthree,\derskeldimenthree>="here"
            *{\box\derskelboxone}**\dir{-};
         "here"+<\derskeldimentwo,0pt>="here"**\dir{-};
         "here"+<1em,0pt>="here"**\dir{-};
         "here"+<\derskeldimeneight,0pt>="here"
            *{\box\derskelboxfour}**\dir{-};
         "here"+<\derskeldimeneight,0pt>="here"**\dir{-};
         "here"+<1em,0pt>="here"**\dir{-};
         "here"+<\derskeldimenfour,0pt>*{\box\derskelboxtwo}**\dir{-};
         0*{\box\derskelboxthree}**\dir{-};
         <-\derskeldimenthree,\derskeldimenthree>**\dir{-}
         \endxy
         \kern\derskeldimenfour\kern\derskeldimenseven}%
      \kern\derskeldimenfive}}%

\newbox\DerivOneBox
\newbox\DerivTwoBox
\newbox\DerivThreeBox
\newbox\DerivFourBox
\newdimen\DerivOneDimen
\newdimen\DerivTwoDimen
\newdimen\DerivThreeDimen
\newdimen\DerivFourDimen
\def\Derivationleaf #1#2#3#4#5{\global\setbox\derboxone=\hbox{\strut
                                    $\DerivationFactors{#1}{#2}{#3}{#4}{#5}11$}}%
\def\DerivationFactors #1#2#3#4#5#6#7{%
   \setbox\DerivOneBox=\hbox{$#1\strut$}%
      \DerivOneDimen=\wd\DerivOneBox\divide\DerivOneDimen by2
   \setbox\DerivThreeBox=\hbox{$#3\strut$}%
      \DerivThreeDimen=\wd\DerivThreeBox\divide\DerivThreeDimen by2
   \setbox\DerivTwoBox=\hbox{\box\DerivOneBox\hbox{$#2$}\box\DerivThreeBox}%
      \DerivTwoDimen=\wd\DerivTwoBox
   \setbox\DerivFourBox=\hbox{$#4\strut$}%
      \DerivFourDimen=\wd\DerivFourBox
   \ifdim\DerivFourDimen>\DerivTwoDimen
      \global\dercdim=\DerivFourDimen                
      \global\derldim=0pt                            
      \global\derrdim=0pt                            
      \advance\DerivFourDimen by-\DerivTwoDimen
      \divide \DerivFourDimen by2
      \advance\DerivTwoDimen  by-\DerivOneDimen
      \advance\DerivTwoDimen  by-\DerivThreeDimen
      \divide \DerivTwoDimen  by 2
   \else
      \global\dercdim=\DerivFourDimen                
      \DerivFourDimen=0pt
      \advance\DerivTwoDimen  by-\DerivOneDimen
      \advance\DerivTwoDimen  by-\DerivThreeDimen
      \global\derldim=\DerivTwoDimen
         \global\advance\derldim by-\dercdim
         \global\divide\derldim by2
         \global\advance\derldim by\DerivOneDimen    
      \global\derrdim=\DerivTwoDimen
         \global\advance\derrdim by-\dercdim
         \global\divide\derrdim by2
         \global\advance\derrdim by\DerivThreeDimen  
      \divide \DerivTwoDimen  by 2
   \fi
   \vbox{\offinterlineskip\hbox{\kern\DerivFourDimen\box\DerivTwoBox}%
         \hbox{\kern\DerivFourDimen\kern\DerivOneDimen
               \kern\DerivTwoDimen\kern-#6\DerivTwoDimen\hbox{$\xy
               0;<#6\DerivTwoDimen,0pt>:<0pt,#7\DerivTwoDimen>::
               (0,1);(2,1)**\crv{(1.25,1.1875)&(0.75,0.8125)};
               (1,0)**@{-};(0,1)**@{-};
               (1,0.625)*{\scriptstyle #5}
               \endxy$}}%
         \hbox{\kern\DerivFourDimen\kern\DerivOneDimen\kern\DerivTwoDimen
               \hbox to0pt{\hss\box\DerivFourBox\hss}%
               \kern\DerivFourDimen\kern\DerivOneDimen\kern\DerivTwoDimen}}}%

\usepackage{listings}
\begin{document}

\ufmgthesis
{
  title      	= {Aspectos computacionais do cálculo das estruturas}, 
  author     	= {Mário Sérgio Ferreira Alvim Júnior}, 
  university 	= {Universidade Federal de Minas Gerais}, 
  course     	= {Ciência da Computação}, 
  address    	= {Belo Horizonte, Minas Gerais}, 
  advisor    	= {Elaine Gouvêa Pimentel}{Ph.~D}{Departamento de Matemática / UFMG}, 
  coadvisor  	= {Roberto da Silva Bigonha}{Ph.~D}{Departamento de Ciência da Computação / UFMG},
  date        	= {2008-04-04},
  logo			= {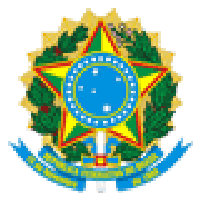},
  member		= {João Marcos}{Ph.~D}{Departamento de Informática e Matemática Aplicada / UFRN},
  member		= {Lucília Camarão de Figueiredo}{Ph.~D}{Departamento de Computação / UFOP},
  abstract		= {resumo.tex},
  englishabstract= {abstract.tex},
}


\chapter{Introdução}
\label{cap:introducao}

\section{Definição do problema}
\label{section:definicao-problema}

\subsection{Introdução}
\label{section:introducao}

A lógica é o estudo dos princípios e critéiros de inferências e demonstrações válidas.

Um sistema lógico é composto por três partes: a sintaxe (ou notação), uma especificação de regras de inferência (ou de argumentação), e o significado das sentenças lógicas (semântica). 

Em geral, para uma dada lógica, existem diversas formas de se definir o sistema de regras de um sistema lógico formal. Para as lógicas mais difundidas como a lógica clássica, a lógica intuicionista e a lógica linear, os sistemas de inferência mais utilizados são
baseados em formalismos como a dedução natural e o cálculo de seqüentes, ambos propostos por Gentzen no século XX. A dedução natural é um formalismo cujas regras possuem uma apresentação mais intuitiva, semelhante ao raciocínio humano. Entretanto, construir demonstrações utilizando dedução natural é
muito difícil, uma vez que não existe, começando da
fórmula que se deseja demonstrar e seguindo de baixo para cima\footnote{Em inglês: \emph{bottom-up}}, uma
estratégia de demonstração. 

Já o cálculo de seqüentes é um formalismo menos intuitivo, mas computacionalmente muito mais poderoso, uma vez que as regras relativas aos conectivos lógicos, quando analisadas de baixo para cima, reduzem o número de conectivos utilizados, reduzindo assim também, a cada passo, o número de opções para se seguir com a demonstração. 

Utilizando este formalismo, foi possível a formulação do conceito de demonstrações \emph{uniformes}. Intuitivamente, um sistema de demonstrações é dito uniforme se existe um método único de demonstrações para qualquer seqüente provável. Este conceito facilita enormemente a construção de demonstrações, uma vez que o não-determinismo é reduzido ao máximo, permitindo uma implementação direta do sistema. O ramo de pesquisa que se preocupa com a busca de demonstrações em sistemas lógicos é conhecido como \emph{Teoria da Demonstração}\footnote{Em inglês: \emph{Proof Theory}. Algumas traduções para o português utilizam o termo \emph{Teoria da Prova}.}.

Atualmente, existe um novo sistema formal conhecido como \emph{cálculo das estruturas}, proposto por Alessio Guglielmi em 2004, que é, sob alguns aspectos, ainda mais poderoso que o cálculo de seqüentes. 

Para algumas lógicas, ocorre que o problema de como tratar com o não-determinismo deste formalismo a fim de produzir uma interpretação computacional eficiente está ainda engatinhando. Discutir caminhos para encontrar uma estratégia de demonstração para este formalismo é o foco principal deste trabalho.

\subsection{Motivação}
\label{section:motivacao}
Desde o surgimento da dedução natural e do cálculo de seqüentes nos anos de 1930, pouco foi feito no sentido de se proporem novos formalismos de representação de regras de inferência. Apenas em 2004
foi proposto um novo sistema de inferência conhecido como \emph{cálculo das estruturas}~\cite{guglielmi07}, que não só revolucionou o conceito de demonstrações, como também permitiu a observação de uma série de simetrias nunca antes apresentadas em sistemas dedutivos. 

Uma \emph{estrutura} é uma expressão intermediária entre um seqüente representado em apenas um lado\footnote{Em inglês: \emph{one-sided sequent}} e uma fórmula. Mais precisamente, é uma fórmula lógica comum, módulo uma teoria equacional normalmente imposta aos seqüentes. Na prática, os conectivos desaparecem e as regras são \emph{estruturais} (em vez de \emph{lógicas}), no sentido de que elas lidam com a posição relativa de uma subestrutura dentro da estrutura mais geral. As estruturas são o único tipo de expressão permitido, e as regras de inferência são simplesmente regras de reescrita de estruturas. Daí o nome de \emph{cálculo das estruturas}.

O cálculo das estruturas apresenta um aspecto único que representa uma grande vantagem sobre os demais sistemas de inferência: a \emph{deep-inference}. \emph{Deep-inference} é um mecanismo que permite que, durante a construção de uma demonstração, as regras de inferência possam ser aplicadas tão profundamente quanto se queira dentro da estrutura. Isso quer dizer que, ao contrário do cálculo de seqüentes, por exemplo, não existe um lugar delimitado em cada passo onde se pode trabalhar na construção de uma demonstração (o conectivo principal). No cálculo das estruturas é possível aplicar regras de inferência em qualquer subestrutura da estrutura principal. Esta característica leva a duas conseqüências muito importantes. Em primeiro lugar, as demonstrações se tornam potencialmente menores, quando comparadas às demonstrações obtidas com o cálculo de seqüentes. Em contrapartida, devido à liberdade propiciada pela \emph{deep-inference}, o não-determinismo na construção de uma demonstração se torna muito maior, uma vez que há muitas maneiras em cada passo de se aplicarem regras de inferência.

O primeiro sistema de cálculo das estruturas proposto, e também o mais simples, é o sistema \BV~\cite{guglielmi07}, que é o fragmento multiplicativo da lógica linear estendido com a regra \mix\ e um operador \emph{auto-dual} e \emph{não comutativo}, o \emph{seq}. Um fato notável é que \BV\ não pode ser formalizado em cálculo de seqüentes~\cite{tiuBVII}. Então, a pergunta que surge é se todos os sistemas formalizados em cálculo de seqüentes podem ser formalizados em cálculo das estruturas. Para responder a esta pergunta, houve inicialmente um grande interesse em formalizar outras lógicas no cálculo das estruturas. A figura~\ref{fig:cos-lk} apresenta um sistema em cálculo das estruturas para lógica clássica, o SKSg.

\begin{figure}[!htb]
$$
\begin{array}{|c c c|}
\hline
                                              &                         & \\
\infer[i\downarrow]{[R,\dual{R}]}{S\{t\}} &                         & \infer[i\uparrow]{S\{f\}}{S(R, \dual{R})} \\															&                         & \\
                              & \infer[s]{S[(R,T),U]}{S([R,U],T)} & \\
															&                         & \\
\infer[w\downarrow]{S\{R\}}{S\{f\}}           &                         & \infer[w\uparrow]{S\{t\}}{S\{R\}} \\
															&                         & \\
\infer[c\downarrow]{S\{R\}}{S[R,R]} &															&  \infer[c\uparrow]{S(R,R)}{S\{R\}} \\
															&                         & \\
\hline																					                              
\end{array}
$$
\label{fig:cos-lk}
\caption{Sistema SKSg}
\end{figure}

Além lógica clássica~\cite{kai06}, outras lógicas já foram formalizadas utilizando o cálculo das estruturas como a lógica intuicionista~\cite{tiu06}, lógica linear~\cite{lutzll} e modal~\cite{stewart_stouppa04}. 

Não pretendemos aqui trabalhar no sentido de formalizar lógicas. O nosso interesse é em formular ao cálculo das estruturas algo que ainda não tem: um significado computacional. Alguns esforços no sentido de diminuir o não-determinismo do formalismo têm sido empreendidos mas, até hoje, nenhum resultado definitivo, como o conceito de \emph{demonstrações uniformes} para o cálculo de seqüentes, foi encontrado.

\subsection{Objetivo}
\label{section:objetivo-intro}

O objetivo principal do trabalho é:

\emph{Discutir caminhos possíveis para uma estratégia de busca por demonstrações em cálculo das estruturas adequada para a implementação computacional. Esta estratégia deve ser, de preferência, teórica e não puramente operacional.}

Ou seja, desejamos aprofundar o conhecimento existente sobre o cálculo das estruturas no sentido de contribuir para, no futuro, conseguir um resultado similar ao obtido através conceito de demonstrações uniformes para o cálculo de seqüentes.

\subsection{Contribuição}
\label{section:contribuicao}

O cálculo das estruturas é um formalismo poderoso e vem sendo estendido para diversas lógicas. Embora seu uso permita a construção de demonstrações menores (em comparação com o cálculo de seqüentes), simétricas e elegantes, seu aspecto computacional ainda não foi discutido a fundo. O não-determinismo na construção de demonstrações por este formalismo ainda é um limitador na sua utilização prática em implementações para computador. Embora algumas tentativas de diminuir este não-determinismo tenham sido empreendidas, elas ainda têm um caráter operacional, e não teórico. 

As principais contribuições deste trabalho são: \textbf{(i)} apresentar de forma sucinta o cálculo das estruturas em português, assim como uma breve compilação do estado da arte e \textbf{(ii)} propor uma tentativa de estratégia \textit{teórica} que permita a implementação computacional do cálculo das estruturas para um subconjunto do sistema \FBV\ (\emph{flat} \BV).

\section{Revisão da literatura}
\label{section:revisao_literatura}

O formalismo de cálculo das estruturas (CoS, \emph{do inglês Calculus of Structures}) foi apresentado por Alessio Guglielmi em 2004 e publicado em 2007~\cite{guglielmi07}. Neste artigo foi introduzido o sistema lógico \BV, que estende o fragmento multiplicativo da lógica linear com a regra \mix\ e um operador dual a si mesmo e não comutativo. Até então não se tinha conseguido realizar esta extensão para o cálculo de seqüentes, mas no cálculo das estruturas ela ocorreu de forma natural. O cálculo das estruturas foi obtido como uma generalização do cálculo de seqüentes de tal forma que existisse uma \emph{simetria top down de derivações}. No artigo também se discute o conceito de \emph{deep-inference} presente no CoS, que permite que as regras de inferência sejam aplicadas tão profundamente quanto se queira em uma estrutura, havendo, portanto, muito maior liberdade na construção de uma demonstração. Um resultado importante do artigo é o Teorema de \emph{Splitting}, que, em linhas gerais, enuncia que derivações complexas podem ser divididas em derivações mais simples e então o resultado dessas subderivações pode ser utilizado na composição da derivação original mais complexa. Este é um resultado importante porque mostra a existência de caminhos alternativos de demonstração para estruturas complicadas, e a quebra em demonstrações mais simples pode reduzir o espaço de busca por derivações. Apesar de ser o artigo seminal em CoS, o trabalho não chegou a abordar os aspectos computacionais do formalismo, no sentido de que não se discutiram maneiras de lidar com o não-determinismo provocado pela \emph{deep-inference}.

Este artigo causou um grande movimento na comunidade de Teoria da Demonstração ao redor do mundo, e motivou vários pesquisadores a escrever sistemas baseados em CoS para outras lógicas que não o fragmento da lógica linear coberto pelo sistema \BV. Em~\cite{kai06} foram propostos sistemas dedutivos para a lógica clássica proposicional. Como no cálculo de seqüentes, estes sistemas possuem uma regra \emph{Cut} admissível\footnote{Uma regra $\rho$ é dita \emph{adimissível} em um sistema dedutivo se $\rho$ sua ocorrência explícita pode ser retirada do sistema sem afetar a sua completude. Em outras palavras, os conjuntos de fórmulas prováveis em um sistema e no sistema sem $\rho$ explicitamente são idênticos.}. As regras \emph{Cut}, \emph{weakening} e \emph{contraction} são apresentadas em sua forma atômica.

Embora já houvesse sistemas de CoS para a lógica linear~\cite{lutzll}, para a lógica clássica~\cite{kai06}, e para várias lógicas modais~\cite{stewart_stouppa04}, ainda era um desafio a criação de um sistema assim para a lógica intuicionista. A dificuldade principal se encontrava em criar um conjunto de regras de inferência que não permitisse a demonstrabilidade do princípio do terceiro excluído\footnote{O \emph{princípio do terceiro excluído} afirma que sempre ou uma fórmula é verdadeira, ou então é falsa. Trata-se de um princípio essencialmente não-construtivo, uma vez que não exige a construção de uma demonstração para a fórmula. É um princípio característico da lógica clássica, mas que não é válido para a lógica intuicionista}. Este problema foi resolvido por Tiu em~\cite{tiu06}, onde sitemas de inferência baseados em CoS para a lógica intuicionista de primeira ordem e algumas extensões (a lógica intermediária de Dummett - LC, a lógica de Göedel e a lógica clássica) foram propostos. Os sistemas propostos utilizam o conceito de contextos positivos (lado direito de um seqüente) e negativos (lado esquerdo de um seqüente). As regras de inferência são positivas ou negativas e só podem ser aplicadas no contexto adequado (regras positivas para contextos positivos e regras negativas para contextos negativos). Esta necessidade de checagem de contexto para aplicação de regras compromete a noção pura de localidade, mas ainda a preserva no sentido de que as regras são todas atômicas. A demonstração de que a regra \emph{Cut} é admissível nestes sistemas foi feita de forma indireta, através da equivalência a outros sistemas de inferência. A grande contribuição do trabalho é o pioneirismo na aplicação de CoS para sistemas lógicos assimétricos\footnote{Aqui o termo \emph{simetria} equivale à presença de negação involutiva e da possibilidade de representação de seqüentes em um só lado.}. 

Embora o cálculo das estruturas viesse sendo empregado com sucesso na criação de sistemas de inferência para várias lógicas, o problema de seu não-determinismo conseqüente da \emph{deep-inference} ainda era uma questão pouco explorada. As demonstrações criadas em sistemas baseados em CoS eram menores que as criadas em sistemas baseados em cálculo de seqüentes, mas o espaço de buscas de demonstrações no primeiro formalismo é muito maior. Em~\cite{ozan04}, Kahramano$\breve{g}$ullari introduz uma nova técnica que reduz este não-determinismo para o sistema \BV, sem quebrar as propriedades teóricas do sistema. Esta técnica propicia um acesso mais imediato a demonstrações menores. No artigo argumenta-se que a técnica, por se basear em aspectos gerais do CoS, pode ser facilmente estendida para outros sistemas que não o \BV. Entretanto, a técnica proposta é basicamente operacional, por tratar-se da introdução de regras de inferência que checam previamente se a decisão tomada em um passo da derivação conduzirá a um estado sobre o qual não será mais possível avançar. 

Portanto, o problema \emph{teórico} de se definir com clareza o comportamento computacional do cálculo das estruturas, assim como o conceito de demonstrações uniformes fez com o cálculo de seqüentes~\cite{miller91}, continua em aberto.

\section{Sumário da dissertação}
\label{section:sumario-dissertacao}

O Capítulo \ref{cap:conceitos-fundamentais} apresenta os conceitos fundamentais de Teoria da Demonstração utilizados nesta dissertação, incluindo o conceito de lógica clássica, intuicionista e linear, cálculo de seqüentes e demonstrações uniformes (em cálculo de seqüentes). O Capítulo \ref{cap:calculo-das-estruturas} apresenta o formalismo do cálculo das estruturas, com sua definição formal, o conceito de teias de interação, o teorema de \emph{splitting} e uma estratégia já proposta na literatura para reduzir seu não-determinismo. O Capítulo \ref{cap:aspectos-computacionais-cos} contém as principais contribuições desta dissertação: uma discussão sobre as possíveis abordagens para o problema de se encontrar uma estratégia de demonstração em cálculo das estruturas, algumas definições e teoremas importantes e o resultado principal deste trabalho, que consiste numa tentativa de estratégia de demonstração para um subconjunto do sistema \FBV. O Capítulo \ref{cap:implementacao} apresenta uma implementação do sistema \FBV\ segundo a tentativa de estratégia proposta nesta dissertação. Por fim, o Capítulo \ref{cap:conclusao} revisa os conceitos apresentados e discute trabalhos futuros.

\section{Consideração relevante}
\label{section:consideracoes}

Apesar de este trabalho ser uma dissertação na área de \emph{teoria da computação}, ele possui um caráter prático bastante relevante, uma vez que o interesse da comunidade acadêmica da área está cada vez mais voltado para a implementação eficiente de sistemas dedutivos.
%
\chapter{Teoria da Demonstração: conceitos fundamentais}
\label{cap:conceitos-fundamentais}

\section{Lógica clássica e sistemas formais}
\label{section:lk}

A lógica clássica lida com a formalização e a análise de tipos de argumentação utilizados em matemática.
Desta forma, os sistemas lógicos formais utilizados para esse fim devem ser ferramentas adequadas para demonstraçãor proposições. Parte do problema com a formalização da argumentação matemática é a necessidade de se especificar de maneira precisa uma linguagem matemática formal. Linguagens naturais, tais como o português ou inglês, não servem a este propósito: elas são muito complexas, ambíguas e em constante modificação. Por outro lado, linguagens formais como (algumas) linguagens de programação, que também são rigidamente definidas, são muito mais simples e menos flexíveis que as linguagens naturais. Utilizamos, então, um \emph{sistema lógico formal} que apresente um balanceamento entre expressividade e precisão, sem, entretanto, apresentar a ambigüidade das linguagens naturais ou ser demasiado simplista como uma linguagem de programação.

Um sistema lógico formal é composto por três partes: a sintaxe (ou notação), uma especificação cuidadosa de regras de argumentação (regras de inferência), e alguma noção de como interpretar e dar um significado a sentenças (ou proposições) da linguagem adotada (semântica). Apresentaremos rapidamente os conceitos citados acima para a lógica clássica de primeira ordem:

\begin{itemize}

\item Sintaxe

A sintaxe consiste nos símbolos utilizados na representação do sistema lógico. Na sintaxe há um \emph{alfabeto}, que é um conjunto de símbolos pré-definidos sobre os quais se pode construir sentenças e uma \emph{gramática}, que define como as sentenças podem ser construídas.

A gramática para a lógica clássica de primeira ordem é:

$$ E ::= \ A \ | \ \top \ | \ \bot \ | \ (E \wedge E) \ | \ (E \vee E) \ | \ (E \cimp E) \ | \ \forall x.E | \ \exists x.E $$

que significa que uma fórmula $E$ pode ser uma proposição atômica (sem conectivos), verdadeiro, falso, a conjunção de duas fórmulas, a disjunção de duas fórmulas, uma implicação ou fórmulas quantificadas (quantificação de universalidade ou de existência).

Não acrescentamos a negação como um conectivo primitivo da lógica, mas representamos $E\cimp \bot$ por $\neg E$. De fato, ao longo do texto utilizaremos uma ou outra representação da maneira que for mais conveniente. 

\item Regras de inferência

As regras de inferência determinam como se podem obter fórmulas a partir de outras fórmulas, num processo equivalente a tirar \emph{conclusões} válidas a partir de \emph{premissas}. Existem diversas formas de se definir regras e axiomas de um sistema
lógico formal. Este trabalho tem como foco apenas sistemas apresentados em \emph{cálculo de seqüentes}
e \emph{cálculo das estruturas}.

Um exemplo de regra de inferência (em cálculo de seqüentes) é a regra \emph{Cut}:

$$\infer[Cut]{\Gamma_1,\Gamma_2\vdash\Delta_1,\Delta_2}{\Gamma_1\vdash\Delta_1,A\quad
A,\Gamma_2\vdash\Delta_2}$$ 

Basicamente, essa regra formaliza o conceito de demonstrações matemáticas utilizando lemas auxiliares. Ou
seja, se podemos demonstraçãor um lema $A$ (ou outros resultados $\Delta_1$) a partir de um conjunto de hipóteses $\Gamma_1$ e, a partir de $A$ (e possivelmente algumas outras hipóteses $\Gamma_2$) é possível demonstraçãor outro conjunto de resultados ($\Delta_2$), então podemos demonstraçãor $\Delta_1,\Delta_2$ diretamente a partir de $\Gamma_1,\Gamma_2$.

\item Semântica

A semântica de um sistema formal dá o significado dos símbolos utilizados. No caso da lógica clássica
proposicional (sem quantificadores), o significado das fórmulas pode ser trivialmente expresso através de uma \emph{tabela de verdade}. Nessa tabela, a cada símbolo é atribuído um valor (\emph{V} para verdadeiro ou \emph{F} para falso). A partir desses valores, podemos calcular os valores das fórmulas compostas. Veja a tabela~\ref{tab:tabela-verdade}.

\begin{table}[!htb]
$$ 
\begin{array}{|c|c||c|c|c|c|c|c|}
\hline
A & B & \neg A & A \wedge B & A \vee B & A \cimp B & \top & \bot \\ \hline
V & V & F      & V          & V        & V            & V    & F    \\ \hline
V & F & F      & F          & V        & V            & V    & F    \\ \hline
F & V & V      & F          & V        & F            & V    & F    \\ \hline
F & F & V      & F          & F        & V            & V    & F    \\ \hline
\end{array}
$$
\label{tab:tabela-verdade}
\caption{Tabela de verdade para a lógica clássica proposicional}
\end{table}

\end{itemize}

\subsection{Cálculo de seqüentes}
\label{section:calculo-sequentes}

O cálculo de seqüentes, introduzido nos anos 1930 por Gerhard
Gentzen~\cite{gentzen35}, consiste em um formalismo para tratar da \emph{verdade lógica}~\footnote{A verdade do conhecimento, ou verdade lógica, é a conformidade da inteligência com o que é, isto é, com o objeto.}
considerando apenas a forma da dedução.

\begin{Definition}\label{sequente}
Um \emph{seqüente} possui a forma $\Gamma\vdash\Delta$, onde $\Gamma$
e $\Delta$ são multiconjuntos finitos (possivelmente vazios) de
fórmulas. Chamamos $\Gamma$ de \emph{antecedente} e $\Delta$ de \emph{
sucedente} do seqüente.

Uma \emph{demonstração} para o seqüente $\Gamma\vdash\Delta$ é uma árvore
finita, construída utilizando as regras de inferência do sistema tal
que a raiz é $\Gamma\vdash\Delta$.
\end{Definition}

As principais características do cálculo de seqüentes são:
\begin{itemize}
\item possui apenas regras de introdução de conectivos lógicos\footnote{Outro formalismo proposto por Gentzen
é a \emph{dedução natural}, que possui regras de introdução e
eliminação de conectivos lógicos.};
\item antecedentes e sucedentes são tratados da
mesma forma e são construídos simultaneamente;
\item é tecnicamente simples: quando lidas
de baixo pra cima, as regras no cálculo de
seqüentes simplificam o processo de construção de demonstrações, com
exceção da regra de \emph{contraction} (\emph{cont}) e da regra \emph{Cut}.
\end{itemize}

As regras do cálculo de seqüentes para a lógica clássica
$\LK$ estão listadas na Figura~\ref{cs}.

\begin{figure}[!htb]
\emph{Axioma inicial e a regra \emph{Cut}}
$$\frac{}{\Seq {\Gamma,A} {\Delta,A}}\ \hbox{\emph{Initial}}
\qquad \frac{\Seq{\Gamma_1}{\Delta_1,A}\quad
      \Seq{A,\Gamma_2}{\Delta_2}}
     {\Seq{\Gamma_1,\Gamma_2}{\Delta_1,\Delta_2}}
            \  \hbox{\emph{Cut}}
$$
\emph{Regras à direita}
$$\frac{}
       {\Seq\Gamma{\top,\Delta}}
            \ \top\hbox{\emph{R}}
\qquad \frac{\Seq\Gamma{A,\Delta}\quad\Seq\Gamma{B,\Delta}}
     {\Seq\Gamma{A\wedge B,\Delta}}\ \wedge\hbox{\emph{R}}\qquad
\qquad \frac{\Seq{\Gamma,A}{B,\Delta}}
     {\Seq\Gamma{A\cimp B,\Delta}}\ \cimp\hbox{\emph{R}}
$$
$$\frac{\Seq\Gamma{A,\Delta}}
     {\Seq\Gamma{A\vee B,\Delta}}\ \vee R_1 \qquad
\frac{\Seq\Gamma{B,\Delta}}
     {\Seq\Gamma{A\vee B,\Delta}}\ \vee R_2
$$
$$\frac{\Seq\Gamma{A[x/y],\Delta}}
    {\Seq\Gamma{\forall xA,\Delta}}\ \forall\hbox{\emph{R}}\qquad
\frac{\Seq\Gamma{A[x/t],\Delta}}
    {\Seq\Gamma{\exists xA,\Delta}}\ \exists\hbox{\emph{R}}
$$
\emph{Regras \`a esquerda}
$$
\frac{}
    {\Seq{\Gamma,\bot}{\Delta}}\ \bot\hbox{\emph{L}}
\qquad \frac{\Seq{\Gamma,A}\Delta}
     {\Seq{\Gamma,A\wedge B}\Delta}\ \wedge L_1 \qquad
\frac{\Seq{\Gamma,B}\Delta}
     {\Seq{\Gamma,A\wedge B}\Delta}\ \wedge L_2
$$
$$\frac{\Seq{\Gamma,A}\Delta\quad \Seq{\Gamma,B}\Delta}
     {\Seq{\Gamma,A\vee B}\Delta}\ \vee\hbox{\emph{L}} \qquad
\qquad
\frac{\Seq{\Gamma_1}{A,\Delta_1}\quad\Seq{\Gamma_2,B}{\Delta_2}}
     {\Seq{\Gamma_1,\Gamma_2,A\cimp B}{\Delta_1,\Delta_2}}\ \cimp\hbox{\emph{L}}
$$
$$\frac{\Seq{\Gamma,A[x/t]}\Delta}
    {\Seq{\Gamma,\forall xA}\Delta}\ \forall\hbox{\emph{L}}\qquad
\frac{\Seq{\Gamma,A[x/y]}\Delta}
    {\Seq{\Gamma,\exists xA}\Delta}\ \exists\hbox{\emph{L}}
$$
\emph{Regras estruturais}
$$\frac{\Seq\Gamma\Delta}
    {\Seq{\Gamma,A}\Delta}\ \hbox{\emph{weak L}}\qquad
\frac{\Seq{\Gamma}\Delta}
    {\Seq{\Gamma}{\Delta,A}}\ \hbox{\emph{weak R}}
$$
$$
\frac{\Seq{\Gamma,A,A}\Delta}
    {\Seq{\Gamma,A}\Delta}\ \hbox{\emph{cont L}}\qquad
\frac{\Seq{\Gamma}{\Delta,A,A}}
    {\Seq{\Gamma}{\Delta,A}}\ \hbox{\emph{cont R}}
$$
\caption{Cálculo de seqüentes para a lógica clássica
$\LK$}\label{cs}
\end{figure}

As regras à direita e à esquerda são chamadas \emph{regras
lógicas}, uma vez que definem o significado dos conectivos
lógicos.

\begin{Example}
Na lógica clássica vale o \emph{princípio do
terceiro excluído}. Ou seja, a proposição
$$p\vee\neg p$$
é sempre válida. Isso significa que uma fórmula é sempre ou
verdadeira, ou falsa.

Essa afirmação é extremamente não construtiva, uma vez que
nada se pode dizer de qual das opções é valida.

A demonstração do princípio do terceiro excluído no sistema $\LK$ é
dada abaixo.
$$\infer[cont R]{\vdash p\vee\neg p}{\infer[\vee R_1]{\vdash p\vee\neg p,p\vee\neg p}
{\infer[\vee R_2]{\vdash p,p\vee\neg p}{\infer[\cimp R]{\vdash p,\neg
p} {\infer[Initial]{p\vdash\bot,p}{}}}}}$$
\end{Example}

Chamaremos de \textbf{ C}-demonstração qualquer demonstração no sistema $\LK$.

\section{Lógica intuicionista}
\label{section:lj}

Como descrito na Seção~\ref{section:lk}, o entendimento clássico de
lógica é baseado na noção de verdade. Ou seja, a veracidade
de uma afirmativa é ``absoluta'' e independente de qualquer
argumentação, crença ou ação.

Desta forma, afirmativas são ou verdadeiras ou falsas, onde falso
é a mesma coisa que não verdadeiro. De fato, é fácil demonstraçãor em
$\LK$ as seguintes equivalências: 

$$\top\equiv\bot\cimp\bot$$ 
e
$$\bot\equiv\top\cimp\bot$$,

onde $\equiv$ é o símbolo de relação de equivalência entre fórmulas.

Claro que essa abordagem de pensamento é muito intuitiva e baseada
em experiência e observação. Para um matemático preocupado
em demonstraçãor um teorema, é importante a idéia de que toda
afirmativa pode ser demonstraçãoda verdadeira se uma demonstração é apresentada
ou falsa se existe um contra-exemplo. Além disso, várias
técnicas de demonstração utilizam implicitamente o princpípio do terceiro excluído.

A lógica intuicionista abandona a idéia de verdade absoluta, e
afirmativas são consideradas válidas se, e somente se, existe uma
demonstração construtiva das mesmas.

Em cálculo de seqüentes, o sistema de demonstrações mais conhecido para a lógica intuicionista é o sistema $\LJ$, onde os seqüentes válidos possuem exatamente uma
fórmula como sucedente e as regras de \emph{weakening} e \emph{contraction} não são válidas à direita. O sistema $\LJ$ é
apresentado na Figura~\ref{LJ_proofsystem}.

\begin{figure}[!htb]
$$\frac{}{\Seq {\Gamma,A} {A}}\ \hbox{\emph{Initial}}
\qquad \frac{\Seq{\Gamma_1}{A}\quad
      \Seq{A,\Gamma_2}{C}}
     {\Seq{\Gamma_1,\Gamma_2}{C}}
            \  \hbox{\emph{Cut}}
$$
$$\frac{}
       {\Seq\Gamma{\top}}
            \ \top\hbox{\emph{R}}
\qquad \frac{\Seq\Gamma{A}\quad\Seq\Gamma{B}}
     {\Seq\Gamma{A\wedge B}}\ \wedge\hbox{\emph{R}}\qquad
\qquad \frac{\Seq{\Gamma,A}{B}}
     {\Seq\Gamma{A\cimp B}}\ \cimp\hbox{\emph{R}}
$$
$$\frac{\Seq\Gamma{A}}
     {\Seq\Gamma{A\vee B}}\ \vee R_1 \qquad
\frac{\Seq\Gamma{B}}
     {\Seq\Gamma{A\vee B}}\ \vee R_2
$$
$$\frac{\Seq\Gamma{A[x/y]}}
    {\Seq\Gamma{\forall xA}}\ \forall\hbox{\emph{R}}\qquad
\frac{\Seq\Gamma{A[x/t]}}
    {\Seq\Gamma{\exists xA}}\ \exists\hbox{\emph{R}}
$$
$$
\frac{}
    {\Seq{\Gamma,\bot}{A}}\ \bot\hbox{\emph{L}}
\qquad \frac{\Seq{\Gamma,A}C}
     {\Seq{\Gamma,A\wedge B}C}\ \wedge L_1 \qquad
\frac{\Seq{\Gamma,B}C}
     {\Seq{\Gamma,A\wedge B}C}\ \wedge L_2
$$
$$\frac{\Seq{\Gamma,A}C\quad \Seq{\Gamma,B}C}
     {\Seq{\Gamma,A\vee B}C}\ \vee\hbox{\emph{L}} \qquad
\qquad \frac{\Seq{\Gamma_1}{A}\quad\Seq{\Gamma_2,B}{C}}
     {\Seq{\Gamma_1,\Gamma_2,A\cimp B}{C}}\ \cimp\hbox{\emph{L}}
$$
$$\frac{\Seq{\Gamma,A[x/t]}C}
    {\Seq{\Gamma,\forall xA}C}\ \forall\hbox{\emph{L}}\qquad
\frac{\Seq{\Gamma,A[x/y]}C}
    {\Seq{\Gamma,\exists xA}C}\ \exists\hbox{\emph{L}}
$$
$$\frac{\Seq{\Gamma}C}
    {\Seq{\Gamma,A}C}\ \hbox{\emph{weak L}}\qquad
\frac{\Seq{\Gamma,A,A}C}
    {\Seq{\Gamma,A}C}\ \hbox{\emph{cont L}}
$$
\caption{Cálculo de seqüentes para a lógica intuicionista
$\LJ$}\label{LJ_proofsystem}
\end{figure}

\begin{Example}
Todos os seqüentes abaixo são prováveis em lógica
clássica:
\begin{enumerate}
\item $\neg{(p\vee q)}\vdash (\neg p\wedge\neg q)$
\item $(p\vee q)\vdash \neg{(\neg p\wedge \neg q)}$
\item $(p\vee q)\vdash (\neg p\cimp q)$
\item $\neg{(p\wedge q)}\vdash (\neg p\vee\neg q)$
\item $((p\cimp q)\cimp p)\vdash p$
\item $\vdash (p\cimp q)\vee (q\cimp p)$
\end{enumerate}
Mas apenas $(1),(2)$ e $(3)$ apresentam demonstrações construtivas, isto
é, são prováveis intuicionisticamente.
\end{Example}

Chamaremos de \textbf{I}-demonstração qualquer demonstração no sistema $\LJ$.

\section{Lógica linear}
\label{section:ll}

Na lógica clássica (seção ~\ref{section:lk}), matemáticos começam de um conjunto de axiomas, demonstraçãom alguns lemas e então os utilizam para demonstraçãor teoremas. Algumas das demonstrações utilizadas não são construtivas, e o uso da estratégia conhecida como \emph{redução ao absurdo} é muito comum.

Uma vez que um lema é demonstraçãodo, ele pode ser usado de novo para demonstraçãor outras proposições ou teoremas, uma vez que um lema demonstraçãodo verdadeiro será verdadeiro para sempre. Portanto, matemáticos trabalham com a lógica clássica, a lógica da \emph{verdade estável} e de \emph{recursos e conclusões infinitos}.

Já a lógica intuicionista (Seção~\ref{section:lj}) descarta essa noção de verdade absoluta e a veracidade de uma afirmativa depende da existência de uma demonstração (ou construção) da afirmativa. Alguns matemáticos preferem a lógica intuicionista à clássica, justamente por acreditarem na importância de se construir uma demonstração, em vez de se contentar em apenas saber que ela existe. A lógica intuicionista ainda é muito utilizada em ciência da computação, sendo a base de linguagens de programação como Prolog. Mas a lógica intuicionista ainda é uma lógica de \emph{infinitos recursos} -- mas não infinitas conclusões, uma vez que
permitir a demonstração de todos os resultados possíveis em lógica clássica implicaria em
permitir o princípio do terceiro excluído.

A lógica linear (desenvolvida por Girard~\cite{girard87tcs}) é
uma lógica de \emph{recursos conscientes}. É, portanto, um
refinamento das lógicas clássica e intuicionista, no sentido que
substitui a ênfase em \emph{verdade} ou {em demonstraçãobilidade} por \emph{recursos}.

Em lógica linear, afirmativas não podem ser livremente copiadas
(\emph{contraction}) ou descartadas (\emph{weakening}), apenas em
situações especiais, onde aparece um tipo muito particular de
conectivos: os exponenciais ``$?$'' e ``$!$''. Intuitivamente, $!B$
significa que o recurso $B$ pode ser usado tantas vezes quanto
necessárias. De maneira dual, $?B$ indica a possibilidade de
produção de uma quantidade infinita da conclusão $B$.

A implicação linear é representada pelo símbolo
``$\limp$'' e o significado de $A\limp B$ é:
$$\mbox{consome-se }A\mbox{ dando origem a }B$$
Isto significa que, a partir do ponto em que $B$ é produzido, o
predicado $A$ deixa de ser válido. A implicação intuicionista
``$\Rightarrow$'' então significa:
$$A\Rightarrow B\equiv \;!A\limp B$$
ou seja, um predicado $A$ implica $B$ intuicionisticamente se e
somente se existe uma quantidade infinita de $A$ que linearmente
implica $B$.

A ausência de \emph{contraction} e \emph{weakening} 
muda a natureza
dos conectivos lógicos. 
De fato, poderíamos propor uma variante das regras do sistema $\LK$, considerando as regras para a conjunção como:
$$\infer[\wedge L']{\Gamma,A\wedge B\vdash\Delta}{\Gamma,A, B\vdash\Delta}\qquad
\infer[\wedge R']{\Gamma_1,\Gamma_2\vdash A\wedge B,\Delta_1,\Delta_2}{\Gamma_1\vdash A,\Delta_1\quad
\Gamma_2\vdash B,\Delta_2}$$
As regras acima são chamadas de \emph{multiplicativas}, enquanto que as regras apresentadas na Figura~\ref{cs}
são chamadas de \emph{aditivas}. Ocorre que, na presença das regras \emph{cont} e \emph{weak}, esses dois formatos
são equivalentes. De fato, se supusermos a regra $\wedge R$, podemos derivar a regra $\wedge R'$:
$$\infer[\wedge R]{\Gamma_1,\Gamma_2\vdash A\wedge B,\Delta_1,\Delta_2}{
\infer=[weak\,R,weak\,L]{\Gamma_1,\Gamma_2\vdash A,\Delta_1,\Delta_2}{
\deduce{\Gamma_1\vdash A,\Delta_1}{}}&
\infer=[weak\,R,weak\,L]{\Gamma_1,\Gamma_2\vdash B,\Delta_1,\Delta_2}
{\deduce{\Gamma_2\vdash B,\Delta_2}{}}}$$
e vice versa:
$$\infer=[cont\,R,cont\,L]{\Gamma\vdash A\wedge B,\Delta}{
\infer[\wedge R']{\Gamma,\Gamma\vdash A\wedge B,\Delta,\Delta}
{\deduce{\Gamma\vdash A,\Delta}{}&
\deduce{\Gamma\vdash B,\Delta}{}}}$$
Se removermos \emph{contraction} e \emph{weakening}, então as regras não são mais equivalentes,
e a conjunção
(assim como a disjunção) é separada em dois conectivos
diferentes. Portanto, existem duas maneiras distintas de formular a
conjunção, correspondendo a dois conectivos distintos em
lógica linear: o conectivo multiplicativo ``$\otimes$'' ($A\otimes
B$ significa ambos $A$ e $B$) e o aditivo ``$\&$'' ($A\& B$ =
escolha entre $A$ e $B$). O mesmo para a disjunção:
multiplicativo ``$\lpar$'' ($A\lpar B$ é igual a $A$ paralelo a
$B$) e aditivo ``$\oplus$'' ($A\oplus B$ significa ou $A$ ou $B$).

A lógica linear utiliza ainda os seguintes conectivos: $\bot$, e $1$
para a versão multiplicativa de falso e verdadeiro
respectivamente; $0$, $\top$ para a versão aditiva desses
conectivos; e $\forall$ e $\exists$ para quantificações
universal e existencial.

Na Figura~\ref{LL proof system} apresentamos as regras de inferência
para a lógica linear.

\begin{figure}[!htb]
$$\frac{}{\Seq A A}\ \hbox{\emph{Initial}}
\qquad \frac{\Seq{\Gamma_1}{\Delta_1,A}\quad
      \Seq{A,\Gamma_2}{\Delta_2}}
     {\Seq{\Gamma_1,\Gamma_2}{\Delta_1,\Delta_2}}
            \  \hbox{\emph{Cut}}
$$
$$\frac{\Seq{\Gamma_1}{\Delta_1,\bang A}\quad
      \Seq{(\bang A)^{n},\Gamma_2}{\Delta_2}}
     {\Seq{\Gamma_1,\Gamma_2}{\Delta_1,\Delta_2}}
            \  \hbox{\emph{Cut!}} \qquad
\frac{\Seq{\Gamma_1}{\Delta_1,(\quest A)^n}\quad
      \Seq{\quest A,\Gamma_2}{\Delta_2}}
     {\Seq{\Gamma_1,\Gamma_2}{\Delta_1,\Delta_2}}
            \  \hbox{\emph{Cut?}}\qquad n>1
$$
\emph{Regras à direita}
$$\frac{}
       {\Seq\Gamma{\top,\Delta}}
            \ \top\hbox{\emph{R}}
\qquad \frac{\Seq\Gamma\Delta}
    {\Seq\Gamma{\bottom,\Delta}}\ \bottom\hbox{\emph{R}}
\qquad \frac{}
       {\Seq{  }\one}
            \ \one\hbox{\emph{R}}
$$
$$\frac{\Seq\Gamma{A,\Delta}\quad\Seq\Gamma{B,\Delta}}
     {\Seq\Gamma{A\with B,\Delta}}\ \with\hbox{\emph{R}}\qquad
\frac{\Seq\Gamma{A,B,\Delta}}
     {\Seq\Gamma{A\lpar B,\Delta}}\ \lpar\hbox{\emph{R}}
$$
$$\frac{\Seq\Gamma{A,\Delta}}
     {\Seq\Gamma{A\oplus B,\Delta}}\ \oplus\hbox{\emph{R}} \qquad
\frac{\Seq\Gamma{B,\Delta}}
     {\Seq\Gamma{A\oplus B,\Delta}}\ \oplus\hbox{\emph{R}}
$$
$$\frac{\Seq{\Gamma_1}{A,\Delta_1}\quad\Seq{\Gamma_2}{B,\Delta_2}}
     {\Seq{\Gamma_1,\Gamma_2}{A\otimes B,\Delta_1,\Delta_2}}\
\otimes\hbox{\emph{R}} \qquad \frac{\Seq{\Gamma,A}{B,\Delta}}
     {\Seq\Gamma{A\limp B,\Delta}}\ \limp\hbox{\emph{R}}
$$
$$\frac{\Seq\Gamma{A[x/y],\Delta}}
    {\Seq\Gamma{\lforall xA,\Delta}}\ \lforall\hbox{\emph{R}}\qquad
\frac{\Seq\Gamma{A[x/t],\Delta}}
    {\Seq\Gamma{\lexists xA,\Delta}}\ \lexists\hbox{\emph{R}}
$$
\emph{Regras à esquerda}
$$\frac{}
       {\Seq{\zero,\Gamma}\Delta}
            \ \zero\hbox{\emph{L}}
\qquad \frac{}
    {\Seq\bottom{  }}\ \bottom\hbox{\emph{L}}
\qquad \frac{\Seq\Gamma\Delta}
       {\Seq{\one,\Gamma}\Delta}
            \ \one\hbox{\emph{L}}
$$
$$\frac{\Seq{\Gamma,A}\Delta}
     {\Seq{\Gamma,A\with B}\Delta}\ \with\hbox{\emph{L}} \qquad
\frac{\Seq{\Gamma,B}\Delta}
     {\Seq{\Gamma,A\with B}\Delta}\ \with\hbox{\emph{L}}
$$
$$\frac{\Seq{\Gamma,A}\Delta\quad \Seq{\Gamma,B}\Delta}
     {\Seq{\Gamma,A\oplus B}\Delta}\ \oplus\hbox{\emph{L}} \qquad
\frac{\Seq{\Gamma_1,A}{\Delta_1}\quad \Seq{\Gamma_2,B}{\Delta_2}}
     {\Seq{\Gamma_1,\Gamma_2,A\lpar B}{\Delta_1,\Delta_2}}\ \lpar\hbox{\emph{L}}
$$
$$\frac{\Seq{\Gamma,A,B}\Delta}
     {\Seq{\Gamma,A\otimes B}\Delta}\ \otimes\hbox{\emph{L}} \qquad
\frac{\Seq{\Gamma_1}{A,\Delta_1}\quad\Seq{\Gamma_2,B}{\Delta_2}}
     {\Seq{\Gamma_1,\Gamma_2,A\limp B}{\Delta_1,\Delta_2}}\ \limp\hbox{\emph{L}}
$$
$$\frac{\Seq{\Gamma,A[x/t]}\Delta}
    {\Seq{\Gamma,\lforall xA}\Delta}\ \lforall\hbox{\emph{L}}\qquad
\frac{\Seq{\Gamma,A[x/y]}\Delta}
    {\Seq{\Gamma,\lexists xA}\Delta}\ \lexists\hbox{\emph{L}}
$$
\emph{Regras para os exponenciais}
$$\frac{\Seq\Gamma\Delta}
    {\Seq{\Gamma,\bang A}\Delta}\ \bang\hbox{\emph{W}}\qquad
\frac{\Seq{\Gamma,A}\Delta}
    {\Seq{\Gamma,\bang A}\Delta}\ \bang\hbox{\emph{D}}\qquad
\frac{\Seq{\bang\Gamma}{A,\quest\Delta}}
    {\Seq{\bang\Gamma}{\bang A,\quest\Delta}}\ \bang\hbox{\emph{R}}\qquad
\frac{\Seq{\Gamma,\bang A,\bang A}\Delta}
    {\Seq{\Gamma,\bang A}\Delta}\ \bang\hbox{\emph{C}}
$$
$$\frac{\Seq\Gamma\Delta}
    {\Seq{\Gamma}{\quest A,\Delta}}\ \quest\hbox{\emph{W}}\qquad
\frac{\Seq{\bang\Gamma,A}{\quest\Delta}}
    {\Seq{\bang\Gamma,\quest A}{\quest\Delta}}\ \quest\hbox{\emph{L}}\qquad
\frac{\Seq{\Gamma}{A,\Delta}}
    {\Seq{\Gamma}{\quest A,\Delta}}\ \quest\hbox{\emph{D}}\qquad
\frac{\Seq\Gamma{\quest A,\quest A,\Delta}}
    {\Seq\Gamma{\quest A,\Delta}}\ \quest\hbox{\emph{C}}
$$
\caption{Cálculo de seqüentes para a lógica linear clássica $\LL$}
\label{LL proof system}
\end{figure}

\subsection{$\MLL$: fragmento multiplicativo da lógica linear}\label{section:MLL}
Observe que as regras da Figura~\ref{LL proof system} apresentam uma simetria
fantástica. De fato, a regra à direita para o conectivo $\lpar$ possui o
mesmo formato da regra à esquerda para o conectivo $\otimes$. O mesmo ocorre com os conectivos
$\with$ e $\oplus$. Este fato, juntamente com 
o fato de que, na lógica linear, a negação é involutiva\footnote{A negação é dita \emph{involutiva} se $\neg{\neg{A}}\equiv A$.} permitem a representação da lógica linear através de seqüentes que possuem apenas o sucedente. Desta forma, os conectivos possuem apenas regras à direita. 

Na Figura~\ref{MLL_proof_system} apresentamos as regras utilizando o cálculo de seqüentes
de um lado só\footnote{Em inglês: \emph{one sided sequent calculus}} para \MLL, que é o fragmento da lógica linear
contendo apenas os conectivos multiplicativos.

Tal sistema será importante para a definição do sistema $\BV$, no Capítulo\ref{section:cos}.

\begin{figure}[!htb]
$$\frac{}{\vdash A, \neg{A}}\ \hbox{\emph{Initial}}
\qquad \frac{\vdash\Gamma_1,A\quad
      \vdash \neg{A},\Gamma_2}
     {\vdash\Gamma_1,\Gamma_2}
            \  \hbox{\emph{Cut}}
$$
$$
\frac{\vdash\Gamma,A,B}
     {\vdash\Gamma,A\lpar B}\ \lpar
\qquad\frac{\vdash\Gamma_1,A\quad\vdash\Gamma_2,B}
     {\vdash\Gamma_1,\Gamma_2,A\otimes B}\
\otimes
$$
\caption{Cálculo de seqüentes de um lado só para a lógica linear multiplicativa $\MLL$}
\label{MLL_proof_system}
\end{figure}

\section{Regra $\mix$}

Outra regra que vai ser de fundamental importância na Seção~\ref{section:sistema-bv} é a regra
$\mix$:
$$\infer[\mix]{\vdash\Gamma,\Delta}{\vdash\Gamma\quad\vdash\Delta}$$

Tal regra se parece com a regra $Cut$, a diferença está no fato de que nenhuma fórmula é eliminada, elas são na verdade todas ``misturadas''. A presença da regra $\mix$ muda completamente a demonstraçãobilidade de um
sistema lógico. Por exemplo, em $\MLL+\mix$ é possível demonstraçãor o seqüente $\vdash A\lpar B,\neg{A}\lpar\neg{B}$ (que em cálculo de seqüentes de dois lados corresponde a $A\otimes B\vdash A\lpar B$):
$$\infer[\lpar]{\vdash A\lpar B,\neg{A}\lpar\neg{B}}{
\infer[\lpar]{\vdash A,B,\neg{A}\lpar\neg{B}}{
\infer[\mix]{\vdash A, B,\neg{A},\neg{B}}{
\infer[Initial]{\vdash A,\neg{A}}{}&
\infer[Initial]{\vdash B,\neg{B}}{}}}}$$
Observe que, em lógica linear, a demonstração poderia ser construída de baixo para cima apenas até o seqüente $\vdash A, B,\neg{A},\neg{B}$. Este seqüente não é provável, pois não há nenhuma regra de inferência que possa ser aplicada, e, além disso, o axioma inical não pode ser empregado (em lógica linear o axioma inicial é \emph{relevante}, no sentido de que só admite uma única fórmula e sua negação). Porém, com a regra \mix, podem-se separar as fórmulas de modo que o axioma inicial se aplique, e o seqüente passa a ser provável.

Em resumo, a regra \mix\ estende o conjunto de seqüentes prováveis em \MLL\ , uma vez que permite a simulação de um axioma inical do tipo:

$$\infer[Initial']{\Seq{}{A_{1},\neg{A_{1},\hdots,A_{h},\neg{A_{n}}}}}{}$$

\section{Demonstrações uniformes}
\label{section:demonstracoes-uniformes}

Considere o seqüente $A\vee B\vdash B\vee A$. Tal seqüente é sabidamente demonstrável em lógica clássica e em lógica intuicionista, por exemplo, uma vez que ele corresponde à demonstração de que a disjunção é comutativa.

A princípio, poderia-se pensar que seria uma tarefa simples demonstraçãor
tal seqüente. Entretanto, observe que se decidirmos começar a demonstração
eliminando a disjunção da direita

$$\infer[\vee R_1]{A\vee B\vdash B\vee A}{A\vee B\vdash B}\qquad
\mbox{ou}\qquad\infer[\vee R_2]{A\vee B\vdash B\vee A}{A\vee
B\vdash A}$$ temos necessariamente que fazer primeiro a escolha de qual fórmula
demonstraçãor $(A$ ou $B)$, para depois proceder com a demonstração.

Entretanto, os seqüentes das premissas de ambas derivações não são prováveis: a única regra aplicável é $\vee L$, $weak \ L$  ou $cont \ L$, e nenhuma delas conduz a uma demonstração. Desta
forma, devemos obrigatoriamente \emph{adiar} a escolha, o que
significa, na lógica intuicionista, que devemos começar a redução
pela disjunção da esquerda.
$$\infer[\vee L]{A\vee B\vdash B\vee A}{
\infer[\vee R_2]{A\vdash B\vee A}{\infer[Initial]{A\vdash A}{}}&
\infer[\vee R_1]{B\vdash B\vee A}{\infer[Initial]{B\vdash
B}{}}}$$

Essa é uma situação que se deseja evitar na prática, se estamos preocupados com a
automatização completa de sistemas lógicos: a implementação teria
que tentar \emph{todas} as possibilidades até decidir o caminho
certo, o que tornaria o tempo de execução exponencial, no pior caso. Desta forma,
é importante saber quando um sistema lógico possui uma \emph{estratégia de redução}. Ou seja, um caminho único e bem determinado
para se demonstraçãor um seqüente qualquer.

Ainda analisando o seqüente $A\vee B\vdash B\vee A$, vemos que, no sistema $\LK$,
existe ainda outra possibilidade de demonstração, duplicando a fórmula da direita:
$$\infer[cont\, R]{A\vee B\vdash B\vee A}{
\infer[\vee R_1]{A\vee B\vdash B\vee A,B\vee A}{\infer[\vee
R_2]{A\vee B\vdash B,B\vee A}{ \infer[\vee L]{A\vee B\vdash B,A}{
\infer[Initial]{A\vdash B,A}{}&\infer[Initial]{B\vdash
B,A}{}}}}}$$ Tal derivação segue a estratégia de começar a demonstração
pelo lado direito do seqüente, mas não pode ser chamada de uma
estratégia \emph{de redução}. Isto porque, ao duplicar a fórmula
$B\vee A$, estamos na verdade aumentando o número de fórmulas no
seqüente, em vez de diminuir.

De fato, quando derivações são analisadas de baixo para cima, as
regras estruturais de contração da Figura~\ref{cs} podem sempre ser
aplicadas, duplicando fórmulas e, consequentemente, aplicação de
regras de inferência. Outra regra que pode sempre ser aplicada é a
regra \emph{Cut}, que ``cria'' uma fórmula na derivação. Este é um
problema extremamente sério, uma vez que o processo de criação
requer a presença de inteligência, que um programa de computador não
possui. Um problema semelhante ocorre com as regras $\forall L$ e
$\exists R$, pois a aplicação de tais regras requer que se saiba, \emph{a
priori}, o termo $t$ que deve ser usado como instância.

Claro que tais dificuldades podem ser superadas parcialmente. Por
exemplo, as regras estruturais podem vir implícitas em algumas
regras (dando origem ao sistema $\Gtc$ -- veja a Figura~\ref{G3c}),
a regra \emph{Cut} pode ser eliminada~\cite{vonPlato01} nas lógicas
clássica, intuicionista e linear e o problema com a aplicação de
regras para os quantificadores pode ser parcialmente resolvido
utilizando unificação~\cite{miller83}.

\begin{figure}[!htb]
$$\infer[Initial]{\Seq {\Gamma,A} {\Delta,A}}{}
$$
$$\infer[\top\; R]
       {\Seq\Gamma{\top,\Delta}}{}
\qquad \infer[\wedge R]{\Seq{\Gamma}{A\wedge B,\Delta}}
{\Seq{\Gamma}{A,\Delta}\quad\Seq{\Gamma}{B,\Delta}}
$$
$$
\infer[\cimp R]{\Seq\Gamma{A\cimp B,\Delta}}{\Seq{\Gamma,A}{B,\Delta}}
\qquad\infer[\vee R]{\Seq\Gamma{A\vee B,\Delta}}{\Seq\Gamma{A,B,\Delta}}
$$
$$\infer[\forall R]{\Seq\Gamma{\forall xA,\Delta}}{\Seq\Gamma{A[x/y],\Delta}}
\qquad
\infer[\exists R]{\Seq\Gamma{\exists xA,\Delta}}{\Seq\Gamma{\exists xA, A[x/t],\Delta}}
$$

$$
\infer[\bot\; L]
    {\Seq{\Gamma,\bot}{\Delta}}{}
$$
$$
\infer[\wedge L]{\Seq{\Gamma,A\wedge B}\Delta}{\Seq{\Gamma,A,B}\Delta}
$$
$$\infer[\vee L]{\Seq{\Gamma,A\vee B}{\Delta}}{\Seq{\Gamma,A}{\Delta}\quad \Seq{\Gamma,B}{\Delta}}
\qquad
\infer [\cimp L]{\Seq{\Gamma,A\cimp B}{\Delta}}{\Seq{\Gamma}{A,\Delta}\quad\Seq{\Gamma,B}{\Delta}}
$$
$$\infer[\forall L] {\Seq{\Gamma,\forall xA}\Delta}{\Seq{\Gamma,A[x/t], \forall xA}\Delta}
\qquad
\infer[\exists L]{\Seq{\Gamma,\exists xA}\Delta}{\Seq{\Gamma,A[x/y]}\Delta}
$$
\caption{Sistema $\Gtc$ para a lógica clássica de primeira
ordem}\label{G3c}
\end{figure} %

Substituir o sistema $\LK$ pelo sistema $\Gtc$ pode resolver alguns
problemas relativos à implementação da lógica clássica, mas introduz
outros.

\begin{Example}
O seqüente $p(a)\vee p(b)\vdash \exists x.p(x)$ é provável em lógica
clássica. No sistema $\LK$, uma possibilidade de demonstração é começando
pelo antecedente:
$$
\infer[\vee L]{p(a)\vee p(b)\vdash \exists x.p(x)}{
\infer[\exists R]{p(a)\vdash \exists x.p(x)}{
\infer[Initial]{p(a)\vdash p(a)}{}}& \infer[(\exists
R)]{p(b)\vdash \exists x.p(x)} {\infer[Initial]{p(b)\vdash
p(b)}{}}}
$$

Claro que podemos sempre começar pela contração da fórmula
quantificada, desta forma começando pelo sucedente. Em $\Gtc$, a
demonstração correspondente seria:
$$
\infer[\exists R]{p(a)\vee p(b)\vdash \exists x.p(x)}{
\infer[\exists R]{p(a)\vee p(b)\vdash \exists x.p(x), p(a)}{
\infer[\vee L]{p(a)\vee p(b)\vdash \exists x.p(x), p(a), p(b)}{
\infer[Initial]{p(a)\vdash \exists x.p(x), p(a), p(b)}{}&
\infer[Initial]{p(b)\vdash \exists x.p(x), p(a), p(b)}{}}}}
$$
O problema que surge é que a aplicação de $\exists R$ gera a
duplicação da fórmula $\exists x.p(x)$ o que, na prática, cria um
problema de implementação. Como a regra pode ser aplicada
indefinidamente, torna-se mais complicado controlar o seu uso, evitando que o programa entre em \emph{loop}.
\end{Example}

Outros problemas podem surgir com a procura por demonstrações em seqüentes
que contenham mais de uma fórmula no sucedente. O principal, no caso da lógica clássica, é o de \emph{scope
extrusion}. Para entender o que significa este conceito, considere o
seqüente $\Gamma\vdash D\cimp G$. Tal seqüente é provável se e
somente se $\Gamma,D\vdash G$ é provável\footnote{A regra ($\cimp
R$) é \emph{inversível}, veja~\cite{vonPlato01}.}. Fazendo a relação
entre lógica e programação, a fórmula $D$ pode ser vista como uma
unidade de programa que é adicionada ao programa $\Gamma$ durante a
computação~\footnote{Ou seja, \emph{modularização}.}. Para garantir
que tal procedimento irá obedecer a noção correta de \emph{escopo}
(ou seja, que as ações em $D$ terão reflexo apenas sobre $G$),
precisamos de um cálculo que permita apenas uma fórmula no
sucedente.

De fato, se $G=G_1\vee(D\cimp G_2)$, uma redução no sistema $\LJ$
começando pelo sucedente resultaria em uma busca por demonstrações de um
dos seqüentes: $\Gamma\vdash G_1$ ou $\Gamma,D\vdash G_2$. Em
particular, a fórmula $D$ está disponível \emph{apenas} para demonstraçãor a
fórmula $G_2$. Entretanto, a fórmula $G_1\vee(D\cimp G_2)$ é
classicamente equivalente a $(D\cimp G_1)\vee G_2$ e $D\cimp
(G_1\vee G_2)$. Em particular, $p\vee (p\cimp q)$ não possui uma
\textbf{ I}-demonstração, mas possui uma \textbf{ C}-demonstração. De fato, observe que na
demonstração abaixo o ``módulo'' (a fórmula $p$) é utilizado para demonstraçãor um programa que
não está no seu escopo:
$$\infer[\vee R]{\vdash p\vee(p\cimp q)}{
\infer[\cimp R]{\vdash p,(p\cimp q)}{ \infer[(Initial)]{p\vdash
p,q}{}}}$$

Por este motivo, para determinar uma estratégia de redução bem
definida, que suporte esta noção de programação modular com uma
disciplina correta para escopo, é necessário, no presente momento,
restringir as demonstrações para \textbf{ I}-demonstrações. Iremos generalizar tal
conceito para o caso da lógica linear, em~\ref{section:demonstrações-uniformes-ll}.

Em~\cite{miller91} foi apresentada uma fundamentação teórica para
caracterizar lógicas como linguagens abstratas de programação. Tal
fundamentação é baseada no conceito de \emph{demonstrações uniformes},
restrito ao caso de \textbf{ I}-demonstrações.

\begin{Definition}\label{uniformeI}
Uma \textbf{ I}-demonstração livre da regra \emph{Cut}\footnote{Em inglês: \emph{cut-free proof}} $\Xi$ é \emph{uniforme} se para
toda subdemonstração $\Xi'$ de $\Xi$ e para toda ocorrência de uma fórmula
não atômica $B$ no lado direito de um sequente em $\Xi'$, existe uma
demonstração $\Xi''$ que é igual a $\Xi'$ a menos de permutação de regras
de inferência e tal que a última regra de inferência de $\Xi''$
introduz o conectivo principal de $B$.
\end{Definition}

Em outras palavras: 

\begin{center}
\emph{uma demonstração uniforme pode ser construída utilizando um algoritmo determinístico}.
\end{center}

Desta forma, a estratégia de redução é baseada em \emph{busca
direcionada por objetivo}\footnote{Em inglês, \emph{goal-directed
search}.}. Ou seja, o seqüente $\mathcal{P}\vdash G$ denota um 
\emph{estado} de um interpretador, onde $\mathcal {P}$ denota um programa
e $G$ o objetivo que queremos demonstraçãor a partir de $\mathcal {P}$. Além
disso, uma demonstração de tal seqüente representa uma seqüência de
transições de estado determinadas pelo interpretador, que deve
necessariamente reduzir o conectivo principal da fórmula $G$, caso
exista.

Com esta estratégia, podemos definir \emph{linguagens lógicas de
programação}:

\begin{Definition}
Uma lógica com um sistema de demonstrações baseado em cálculo de seqüentes
é uma \emph{linguagem lógica de programação abstrata} se, restrita a demonstrações
uniformes, não perde a completude~\footnote{Neste contexo, entende-se que não perder a \emph{completude} signifca que o conjunto de fórmulas prováveis continua sendo exatamente o mesmo}.
\end{Definition}

Pelos exemplos apresentados anteriormente, vimos que as lógicas
clássica e intuicionista não são linguagens lógicas de programação, uma vez que há seqüentes que são prováveis, mas não há uma demonstração uniforme para eles, como $\Seq{A \vee B}{B \ vee A}$.

Para obter completude de linguagens lógicas de programação, devemos
ou restingir a gramática e as fórmulas que podem ser inseridas nos
antecedentes e sucedentes, ou escolher a lógica base com muito
cuidado. No que se segue, faremos ambos.

\subsection{Cláusulas de Horn e fórmulas de Harrop}
\label{section:horn-harrop}

Comecemos por limitar a gramática da lógica de primeira ordem. Pelos exemplos que
vimos anteriormente, fica claro que a disjunção e o quantificador
existencial, quando presentes no sucedente, trazem sérios problemas
para uniformidade.

Baseado nessa observação nasceu o Prolog, baseado em \emph{Cláusulas
de Horn}:

\begin{center}
$\begin{array}[c]{lcl} G&::=&A|(G\wedge G)\\
D&::=&A|G\cimp A|\forall x. D
\end{array}$
\end{center}
onde $G$ é o objetivo e $D$ é o programa. Objetivos (sucedentes) são conjunções de fórmulas atômicas e
programas (antecedentes) possuem a forma
$$\forall x_1\ldots x_m[A_1\wedge\ldots\wedge A_n\cimp A_0]$$
com $m,n\geq 0$.

Pode-se observar que, restrito às cláusulas de Horn, \textbf{ I}-demonstrações e \textbf{ C}-demonstrações coincidem.

Entretanto, as cláusulas de Horn dão origem a um sistema lógico muito fraco, que não comporta, por exemplo, modularização ou mecanismos de abstração de dados.

Com esta motivação, em \cite{miller91} as Cláusulas de Horn foram
generalizadas para \emph{fórmulas de Harrop}\footnote{Em inglês, \emph{hereditary Harrop formulas}.}:

\begin{center}
$\begin{array}[c]{lcl} G&::=&A|(G\wedge G)|D\cimp G|\forall x.G\\
D&::=&A|G\cimp A|\forall x. D
\end{array}$
\end{center}

Nasceu assim a linguagem de programação $\lambda$-Prolog~\cite{miller87lp}.

Não há meio de estender mais a gramática de modo a obter um
subconjunto da lógica de primeira ordem que seja uniforme e
completo. Ou seja, a gramática acima é \emph{maximal}. Desta forma, a
única possibilidade de obtermos maior expressividade, ou seja, a demonstrabilidade de fórmulas mais interessantes, é mudando o sistema lógico.

\subsection{demonstrações uniformes em lógica linear}
\label{section:demonstrações-uniformes-ll}

É fácil ver que a lógica linear clássica, com regras de inferência apresentadas no sistema $\LL$, não é, segundo a Definição~\ref{uniformeI}, uma linguagem lógica de programação. Por exemplo, os seqüentes
$$\begin{array}{rcl}
a\otimes b&\vdash& b\otimes a\\
\bang a&\vdash& \bang a\otimes\bang a\\
\bang a\with b&\vdash& \bang a\\
b\otimes (b\limp \bang a)&\vdash& \bang a\\
\one&\vdash&\one
\end{array}$$
são prováveis em $\LL$ mas não possuem demonstrações uniformes\footnote{O
problema é que as regras ($\one R$), ($\otimes R$) e ($\bang R$) não
permutam com regras à esquerda.}.

Uma alternativa seria restringir a lógica linear de modo a não
permitir os conectivos $\oplus$, $\bang$ e $\one$ no sucedente,
restrição parecida com a que foi feita na lógica de primeira ordem.
Mas isso resultaria em uma lógica muito restrita, e a linguagem que
obteríamos não seria tão mais interessante que $\lambda$-Prolog.

Ocorre que não há necessidade de restringir a lógica linear pois
ela é, completa~\footnote{Ou seja, não estamos lidando com fragmentos, mas com a lógica inteira.}, uma linguagem lógica de programação. Basta que tomemos o cuidado de apresentar suas regras de inferênciamaneira adequada a este objetivo.

De fato, os conectivos da lógica linear podem ser classificados em
\emph{síncronos} e \emph{assíncronos}~\cite{andreoli92jlc},
dependendo de se a regra de introdução à direita para aquele
conectivo depende ou não do seu contexto. O dual De Morgan de um
conectivo em uma dessas classes é um conectivo na outra classe.

Dada essa divisão de conectivos, Miller propôs
em~\cite{miller96tcs} a presentação \emph{Forum} de lógica
linear na qual fórmulas são construídas utilizando apenas os
conectivos assíncronos, a saber: $\quest$, $\lpar$, $\bottom$,
$\with$, $\top$, $\limp$ e $\forall$, junto com a versão
intuicionista da implicação $B\cimp C$.\footnote{Utilizamos
aqui o símbolo $\cimp$ ao invés de $\iimp$ para a implicação
intuicionista para seguir a notação em~\cite{miller96tcs}.} Os
conectivos síncronos da lógica linear estão disponíveis
implicitamente, uma vez que conectivos que aparecem no antecedente do seqüente de cada regra
comportam-se de modo síncrono. Como a negação em lógica linear é
involutiva, podemos usar este fato para simluar os conectivos sícronos, o que faz de Forum um sistema completo para a lógica linear.

O fato impressionante é que Forum é uma linguagem lógica de
programação, uma vez que a busca por demonstrações em seqüentes com
conectivos assíncronos no sucedente corresponde à busca \emph{dirigida por objetivo}, ao mesmo tempo que conectivos assíncronos no
antecedente correspondem ao procedimento de \emph{backchaining} sobre
cláusulas de programas~\cite{miller91}.

O sistema de demonstrações de Forum é apresentado na Figura ~\ref{forum proof
system}.
\begin{figure}[!htb]
$$\frac{}{\Four\Sigma\Psi\Delta{\top,\Gamma}\Upsilon}\ \top\hbox{\emph{R}}
$$
$$
\frac{\Four\Sigma\Psi\Delta{B,\Gamma}\Upsilon\quad
      \Four\Sigma\Psi\Delta{C,\Gamma}\Upsilon}
     {\Four\Sigma\Psi\Delta{B\with C,\Gamma}\Upsilon}
            \  \with\hbox{\emph{R}}$$
$$\frac{\Four\Sigma\Psi\Delta{\Gamma}\Upsilon}
       {\Four\Sigma\Psi\Delta{\bot,\Gamma}\Upsilon}
            \ \bot\hbox{\emph{R}}
\qquad \frac{\Four\Sigma\Psi\Delta{B,C,\Gamma}\Upsilon}
     {\Four\Sigma\Psi\Delta{B\lpar C,\Gamma}\Upsilon}\ \lpar\hbox{\emph{R}}
$$
$$\frac{\Four\Sigma\Psi{B,\Delta}{C,\Gamma}\Upsilon}
       {\Four\Sigma\Psi\Delta{B\limp C,\Gamma}\Upsilon}
   \ \limp\hbox{\emph{R}}
\qquad \frac{\Four\Sigma{B,\Psi}{\Delta}{C,\Gamma}\Upsilon}
       {\Four\Sigma\Psi\Delta{B\cimp C,\Gamma}\Upsilon}
   \ \cimp\hbox{\emph{R}} $$
$$\frac{\Four{y\colon\tau,\Sigma}\Psi\Delta{B[y/x],\Gamma}\Upsilon}
       {\Four\Sigma\Psi\Delta{\forall_\tau x.B,\Gamma}\Upsilon}
   \ \forall\hbox{\emph{R}}
\qquad
  \frac{\Four{\Sigma}\Psi\Delta{\Gamma}{B,\Upsilon}}
       {\Four\Sigma\Psi\Delta{\quest B,\Gamma}\Upsilon}
   \ \quest\hbox{\emph{R}}
$$
$$
\frac{\Five\Sigma{B,\Psi}\Delta B\Ascr\Upsilon}
       {\Four\Sigma{B,\Psi}{\Delta}\Ascr\Upsilon}
   \ decide\bang
\qquad \frac{\Four\Sigma{\Psi}\Delta{\Ascr,B}{B,\Upsilon}}
       {\Four\Sigma{\Psi}{\Delta}{\Ascr}{B,\Upsilon}}
   \ decide\quest$$
$$
\frac{\Five\Sigma\Psi\Delta
B\Ascr\Upsilon}{\Four\Sigma\Psi{B,\Delta}\Ascr\Upsilon}
   \ decide$$
$$\frac{}{\Five\Sigma\Psi{\cdot} A{A}\Upsilon}\ \hbox{\emph{initial}}
\quad
  \frac{}{\Five\Sigma\Psi{\cdot} A{\cdot}{A,\Upsilon}}\ \hbox{\emph{initial}}\quest
$$
$$\frac{}{\Five\Sigma\Psi{\cdot}\bot{\cdot}\Upsilon}\ \bot\hbox{\emph{L}}
\qquad \frac{\Five\Sigma\Psi\Delta{B_i}\Ascr\Upsilon}
     {\Five\Sigma\Psi\Delta{B_1\with B_2}\Ascr\Upsilon}
   \ \with\hbox{\emph{L}}_i
\qquad \frac{\Four\Sigma{\Psi}{B}{\cdot}{\Upsilon}}
     {\Five\Sigma{\Psi}{\cdot}{\quest B}{\cdot}{\Upsilon}}
   \ \quest\hbox{L}
$$
$$\frac{\Five\Sigma\Psi{\Delta_1} B{\Ascr_1}\Upsilon\quad
        \Five\Sigma\Psi{\Delta_2} C{\Ascr_2}\Upsilon}
       {\Five\Sigma\Psi{\Delta_1,\Delta_2}{B\lpar C}{\Ascr_1,\Ascr_2}\Upsilon}
   \ \lpar\hbox{\emph{L}}
\qquad
  \frac{\Five\Sigma\Psi{\Delta}{B[t/x]}{\Ascr}\Upsilon}
     {\Five\Sigma\Psi{\Delta}{\forall_\tau x.B}{\Ascr}\Upsilon}
   \ \forall\hbox{\emph{L}}$$
$$\frac{\Four\Sigma\Psi{\Delta_1}{\Ascr_1,B}\Upsilon\quad
        \Five\Sigma\Psi{\Delta_2} C{\Ascr_2}\Upsilon}
       {\Five\Sigma\Psi{\Delta_1,\Delta_2}{B\limp C}{\Ascr_1,\Ascr_2}\Upsilon}
   \ \limp\hbox{\emph{L}}$$
$$\frac{\Four\Sigma\Psi{\cdot}{B}\Upsilon\quad
        \Five\Sigma\Psi{\Delta} C{\Ascr}\Upsilon}
       {\Five\Sigma\Psi{\Delta}{B\cimp C}{\Ascr}\Upsilon}
   \ \cimp\hbox{\emph{L}}$$
\caption{Sistema de demonstrações de Forum} \label{forum proof system}
\end{figure}

Seqüentes em Forum possuem uma das formas
$\Four\Sigma\Psi\Delta\Gamma\Upsilon\hbox{\quad e \quad}
  \Five\Sigma\Psi\Delta B\Gamma\Upsilon,$
onde $\Sigma$ é uma assinatura, $\Delta$ e $\Gamma$ são
multiconjuntos de fórmulas,  $\Psi$ e $\Upsilon$ são conjuntos
de fórmulas, e $B$ é uma fórmula.  Todas as fórmulas nos
seqüentes são compostas dos conectivos assíncronos listados
anteriormente (juntamente com $\cimp$). Os significados de tais
seqüentes em lógica linear são
$\Seq{\bang\Psi,\Delta}{\Gamma,\quest\Upsilon}\hbox{\quad e \quad}
  \Seq{\bang\Psi,\Delta, B}{\Gamma,\quest\Upsilon},$
respectivamente.

No sistema de demonstrações da Figura~\ref{forum proof system}, as regras
à direita atuam apenas sobre seqüentes da forma
$\Four\Sigma\Psi\Delta\Gamma\Upsilon$. A variável sintática
$\Ascr$ na Figura~\ref{forum proof system} denota um multiconjunto
de fórmulas atômicas. Regras à esquerda são aplicadas apenas
à fórmula $B$, que é o label de $\Five\Sigma\Psi\Delta
B\Ascr\Upsilon$.

O conceito de demonstrações uniformes foi então generalizado
em~\cite{miller96tcs} para seqüentes com mais de uma fórmula no
sucedente.

\begin{Definition}\label{uniformeLL}
Uma demonstração livre da regra \emph{Cut} $\Xi$ é \emph{uniforme} se para toda
subdemonstração $\Xi'$ de $\Xi$ e para toda ocorrência  de uma fórmula não
atômica $B$ no lado direito de um sequente em $\Xi'$, existe uma
demonstração $\Xi''$que é igual a $\Xi'$ a menos de uma permutação de regras de
inferência e tal que a última regra de inferência de $\Xi''$
introduz o conectivo principal de $B$.
\end{Definition}

A busca por demonstrações em Forum é uniforme. De fato, podemos sempre
começar pelo sucedente, passando ao antecedente somente quando
temos apenas átomos do lado direito do seqüente.

\chapter{Cálculo das estruturas}
\label{cap:calculo-das-estruturas}

\section{O Cálculo das estruturas}
\label{section:cos}

\subsection{Visão geral}
\label{section:cos-visao-geral}

O cálculo das estruturas é um formalismo proposto por Alessio Guglielmi~\cite{guglielmi07} que generaliza o cálculo de seqüentes de tal forma que uma nova simetria vertical (\emph{top-down}) é observada nas derivações. Além disso, este formalismo emprega regras de reescrita que podem ser aplicadas dentro da estrutura em qualquer profundidade. 

Uma das motivações para a invenção do cálculo das estruturas foi a dificuldade em se expressar, usando o cálculo de seqüentes, lógicas com operadores auto-duais não comutativos. Estes operadores naturalmente geram uma classe de fórmulas cujas demonstraçãos dependem de um acesso a subfórmulas em profundidades arbitrárias, que é uma característica que o cálculo de seqüentes não suporta. O cálculo das estruturas supre essa necessidade, sendo mais geral que o cálculo de seqüentes para lógicas com negação involutiva, como as lógicas clássica e linear, sem sacrificar com isso a simplicidade.

O cálculo das estruturas é uma generalização do cálculo de seqüentes de um lado só. Muitas lógicas com negação involutiva e leis de De Morgan podem ser definidas no cálculo de seqüentes de um lado só, e a tradução delas para cálculo das estruturas é uma tarefa direta e trivial. O que torna o cálculo das estruturas mais atraente é a possibilidade de definir lógicas utilizando conceitos fundamentalmente diferentes daqueles conceitos utilizados no cálculo de seqüentes. Dois conceitos centrais desse formalismo são:

\begin{itemize}

\item \emph{Deep inference}: regras de inferência no cálculo das estruturas podem operar em qualquer lugar dentro das expressões, e não apenas na subfórmula mais externa em torno da raiz das árvores de fórmulas.

\item \emph{Simetria vertical}: ao contrário do cálculo de seqüentes, da dedução natural e outros formalismos em que as derivações são essencialmente baseadas em árvores, no cálculo das estruturas as derivações podem ser invertidas de cima para baixo e negadas, permanecendo ainda válidas.

\end{itemize}

A implementação dessas idéias é possível utilizando o conceito de "estrutura". Uma \emph{estrutura} é uma expressão intermediária entre um seqüente de um lado só e uma fórmula. Mais precisamente, é uma fórmula lógica comum módulo uma teoria de equações do tipo tipicamente imposta a seqüentes. De um ponto de vista prático, conectivos lógicos desaparecem (em particular conectivos na raiz das árvores de fórmulas) e as regras lógicas se tornam \emph{estruturais} (em oposição a regras \emph{lógicas}), no sentido de que elas lidam com a posição relativa de subestruturas dentro de uma estrutura. Estruturas são a única forma de expressão permitida, e regras de inferência são simplesmente regras de reescrita sobre estruturas, de onde vem o nome "cálculo das estruturas".

O cálculo das estruturas é um formalismo adequado à Teoria da demonstração. Pode-se demonstrar propriedades importantes como uma regra Cut e, assim como no cálculo de seqüentes, a eliminação da regra Cut faz sentido. Além disso, há uma propriedade análoga à propriedade da subfórmula, que garante que qualquer regra dada (exceto pela regra Cut e excepcionalmente outras regras especiais) tem aplicação finita. Com essas características, o cálculo das estruturas está mais próximo do cálculo de seqüentes do que de qualquer outro formalismo. Entretanto, a eliminação da regra Cut nos dois formalismos é bastante distinta, ao menos quando se utiliza a característica de \emph{deep inference}. No cálculo de seqüentes, a eliminação da regra Cut depende crucialmente da existência de um conectivo principal, o que não acontece no cálculo das estruturas. Além disso, no cálculo das estruturas a regra Cut pode ser dividida em várias regras.

Todos esses conceitos de estrutura, "deep inference" e simetria vertical se baseiam em uma representação gráfica de estruturas que se assemelha a grafos. Essa representação é chamada de redes de interação\footnote{Em inglês:\emph{Relation webs}}. As redes de interação podem ser usadas tanto como uma semântica para as estruturas quanto para um modelo abstrato de computação~\cite{guglielmi07}.

\subsection{Estruturas}
\label{section:estruturas}

Apesar de o conceito de estrutura não ser complicado, começar a entendê-lo através da relação com conceitos já tradicionais da Teoria da demonstração pode ser útil. Simplificando bastante, uma estrutura é, ao mesmo tempo, uma fórmula e um seqüente, além de capturar alguns aspectos de redes de demonstraçãos.

Justamente porque estruturas capturam aspectos de redes de demonstraçãos, é interessante começar a introduzir seu conceito através da construção de uma demonstração na lógica linear multiplicativa. Na representação de seqüentes de um lado só, a conjunção multiplicativa $\tensor$ (\emph{times}) é definida como:

$$\ddernote{}{\tensor}{\Seq{}{A\tensor B, \Phi, \Psi}}{\leaf{\Seq{}{A,\Phi}}}{\leaf{\Seq{}{B,\Psi}}}$$

Instâncias de $\tensor$ podem ser vistas como passos elementares em uma computação que correspondem à construção de uma demonstração (de baixo para cima). Esta perspectiva é chamada de paradigma de \emph{busca por demonstraçãos como computação} ou de \emph{construção de demonstraçãos}. Sob este ponto de vista, a regra $\tensor$ acima apresenta uma grave falha: quando a regra é aplicada, temos que decidir como dividir o contexto $\Phi, \Psi$ da fórmula principal $A \tensor B$, e se há $n$ fórmulas no multiconjunto $\Phi, \Psi$, então há $2^{n}$ maneiras de dividi-las entre $\Phi$ e $\Psi$. Esta é uma fonte exponencial de não-determinismo indesejável, se estamos interessados em implementar o nosso sistema.

Uma alternativa para resolver o problema é adotar uma abordagem externa ao cálculo de seqüentes, ou seja, uma \emph{implementação} que controle o uso da regra \emph{a posteriori}. Nesta abordagem, na construção da demonstração, o multiconjunto $\Phi$, $\Psi$ não é dividido a princípio, mas é utilizado conforme vai surgindo a necessidade por suas fórmulas ao longo da construção da demonstração. Sempre que necessário, a instância da regra $\tensor$ é atualizada para refletir essa utilização, até que uma demonstração válida seja construída. Esta abordagem ``preguiçosa'' foi adotada por Hodas e Miller no caso da lógica intuicionista linear~\cite{hodas93welp}.

Mas ainda resta a questão de se o problema pode ser resolvido \emph{dentro} de um sistema dedutivo. Uma idéia seria criar \emph{clusters} de fórmulas, correspondendo a contextos a serem divididos, e cada fórmula poderia ser tomada sob demanda. No cálculo de seqüentes tradicional, esta solução não é atingível facilmente. O que realmente conta no cálculo de seqüentes, enquanto construímos uma demonstração de baixo para cima, é a fronteira de hipóteses, cuja natureza é de um multiconjunto. Em vez disso, precisamos manter uma \emph{árvore}, cuja estrutura trata automaticamente do aninhamento de \emph{clusters}. 

Consideremos o seguinte exemplo de demonstração em \MLL:

$$ \infer[\lpar]{\Seq{}{a \lpar (b \lpar (\neg{b} \tensor ( (\neg{a} \tensor \neg{c}) \lpar \neg{c} ) ) ) }}
{\infer[\lpar]{\Seq{}{a,b,\lpar(\neg{b}\tensor((\neg{a} \tensor c)\lpar \neg{c}))}}
{
\infer[\tensor]{\Seq{}{a,b,\neg{b} \tensor ((\neg{a} \tensor c) \lpar \neg{c})}}
{\infer[Inicial]{\Seq{}{b,\neg{b}}}{}&
\infer[\lpar]{\Seq{}{a,(\neg{a} \tensor c)\lpar \neg{c}}}
{\infer[\tensor]{a,\neg{a} \tensor c, \neg{c}}
{\infer[Inicial]{\Seq{}{a,\neg{a}}}{}&
\infer[Inicial]{\Seq{}{c,\neg{c}}}{}}}}}} $$

Neste caso, partições apropriadas do contexto foram achadas, e temos então uma demonstração. Como será visto adiante, no cálculo das estruturas, essas partições de contexto são tratadas internamente no sistema dedutivo, graças às características únicas que este formalismo apresenta.

Na sintaxe de cálculo das estruturas, vamos chamar $\lpar$ de \emph{par} e $\tensor$ de \emph{copar}. A negação de um átomo $a$, ou seu dual, é representado como sendo $\dual{a}$. A fórmula demonstraçãoda acima será representada como $\pars{a,b,\aprs{\dual{b},\pars{\aprs{\dual{a},c},\dual{c}}}}$ e expressões como esta são chamadas de \emph{estruturas}. Em vez de usar conectivos binários para definir relações lógicas, as relações são induzidas pelo contexto, significando, por exemplo, que $\pars{R_{1},\hdots,R_{h}}$ é uma estrutura par, dentro da qual as subestruturas são consideradas unidas por conectivos par. Já que a relação par é comutativa e associativa, as estruturas não são distingüidas com base na ordem ou agrupamento de subestruturas. O mesmo é verdade para copar: $\aprs{R_{1},\hdots,R_{h}}$ tem as mesmas propriedades que uma estrutura par.

\begin{Definition} Há infinitos \emph{átomos positivos} e infinitos \emph{átomos negativos}. Átomos, não importando se positivos ou negativos, são denotados por $a$, $b$, $c$, $\hdots$. \emph{Estruturas} são , denotadas por $S$, $P$, $Q$, $R$, $T$, $U$, $V$ e $X$. A gramática para geração de estruturas é:

$$S \ ::= \ \circ \ | \ a \ | \ \underbrace{\seqs{S;\hdots;S}}_{> 0} \ | \ \underbrace{\pars{S,\hdots,S}}_{> 0} \ | \ \underbrace{\aprs{S,\hdots,S}}_{> 0} \ | \ \dual{S} $$
onde $\circ$, a \emph{unidade}, não é um átomo. $\seqs{S;\hdots;S}$ é chamada de uma \emph{estrutura seq} (ou simplesmente \emph{seq}), $\pars{S,\hdots,S}$ é uma \emph{estrutura par} (ou simplesmente \emph{par}) e $\aprs{S,\hdots,S}$ é uma \emph{estrutura copar} (ou simplesmente \emph{copar}). $\dual{S}$ é a negação da estrutura $S$. Um átomo negado $\dual{a}$ é negativo se $a$ for positivo e positivo se $a$ for negativo. Existe um átomo especial chamado \emph{contexto aberto}~\footnote{Em inglês: \emph{hole}}, denotado por $\{ \ \}$, cujo propósito é indicar um lugar específico dentro de uma estrutura onde eventualmente outras estruturas são encaixadas. Estruturas que possuem um contexto aberto que não aparece no escopo de uma negação são denotadas como $S\{ \ \}$, e são chamadas de \emph{contextos de estruturas}, ou simplesmente \emph{contextos}. A estrutura $R$ é uma subestrutura de $S\{R\}$, e $S\{ \ \}$ é o contexto de $R$.

\begin{Remark}
\emph{Delimitadores} são sinais utilizados para delimitar estruturas, indicando sua natureza. Os delimitadores de uma estrutura seq são $\seqs{}$, de uma estrutura par são $\pars{}$ e de uma estrutura copar são $\aprs{}$.
\end{Remark}

\end{Definition}

\begin{Notation} Quando houver delimitadores de algum tipo em torno do conteúdo de um contexto aberto, as chaves do contexto aberto poderão ser omitidas. Por exemplo, $S\pars{a,b}$ é o mesmo que $S\{ \pars{a,b} \}$.
\end{Notation}

\begin{Notation} Uma letra estabelece implicitamente a classe à qual um objeto pertence. Por exemplo, $S$ é sempre uma estrutura, sem ser preciso explicitar isso.
\end{Notation}

A associatividade é válida para todas as estruturas, mas a comutatividade não é válida para estruturas seq. A negação é involutiva e obedece às leis usuais de De Morgan para par e copar, que são duais. Já seq é auto-dual.

\begin{Definition} Estruturas são consideradas equivalentes módulo $=$, que é a relação de equivalência minimal definida pelos axiomas da figura~\ref{fig:equivalencia-sintatica}. Nessa figura, $\vec{R}$, $\vec{T}$, $\vec{U}$ representam seqüências finitas e não vazias de estruturas. Uma estrutura, ou um contexto, é dito estar na \emph{forma normal} quando as únicas estruturas negadas são átomos, não há nenhuma unidade $\circ$ e nenhum delimitador pode ser eliminado mantendo a equivalência. Se duas estruturas $R$ e $T$ são tais que $ R \neq \circ \neq T$, então a estrutura $\seqs{R;T}$ é um \emph{seq próprio}, a estrutura $\pars{R;T}$ é um \emph{par próprio} e a estrutura $\aprs{R;T}$ é um \emph{copar próprio}. Uma estrutura $S\{ \ \}$ é um \emph{contexto seq próprio} (ou um \emph{contexto par próprio}, ou um \emph{contexto copar próprio}) se, para todo $X \neq \circ$, a estrutura $S\{X\}$ é um seq próprio (ou um par próprio, ou um copar próprio). As estruturas cuja forma normal não contenham estruturas seq são chamadas \emph{flat}. Se uma forma normal é \emph{flat}, então todas as demais também o são.

\end{Definition}

\begin{figure}[!htb]
\center

	\begin{multicols}{2}{

	\textbf{Associatividade}\\ 
	$\seqs{\vec{R};\seqs{\vec{T}};\vec{U}} = \seqs{\vec{R};\vec{T};\vec{U}}$\\
	$\pars{\vec{R},\pars{\vec{T}}} = \pars{\vec{R},\vec{T}}$\\
	$\aprs{\vec{R},\pars{\vec{T}}} = \aprs{\vec{R},\vec{T}}$\\
	\ \\
	
	\textbf{\emph{Singleton}}\\
	$\seqs{R} = \pars{R} = \aprs{R} = R$\\
	\ \\
	
	\textbf{Unidade}\\
	$\seqs{\circ;\vec{R}} = \seqs{\vec{R};\circ} = \seqs{\vec{R}}$\\
	$\pars{\circ,\vec{R}} = \pars{\vec{R}}$\\
	$\aprs{\circ,\vec{R}} = \aprs{\vec{R}}$\\
	
 	\ \\ 
	
	\textbf{Comutatividade}\\
	$\pars{\vec{R},\vec{T}} = \pars{\vec{T},\vec{R}}$\\
	$\aprs{\vec{R},\vec{T}} = \aprs{\vec{T},\vec{R}}$\\	
	\ \\
	
	\textbf{Negação}\\
	$\dual{\circ} = \circ$\\
	$\dual{\seqs{R;T}} = \seqs{\dual{R};\dual{T}}$\\
	$\dual{\pars{R;T}} = \aprs{\dual{R};\dual{T}}$\\
	$\dual{\aprs{R;T}} = \pars{\dual{R};\dual{T}}$\\
	$\dual{\dual{R}} = R$
	\ \\	
	
	\textbf{Fechamento de contexto}\\
	$R = T \cimp S\{R\} = S\{T\} \wedge \dual{R}=\dual{T}$

	}
	
	\end{multicols}
	
	\caption{Equivalência sintática =}
	\label{fig:equivalencia-sintatica}
	
\end{figure}

Por exemplo, as estruturas $\overline{\pars{a,\circ,b}}$, $\aprs{\aprs{\overline{\pars{\circ,b}}},\seqs{\dual{a}}}$ e $\aprs{\dual{a},\circ,\dual{b}}$ são todas equivalentes, mas nenhuma está na forma normal. Já a estrutura $\aprs{\dual{a},\dual{b}}$ é equivalente a elas e está na forma normal, assim como $\aprs{\dual{b},\dual{a}}$. E todas as estruturas são \emph{flat}. 

Uma vez que as estruturas são consideradas equivalentes sob $=$, a estrutura $\pars{\circ,\seqs{a;b}}$ é um seq próprio, mas não é um par próprio ou um copar próprio. $\seqs{a;\pars{\{ \ \}, b}}$ é um contexto seq próprio, enquanto $\pars{\{ \ \},b}$ é um contexto par próprio.

\begin{Remark} Toda estrutura pode ser colocada na forma normal, uma vez que a negação sempre pode ser trazida para os átomos mais internos via os axiomas de negação, e unidades podem ser removidas, assim como delimitadores extras (pelas leis de associatividade e \emph{singleton}). Toda estrutura pode ser equivalente a apenas um dos seguintes: unidade, um átomo, um seq próprio, um par próprio ou um copar próprio. 
\end{Remark}

As regras de negação se assemelham às leis de De Morgan para par e copar, mas, para o conectivo auto-dual não comutativo seq, a ordem das subestruturas é mantida sob a negação. A equação $\dual{\circ} = \circ$ indica que em cálculo das estruturas a unidade é um pouco diferente das constantes lógicas tradicionais como $\top$ (verdadeiro) e $\bot$ (falso). A unidade é apenas uma marca de representação à qual não daremos interpretação semântica.

\begin{Definition} Dada uma estrutura $S$, suas \emph{ocorrências de átomos} são todos átomos em $S$ levando em conta aparições distintas do mesmo átomo (por exemplo indexando-os de forma que dois átomos iguais recebem índices diferentes). Assim, em $\seqs{a,a}$ há duas ocorrências de átomos. A notação $\occ{S}$ indica o conjunto de todos as ocorrências de átomos em $S$. O \emph{tamanho} de $S$ é a cardinalidade do conjunto $\occ{S}$. 

O conjunto $\occ{S}$ poderia ser definido como o multiconjunto de átomos de $S$, ou de $S$ na forma normal. Note que $\occ{\circ} = \emptyset$. Além disso, $\occ{\seqs{S;S'}} = \occ{\pars{S,S'}} = \occ{\aprs{S,S'}} = \occ{S} \ \cup \ \occ{S'}$ é verdadeiro somente se $\occ{S}$ e $\occ{S'}$ são disjuntos, e pode-se sempre assumir isto sem perda de generalidade.
\end{Definition}

\subsection{Redes de interação}
\label{section:redes-interacao}

Estruturas sempre podem ser representadas pela gramática exposta na seção anterior. Existe, porém, uma forma gráfica alternativa que captura toda a essência da estrutura e, por motivos que serão expostos mais adiante, é muito conveniente em certas situações. Essa forma de representação são as \textit{redes de interação}. Um ponto importante é que existe uma única teia de interação para cada classe de equivalência de estruturas sob a relação de igualdade $=$.

Redes de interação podem ser usadas como uma espécie de semântica para regras de inferência.

Considere $\pars{R_{1},\hdots,R_{h}}$. Sejam $1 \leq i,j \leq h$ distintos. Para todo átomo $a$ que aparece em $R_{i}$ e todo átomo $b$ que aparece em $R_{j}$, a relação $a \downarrow b$ é válida. Observa-se que $\downarrow$ é simétrica. Analogamente, duas estruturas distintas $R$ e $T$ numa relação de copar induzem em seus átomos a relação $a \uparrow b$, onde $a$ pertence a $R$ e $b$ pertence a T. 

Por exemplo, seja a estrutura:

$$\pars{a,b,\aprs{\dual{b},\pars{\aprs{\dual{a},c},\dual{c}}}}$$

Nesse caso temos os seguintes pares relacionados: $a \downarrow b$, $a \downarrow \dual{b}$, $a \downarrow \dual{a}$, $a \downarrow c$, $a \downarrow \dual{c}$, $b \downarrow \dual{b}$, $b \downarrow \dual{a}$, $b \downarrow c$, $b \downarrow \dual{c}$, $\dual{b} \uparrow \dual{a}$, $\dual{b} \uparrow c$, $\dual{b} \uparrow \dual{c}$, $\dual{a} \uparrow c$, $\dual{a} \downarrow \dual{c}$, $c \downarrow \dual{c}$. Além desses, há os pares simétricos, que foram omitidos.

Além dessas relações, $\seqs{S_{1},\hdots,S_{h}}$, com $1 \leq i < j \leq h$ induz a relação $a \triangleleft b$ para todo $a$ em $S_{i}$ e $b$ em $S_{j}$. Observe que $\triangleleft$ não é simétrica.

\begin{Definition}

Dada uma estrutura $S$ na forma normal, as quatro \emph{relações estruturais} $\triangleleft_{S}$ (\emph{seq}), $\triangleright_{S}$ (\emph{coseq}), $\downarrow_{S}$ (\emph{par}) e $\uparrow_{S}$ (\emph{copar}) são os conujuntos minimais tais que $\triangleleft_{S}, \triangleright_{S}, \downarrow_{S}, \uparrow_{S} \subset (\occ{S})^{2}$ e, para todo $S'\{ \ \}$, $U$ e $V$ e para todo $a$ em $U$ e $b$ em $V$, as seguintes asserções podem ser verificadas:
	
\begin{enumerate}

	\item se $S = S'\seqs{U;V}$, então $a \triangleleft_{S} b$ e $b \triangleright_{S} a$;
		
	\item se $S = S'\pars{U,V}$, então $a \downarrow_{S} b$;
	
	\item se $S = S'\aprs{U,V}$, então $a \uparrow_{S} b$;

\end{enumerate}

Para uma estrutura que não está em sua forma normal, obtém-se a teia de interação a partir de qualquer de suas formas normais, uma vez todas elas levarão à mesma rede. A quádrupla $(\occ{S}, \triangleleft_{S}, \downarrow_{S}, \uparrow_{S})$ é chamada de \emph{teia de interação} (ou simplesmente \emph{rede}) de $S$, denotada por $\relweb{S}$. Pode-se abolir os subscritos em $\triangleleft_{S}, \triangleright_{S}, \downarrow_{S}, \uparrow_{S}$ quando eles não forem necessários. Dados dois conjuntos de átomos $\mu$ e $\nu$, escrevem-se $\mu \triangleleft \nu, \mu \triangleright \nu, \mu \downarrow \nu, \mu \uparrow \nu$ para indicar situações nas quais, para todo $a$ pertencente a $\mu$ e todo $b$ pertencente a $\nu$, o seguinte vale, respectivamente: $a \triangleleft b, a \triangleright b, a \downarrow b, a \uparrow b$. Representam-se relações estruturais entre ocorrências de átomos desenhando \xy{\ar@{~>} (0,1)*{a};(8,1)*{b}}\endxy \ quando $a \triangleleft b$ (e $b \triangleright a$), \xy{\ar@{<~>} (0,1)*{a};(8,1)*{b}}\endxy \ quando $a \triangleleft b$ ou $a \triangleright b$, \xy{\ar@{-} (0,1)*{a};(8,1)*{b}}\endxy \ quando $a \downarrow b$ e \xy{\ar@{~} (0,1)*{a};(8,1)*{b}}\endxy \ quando $a \uparrow b$. Linhas pontilhadas representam a negação das respectivas relações estruturais.

\end{Definition}

Por exemplo, em $\aprs{\overline{\seqs{\dual{a};b}},\overline{\aprs{c,\dual{d}}}} = \aprs{\seqs{a,\dual{b}},\pars{\dual{c},d}}$ determinam-se as relações $a \triangleleft \dual{b}$, $a \uparrow \dual{c}$, $a \uparrow d$, $\dual{b} \triangleright a$, $\dual{b} \uparrow \dual{c}$, $\dual{b} \uparrow d$, $\dual{c} \uparrow a$, $\dual{c} \uparrow \dual{b}$, $\dual{c} \downarrow d$, $d \uparrow a$, $d \uparrow \dual{b}$, $d \downarrow \dual{c}$. A teia de interação para $\circ$ é $(\emptyset,\emptyset,\emptyset,\emptyset)$

\begin{Remark} Uma estrutura $S$ tal que $\relweb{S} = (\occ{S}, \triangleleft, \downarrow, \uparrow)$ é \emph{flat} se, e somente se, $\triangleleft = \emptyset$.
\end{Remark}

Pode-se perceber pelas definições acima que todos os átomos de uma subestrutura respeitam a mesma relação estrutural em relação aos átomos do contexto que envolve esta subestrutura:

\begin{Proposition} Dada uma estrutura $S\{R\}$ e duas ocorrências de átomos $a$ em $S\{ \ \}$ e $b$ em $R$, se $a \triangleleft b$ (respectivamente, $a \triangleright b$, $a \downarrow b$, $a \uparrow b$) então $a \triangleleft c$ (respectivamente, $a \triangleright c$, $a \downarrow c$, $a \uparrow c$) para todas as ocorrências de átomos $c$ em $R$.
\end{Proposition}

A sintaxe de estruturas apresentada, assim como a relação de equivalência $=$, ajuda a manter o foco do sistema mais no significado do que na sua forma de representação. O que realmente importa são os átomos e suas relações mútuas. Entretanto, não basta tomar um conjunto de átomos e atribuir relações arbitrárias entre eles para se obter uma estrutura. Duas questões importantes são levantadas neste momento:

\begin{enumerate}

\item Em que condições uma atribuição de relações a átomos realmente forma uma estrutura.

\item Se duas estruturas que possuem uma dada teia de interação são equivalentes por $=$ ou não.

\end{enumerate}

Os teoremas seguintes~\cite{guglielmi07} fornecem as respostas.

\begin{Theorem}

Dada uma estrutura $S$ e suas relações estruturais $\triangleleft, \triangleright, \downarrow, \uparrow$, as seguintes propriedades são válidas, onde $a$, $b$, $c$ e $d$ são ocorrências de átomos distintas em $S$:

	\begin{description}

		\item[$s_{1}$] Nenhuma das relações $\triangleleft, \triangleright, \downarrow$ e $\uparrow$ é reflexiva: $\neg{(a \triangleleft a)}$, $\neg{(a \triangleright a)}$, $\neg{(a \downarrow a)}$, $\neg{(a \uparrow a)}$.
		
		\item[$s_{2}$] Uma, e apenas uma, das relações $a \triangleleft b$, $a \triangleright b$, $a \downarrow b$, $a \uparrow b$ é válida.
		
		\item[$s_{3}$] As relações $\triangleleft$ e $\triangleright$ são mutuamente inversas: $a \triangleleft b \iff b \triangleright a$.
		
		\item[$s_{4}$] As relações $\triangleleft$ e $\triangleright$ são transitivas:  $a \triangleleft b \wedge b \triangleleft c \implies a \triangleleft c$ e $a \triangleright b \wedge b \triangleright c \implies a \triangleright c$.
		
		\item[$s_{5}$] As relações $\downarrow$ e $\uparrow$ são simétricas: $a \downarrow b \iff b \downarrow a$ e $a \uparrow b \iff b \uparrow a$.
		
		\item[$s_{6}$] Propriedade triangular: para $\sigma_{1}, \sigma_{2}, \sigma_{3} \in \{\triangleleft \cup \triangleright, \downarrow, \uparrow \}$ pode-se afirmar que:
		
		$$ (a \sigma_{1} b) \wedge (b \sigma_{2} c) \wedge (c \sigma_{3} a) \implies (\sigma_{1} = \sigma_{2}) \vee (\sigma_{2} = \sigma_{3}) \vee (\sigma_{3} = \sigma_{1}) $$
		
		\item[$s_{7}$] Propriedade do quadrado:
		
			\begin{description}
			
				\item[$s_{7}^{\triangleleft}$] $(a \triangleleft b) \wedge (a \triangleleft d) \wedge (c \triangleleft d) \implies (a \triangleleft c) \vee (b \triangleleft c) \vee (b \triangleleft d) \vee (c \triangleleft a) \vee (c \triangleleft b) \vee (d \triangleleft b)$.
				
				\item[$s_{7}^{\downarrow}$] $(a \downarrow b) \wedge (a \downarrow d) \wedge (c \downarrow d) \implies (a \downarrow c) \vee (b \downarrow c) \vee (b \downarrow d)$.
				
				\item[$s_{7}^{\uparrow}$] $(a \uparrow b) \wedge (a \uparrow d) \wedge (c \uparrow d) \implies (a \uparrow c) \vee (b \uparrow c) \vee (b \uparrow d)$.
			
			\end{description}
			
	\end{description}
	
	\begin{proof}
			
				As propriedades de $s_{1}$, $s_{2}$, $s_{3}$, $s_{4}$ e $s_{5}$ seguem diretamente da definição. Será dada a demonstração das propriedades $s_{6}$ e $s_{7}$.
				
				\begin{description}
				
					\item[$s_{6}$] Suponha que $a \triangleleft b$ e $b \downarrow c$. Os únicos casos possíveis são $S\seqs{P\{a\};T\pars{Q\{b\},R\{c\}}}$ (e então tem-se que $a \triangleleft c$) ou $S\pars{T\seqs{P\{a\};Q\{b\}},R\{c\}}$ (e então tem-se que $a \downarrow c$), onde $P\{ \ \}$, $Q\{ \ \}$, $R\{ \ \}$, $S\{ \ \}$ e $T\{ \ \}$ são contextos. Outras combinações de $\sigma_{1}$ e $\sigma_{2}$ geram casos análogos.
					
					\item[$s_{7}$] Por indução estrutural em $S$. Qualquer estrutura com menos do que quatro ocorrências de átomos satisfaz $s_{7}$ trivialmente, já que a hipótese do teorema necessita de quatro ocorrências distintas. Serão consideradas, portanto, estruturas em que haja pelo menos quatro ocorrências. Sejam $U$ e $V$ duas estruturas tais que $U \neq \circ \neq V$ e temos uma das seguintes situações pode ser verificada: ou $S = \seqs{U;V}$, ou $S = \pars{U,V}$, ou $S = \aprs{U,V}$. Escolham-se quatro átomos $a$, $b$, $c$ e $d$ em $S$. Se $a$, $b$, $c$ e $d$ estão todos em $U$ ou em $V$, então pode-se aplicar a hipótese de indução. Serão considerados então os casos em que os quatro átomos não estão todos em $U$ ou em $V$. Considere $s_{7}^{\triangleleft}$. Já que $S = \pars{U,V}$ e  $S = \aprs{U,V}$ falsificam a hipótese de $s_{7}^{\triangleleft}$, a única situação a ser considerada é $S = \seqs{U;V}$. Suponha que a conclusão de $s_{7}^{\triangleleft}$ seja falsa e que $a$ está em $U$, logo $c$ deve estar em $U$ (de outra forma $a \triangleleft c$ seria verdadeiro), e então $b$ e $d$ devem estar em $U$, o que contradiz a hipótese assumida. Analogamente, se $a$ está em $V$ então $c$ deve necessariamente estar em $V$ e logo $b$ e $d$ também devem estar em $V$, o que novamente é uma contradição. Enfim, se a hipótese de $s_{7}^{\triangleleft}$ é verdadeira quando $a$, $b$, $c$, e $d$ estão distribuídos entre $U$ e $V$, então sua conclusão é verdadeira. O mesmo argumento se aplica a $s_{7}^{\downarrow}$ e a $s_{7}^{\uparrow}$.
				
				\end{description}
			
			\end{proof}

\end{Theorem} 

A propriedade do triângulo diz que não há estrutura que contenha uma subestrutura da forma:

$$
	\xymatrix{
		b \ar@{<~>}[dd] \ar@{-}[drr] & & \\
	  	& & c \\
	  a	\ar@{~}[urr] & & \\
	}
$$

Em outras palavras, em todo triângulo pelo menos dois lados devem representar a mesma relação estrutural.

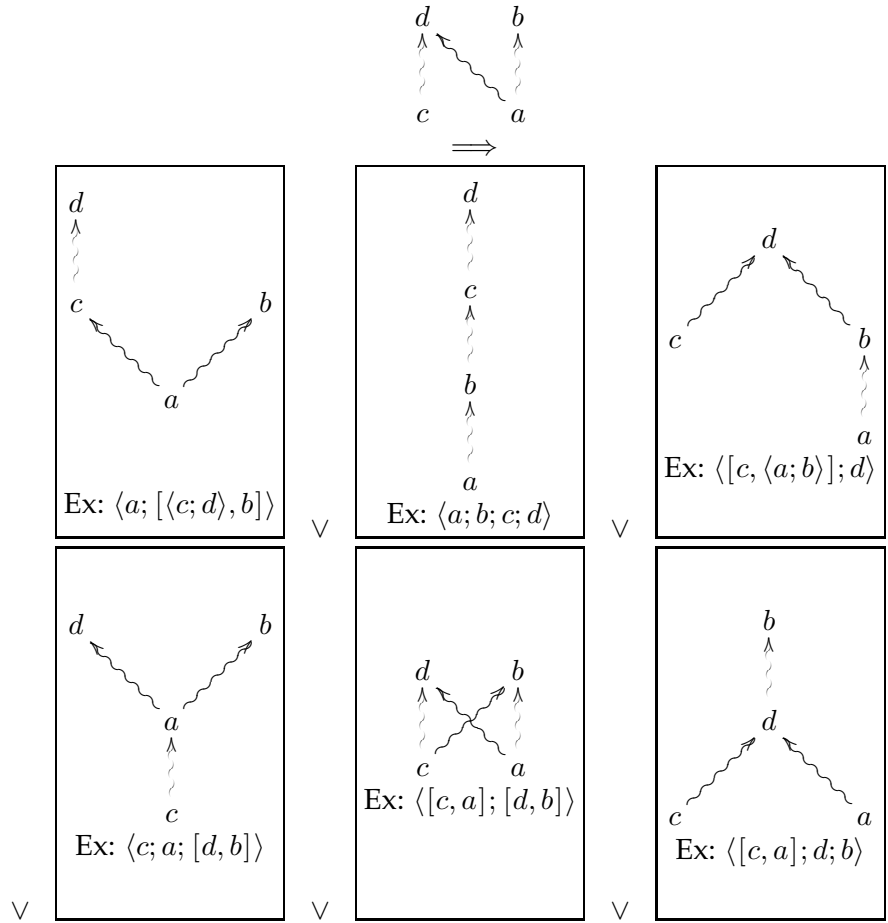
\begin{figure}[!htb]
	\center

	$$
	\begin{array}{cccccc}
	
		& & & 
			\xymatrix{
				d \ar@{<~}[d] \ar@{<~}[dr] & b \ar@{<~}[d] \\
				c & a \\
			}
		& &
		\\
		
		& & & \implies & &
			
		\\
		
		&
		
		\framebox(85,140){
			\begin{tabular}{c}
				\xymatrix{
					d \ar@{<~}[d] & & \\
					c & & b\\
					 & a \ar@{~>}[ul] \ar@{~>}[ur] &
				}
				\\
				\\
				\\
				Ex: $\seqs{a;\pars{\seqs{c;d},b}}$
			\end{tabular}
		}
		
		&
		\vee
		&
		
		\framebox(85,140){
			\begin{tabular}{c}
				\xymatrix{
					d \ar@{<~}[d] \\
					c \ar@{<~}[d] \\
					b \ar@{<~}[d] \\
					a  \\
				}
				\\
				Ex: $\seqs{a;b;c;d}$
			\end{tabular}
		} 
			
		&
		\vee
		&
		
		\framebox(85,140){
			\begin{tabular}{c}
				\xymatrix{
					& d \ar@{<~}[dl] \ar@{<~}[dr] & \\
					c & & b\\
					 & & a \ar@{~>}[u]  \\
				}
				\\
				Ex: $\seqs{\pars{c,\seqs{a;b}};d}$
			\end{tabular}
		}
		\\
		
		\vee
		&
		
		\framebox(85,140){
			\begin{tabular}{c}
				\xymatrix{
					d &   & b \\
				  & a \ar@{<~}[d] \ar@{~>}[ul] \ar@{~>}[ur] &   \\
				  & c &   \\
				}
				\\
				Ex: $\seqs{c;a;\pars{d,b}}$
			\end{tabular}
		}
		
		&
		\vee
		&
		
		\framebox(85,140){
			\begin{tabular}{c}
				\xymatrix{
					d \ar@{<~}[d] \ar@{<~}[dr] & b \ar@{<~}[d] \ar@{<~}[dl] \\
					c & a \\
				}
				\\
				Ex: $\seqs{\pars{c,a};\pars{d,b}}$
			\end{tabular}
		}
		
		&
		\vee
		&
		
		\framebox(85,140){
			\begin{tabular}{c}
				\xymatrix{
					& b \ar@{<~}[d] & \\
		 			& d \ar@{<~}[dl] \ar@{<~}[dr]& \\
					c &  & a \\
				}
				\\
				Ex: $\seqs{\pars{c,a};d;b}$
			\end{tabular}
		}
		
	\end{array}
	$$
	
	\label{fig:propriedade-quadrado-seq}
	\caption{Propriedade do quadrado para $\triangleleft$}
	
\end{figure}

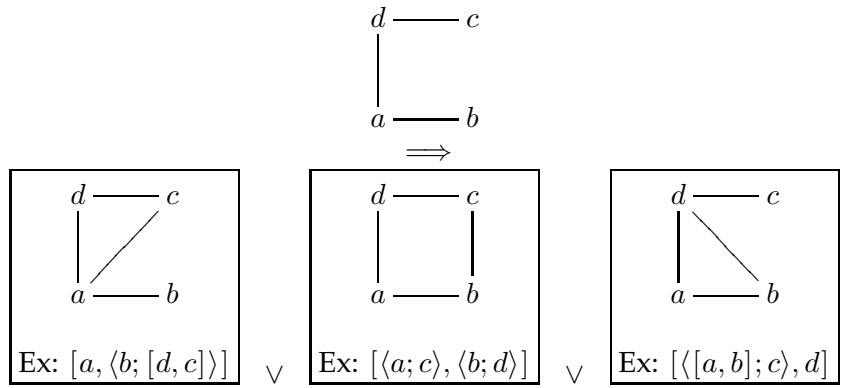
\begin{figure}[!htb]

	\center
	
	$$
	\begin{array}{ccccc}

		& &   
		\xymatrix{
			d \ar@{-}[r] \ar@{-}[d] & c \\
			a \ar@{-}[r] & b \\
		}
		& & 
		
		\\
		
		& & \implies & &
		
		\\
		
		\framebox(85,80){
			\begin{tabular}{c}
				\xymatrix{
					d \ar@{-}[r] \ar@{-}[d] & c \ar@{-}[dl] \\
					a \ar@{-}[r] & b \\
				}
				\\
				\\
				Ex:	$\pars{a,\seqs{b;\pars{d,c}}}$
			\end{tabular}
		}
			
		&
		\vee
		&
		
		\framebox(85,80){
			\begin{tabular}{c}
				\xymatrix{
					d \ar@{-}[r] \ar@{-}[d] & c \ar@{-}[d] \\
					a \ar@{-}[r] & b \\
				}
				\\ 
				\\
				Ex: $\pars{\seqs{a;c},\seqs{b;d}}$
			\end{tabular}
		}
				
		&
		\vee
		&
		
		\framebox(85,80){
			\begin{tabular}{c}
				\xymatrix{
					d \ar@{-}[r] \ar@{-}[d] \ar@{-}[dr] & c  \\
					a \ar@{-}[r] & b \\
				}
				\\ 
				\\
				Ex: $\pars{\seqs{\pars{a,b};c},d}$
			\end{tabular}
		}
	
	\end{array}
	$$
	
	\label{fig:propriedade-quadrado-par}
	\caption{Propriedade do quadrado para $\downarrow$}
	
\end{figure}

A propriedade do quadrado para $\triangleleft$ pode ser representada como na figura~\ref{fig:propriedade-quadrado-seq}, onde a transitividade foi levada em conta e um exemplo de estrutura é mostrado abaixo de cada diagrama. Levando em conta a comutatividade, os casos para par e copar são mais simples. A figura~\ref{fig:propriedade-quadrado-par} mostra o que acontece no caso do par. Informalmente, podemos dizer que nenhum quadrado tem exatamente três lados ou diagonais da mesma natureza (isto é, representando a mesma relação estrutural) e formando um caminho simples (sem levar em conta a orientação).

\begin{Remark}
As relações $\downarrow$ e $\uparrow$ não são transitivas: considere $\pars{\seqs{a;c},b}$ (nesse caso $a \downarrow b$ e $b \downarrow c$ mas $\neg{(a \downarrow c)}$) e $\aprs{\seqs{a;c},b}$ (nesse caso $a \uparrow b$ e $b \uparrow c$ mas $\neg{(a \uparrow c)}$).
\end{Remark}

Até agora foi mostrado que as condições de $s_{1}$ a $s_{7}$ são necessárias para uma estrutura, mas ainda não foi mostrado que são suficientes. Isto será feito em seguida, mas antes são necessárias algumas definições auxiliares.

\begin{Definition}

Uma rede candidata\footnote{Em inglês: \emph{Web candidate}} é uma quádrupla $\zeta = (\xi, \triangleleft, \downarrow, \uparrow)$, onde $\xi$ é um conjunto de ocorrências de átomos e $\triangleleft, \downarrow, \uparrow \subseteq \xi^{2}$. Dadas as redes candidatas $\zeta_{\mu} = (\mu, \triangleleft_{\mu}, \downarrow_{\mu}, \uparrow_{\mu})$ e $\zeta_{\nu} = (\nu, \triangleleft_{\nu}, \downarrow_{\nu}, \uparrow_{\nu})$, tais que $\mu \neq \emptyset \neq \nu$, $\mu \cup \nu = \xi$ e $\mu \cap \nu = \triangleleft_{\mu} \cap \triangleleft_{\nu} = \downarrow_{\mu} \cap \downarrow_{\nu} = \uparrow_{\mu} \cap \uparrow_{\nu} = \emptyset$, o par $(\zeta_{\mu}, \zeta_{\nu})$ pode ser:

	\begin{enumerate}
	
		\item uma partição-$\triangleleft$ de $\zeta$ se, e somente se, $\downarrow = \downarrow_{\mu} \cup \downarrow_{\nu}$, $\uparrow = \uparrow_{\mu} \cup \uparrow_{\nu}$ e 
		
		$$ \triangleleft = \triangleleft_{\mu} \cup \triangleleft_{\nu} \cup \{(a,b)|a \in \mu \wedge b \in \nu\} $$
		
		\item uma partição-$\downarrow$ de $\zeta$ se, e somente se, $\triangleleft = \triangleleft_{\mu} \cup \triangleleft_{\nu}$, $\uparrow = \uparrow_{\mu} \cup \uparrow_{\nu}$ e 
		
		$$ \downarrow = \downarrow_{\mu} \cup \downarrow_{\nu} \cup \{(a,b)|(a \in \mu \wedge b \in \nu)\vee (a \in \nu \wedge b \in \mu) \} $$
		
		\item uma partição-$\uparrow$ de $\zeta$ se, e somente se, $\triangleleft = \triangleleft_{\mu} \cup \triangleleft_{\nu}$, $\downarrow = \downarrow_{\mu} \cup \downarrow_{\nu}$ e 
		
		$$ \uparrow = \uparrow_{\mu} \cup \uparrow_{\nu} \cup \{(a,b)|(a \in \mu \wedge b \in \nu)\vee (a \in \nu \wedge b \in \mu) \} $$
	
	\end{enumerate}
	
	Para toda rede candidata, a relação $\triangleright = \{(a,b) | b \triangleleft a\}$ deve ser definida, e isto é feito de maneira implícita.

\end{Definition}

Evidentemente redes de interação são redes candidatas. Com essas definições já é possível demonstrar que as condições $s_{1}$ a $s_{7}$ são suficientes para uma estrutura.

\begin{Theorem}

	Se as condições de $s_{1}$ a $s_{7}$ valem para uma rede candidata $\zeta$ então existe uma estrutura cuja teia de interação é $\zeta$.

	\begin{proof}
	
		Seja $\zeta = (\xi, \triangleleft, \downarrow, \uparrow)$: vamos proceder por indução sobre a cardinalidade $|\xi|$ de $\xi$ para construir uma estrutura $S$ tal que $\relweb{S} = \zeta$. Se $\xi = \emptyset$ então $S = \circ$. Se $\xi = \{a\}$ então $\triangleleft = \downarrow = \uparrow = \emptyset$ (por $s_{1}$) e $S = a$. Consideremos os casos em que há pelo menos duas ocorrências de átomos em $\xi$. Veremos que as condições de $s_{1}$ a $s_{7}$ reforçam a existência de uma partição -$\triangleleft$, -$\downarrow$ ou -$\uparrow$ de $\zeta$. Suponha que exista uma partição-$\triangleleft$ de $\zeta$, consistindo de $\zeta_{\mu}$ e $\zeta_{\nu}$. As condições de $s_{1}$ a $s_{7}$ valem para $\zeta_{\mu}$ e $\zeta_{\nu}$, então, pela hipótese de indução, existem duas estruturas $U$ e $V$ tais que $\relweb{U}$ = $\zeta_{\mu}$ e $\relweb{V}$ = $\zeta_{\nu}$. Mas então podemos tomar $S = \seqs{U;V}$ e, pela definição e por $s_{2}$ e $s_{3}$, temos que $\relweb{S} = \zeta$. Podemos proceder analogamente quando $\zeta_{\mu}$ e $\zeta_{\nu}$ formarem uma partição-$\downarrow$ (tome $S = \pars{U,V}$) ou uma partição-$\uparrow$ (tome $S = \aprs{U,V}$). O papel de $s_{5}$ é assegurar a correta formação de uma partição.
		
		É preciso mostrar que existe uma partição -$\triangleleft$, -$\downarrow$ ou -$\uparrow$ de $\zeta$, sob as hipóteses dadas, consistindo de  $\zeta_{\mu} = (\mu, \triangleleft_{\mu}, \downarrow_{\mu}, \uparrow_{\mu})$ e $\zeta_{\nu} = (\nu, \triangleright_{\nu}, \downarrow_{\nu}, \uparrow_{\nu})$. Vamos construir os conjuntos de ocorrências de átomos $\mu$ e $\nu$ incrementalmente, começando por $\mu_{2} = \{a\}$ e $\nu_{2} = \{b\}$ para algum $a$ e algum $b$ em $\xi$, e construindo uma família de pares $\{ (\mu_{i},\nu_{i}) \}_{2\leq i \leq |\xi|}$ de tal forma que em cada passo um elemento de $\xi$ que não foi adicionado antes é adicionado à união de $\mu_{i}$ e $\nu_{i}$; em cada passo $\mu_{i} \neq \emptyset \neq \nu_{i}$ e ou $\mu_{i} \triangleleft \nu_{i}$ ou $\mu_{i} \downarrow \nu_{i}$ ou $\mu_{i} \uparrow \nu_{i}$. O passo final dá a partição, ou seja, $\mu = \mu_{|\xi|}$ e $\nu = \nu_{|\xi|}$. Abaixo é apresentado um algoritmo não-determinístico para isso.
		
		\textbf{Primeiro passo}
		
		Tome $\mu_{2} = \{a\}$ e $\nu_{2} = \{b\}$, onde $a$ e $b$ são ocorrências de átomos distintas aleatoriamente escolhidas em $\xi$ e tais que uma das relações $\mu_{2} \triangleleft \nu_{2}$, $\mu_{2} \downarrow \nu_{2}$, ou $\mu_{2} \uparrow \nu_{2}$ é válida (as condições $s_{2}$ e $s_{3}$ se aplicam).
		
		\textbf{Passo iterativo}
		
		Temos dois conjuntos disjuntos e não-vazios de ocorrências $\mu_{i}$ e $\nu_{i}$ tais que todas as ocorrências de átomos em $\mu_{i}$ possuem a mesma relação estrutural $\sigma \in \{ \triangleleft, \downarrow, \uparrow \}$ com as ocorrências de átomos em $\nu_{i}$, ou seja, $\mu_{i} \triangleleft \nu_{i}$, $\mu_{i} \downarrow \nu_{i}$, ou $\mu_{i} \uparrow \nu_{i}$. Escolha um $c$ qualquer em $\xi \backslash (\mu_{i} \cup \nu_{i})$. Se $d \ \sigma \ c$ para todo $d$ em $\mu_{i}$ então faça $\mu_{i+1} = \mu_{i}$ e $\nu_{i+1} = \nu_{i} \cup \{c\}$; se $c \ \sigma \ e$ para todo $e$ em $\nu_{i}$ então faça  $\mu_{i+1} = \mu_{i} \cup \{c\}$ e $\nu_{i+1} = \nu_{i}$; em ambos os casos $\mu_{i+1} \sigma \nu_{i+1}$. Se isso não ocorrer é preciso rearranjar $\mu_{i}$ e $\nu_{i}$ a fim de que eles atendam as nossas restrições. Procedamos por casos:
		
		\begin{enumerate}
		
			\item $\mu_{i} \triangleleft \nu_{i}$ e existem $a$ em $\mu_{i}$ e $b$ em $\nu_{i}$ tais que $\neg{(a \triangleleft c)}$ e $\neg{(c \triangleleft b)}$. Esta situação está representada à esquerda (onde \xy{\ar@{~~>} (0,2)*{a};(10,2)*{c}}\endxy  \ significa $\neg{(a \triangleleft c)}$):
				
			$$
			\begin{array}{ccccc}
			
				\xymatrix{
				\conjunto{\nu_{i}}{b} \ar@{<~}[dd] \ar@{<~~}[drr] & & \\
				                      & & c \ar@{<~~}[dll] \\
				\conjunto{\mu_{i}}{a} & & \\
				}
				
				&
				\implies
				&
				
				\xymatrix{
				\conjunto{\nu_{i}}{b} \ar@{<~}[dd] \ar@{-}[drr] & & \\
				                      & & c \ar@{-}[dll] \\
				\conjunto{\mu_{i}}{a} & & \\
				}
				
				&
				\vee
				&
				
				\xymatrix{
				\conjunto{\nu_{i}}{b} \ar@{<~}[dd] \ar@{~}[drr] & & \\
				                      & & c \ar@{~}[dll] \\
				\conjunto{\mu_{i}}{a} & & \\
				}
				
			\end{array}
			$$
			
			Como $a \triangleleft b$, por transitividade de $\triangleleft$ ($s_{4}$), por simetria de $\downarrow$ e $\uparrow$ ($s_{5}$) e pela propriedade triangular ($s_{6}$), apenas dois casos são possíveis: ou $a \downarrow c$ e $c\downarrow b$, ou $a \uparrow c$ e $c \uparrow b$ (o primeiro caso está representado no diagrama central, o segundo no da direita). Vamos considerar o primeiro caso, o segundo sendo similar. Novamente por $s_{4}$, $s_{5}$ e $s_{6}$, ou $d \downarrow c$ ou $d \triangleleft c$, para cada elemento $d$ em $\mu_{i}$, e ou $c \downarrow e$ ou $c \triangleleft e$ para cada elemento $e$ em $\nu_{i}$. Podemos então particionar $\mu_{i}$ em dois conjuntos disjuntos $\mu_{i}^{\downarrow}$ e  $\mu_{i}^{\triangleleft}$ e particionar $\nu_{i}$ em dois conjuntos disjuntos $\nu_{i}^{\downarrow}$ e  $\nu_{i}^{\triangleright}$ de tal forma que $\mu_{i}^{\downarrow} \downarrow \{c\}$, $\mu_{i}^{\triangleleft} \triangleleft \{c\}$ e $\{c\} \downarrow \nu_{i}^{\downarrow}$, $\{c\} \triangleleft \nu_{i}^{\triangleright}$. Claro que $a \in \mu_{i}^{\downarrow}$ e $b \in \nu_{i}^{\downarrow}$. Essa situação é representada à esquerda:
			
			$$
			\begin{array}{ccc}
			
				\xymatrix{
					\conjunto{\nu_{i}^{\downarrow}}{b}       &   &   & \conjunto{\nu_{i}^{\triangleright}}{} \\
				                                         & c \ar@{-}[ul] \ar@{-}[dl] \ar@{~>}[urr] &   &                                       \\
					\conjunto{\mu_{i}^{\downarrow}}{a} \ar@{~>}[uu] \ar@{~>}[uurrr] &   &   & \conjunto{\mu_{i}^{\triangleleft}}{} \ar@{~>}[uu] \ar@{~>}[ull] \ar@{~>}[uulll]\\
				}
				
				&
				\implies
				&
			
				\xymatrix{
					\conjunto{\nu_{i}^{\triangleright}}{} & & \\
					\conjunto{\nu_{i}^{\downarrow}}{b} \ar@{~>}[u] & & \\
					 & & c \ar@{~>}[uull] \ar@{-}[ull] \ar@{-}[dll] \ar@{<~}[ddll] \\
					\conjunto{\mu_{i}^{\downarrow}}{a} \ar@{~>}[uu] & & \\
					\conjunto{\mu_{i}^{\triangleleft}}{} \ar@{~>}[u] & & \\
				}
				
			\end{array}
			$$			
			
			Devido a $s_{4}$, $s_{5}$, $s_{6}$ e propriedade do quadrado para $\downarrow$ ($s_{7}^{\downarrow}$) entre $\mu_{i}^{\downarrow}$, $\mu_{i}^{\triangleleft}$, $\nu_{i}^{\downarrow}$ e $c$, temos que necessariamente $\mu_{i}^{\triangleleft} \triangleleft \mu_{i}^{\downarrow}$. Analogamente, necessariamente temos que $\nu_{i}^{\downarrow} \triangleleft \nu_{i}^{\triangleright}$. A situação resultante, simplificada por transitividade, é mostrada na figura da direita. Se $\mu_{i}^{\triangleleft} \neq \emptyset$ então tome $\mu_{i+1} = \mu_{i}^{\triangleleft}$ e $\nu_{i+1} = \mu_{i}^{\downarrow} \cup \nu_{i}^{\downarrow} \cup \nu_{i}^{\triangleright} \cup \{c\}$: nesse caso $\mu_{i+1} \triangleleft \nu_{i+1}$. Se $\mu_{i}^{\triangleleft} = \emptyset $, então se $\nu_{i}^{\triangleright} \neq \emptyset$ tome $\mu_{i+1} = \mu_{i}^{\downarrow} \cup \nu_{i}^{\downarrow} \cup \{c\}$ e $\nu_{i+1} = \nu_{i}^{\triangleright}$: de novo $\mu_{i+1} \triangleleft \nu_{i+1}$. Se $\mu_{i}^{\triangleleft} = \emptyset$ e $\nu_{i}^{\triangleright} = \emptyset$, tome $\mu_{i+1} = \mu_{i}^{\downarrow} \cup \nu_{i}^{\downarrow}$ e $\nu_{i+1} = \{c\}$: nesse caso $\mu_{i+1} \downarrow \nu_{i+1}$.
					
			\item $\mu_{i} \downarrow \nu_{i}$ e existem $a$ em $\mu_{i}$ e $b$ em $\nu_{i}$ tais que $\neg{(a \downarrow c)}$ e $\neg{(b \downarrow c)}$: por um argumento análogo ao do caso 1, temos que essa situação, representada à esquerda no diagrama abaixo (onde \xy{\ar@{--} (0,2)*{a};(10,2)*{c}}\endxy \ \  representa $\neg{(a \downarrow c)}$), implica uma das três possibilidades que estão representadas abaixo:
			
			$$
			\begin{array}	{ccccc}
				
				& &
				
				\xymatrix{
				                      & c \ar@{--}[dl] \ar@{--}[dr] &                       \\
				\conjunto{\mu_{i}}{a} \ar@{-}[rr] &   & \conjunto{\nu_{i}}{b} \\
				}
				
				& &
				
				\\
			
				& &
				\implies
				& &
			
				\\
			
				\xymatrix{
				                      & c \ar@{<~}[dl] \ar@{<~}[dr] &                       \\
				\conjunto{\mu_{i}}{a} \ar@{-}[rr] &   & \conjunto{\nu_{i}}{b} \\
				}
				
				&
				\vee
				&
				
				\xymatrix{			                                 
					\conjunto{\mu_{i}}{a} \ar@{-}[rr] &   & \conjunto{\nu_{i}}{b} \\	  					
					 & c \ar@{~>}[ul] \ar@{~>}[ur] &                       \\	
				}
				
				&
				\vee
				&
			
				\xymatrix{
				                                  & c \ar@{~}[dl] \ar@{~}[dr] &                       \\
				\conjunto{\mu_{i}}{a} \ar@{-}[rr] &   & \conjunto{\nu_{i}}{b} \\	                                  
				}
				
			\end{array}
			$$
			
			Vamos considerar o primeiro caso, onde $a \triangleleft c$ e $b \triangleleft c$. Usando a propriedade triangular ($s_{6}$), podemos parcitionar $\mu_{i}$ em $\mu_{i}^{\downarrow}$ e $\mu_{i}^{\triangleleft}$ e $\nu_{i}$ em $\nu_{i}^{\downarrow}$ e $\nu_{i}^{\triangleleft}$ de tal forma que temos a situação representada à esquerda:
			
			$$
			\begin{array}{ccc}
			
				\xymatrix{
				 & c \ar@{<~}[dl] \ar@{<~}[dr] \ar@{-}[ddl] \ar@{-}[ddr] &                                     \\
				\conjunto{\mu_{i}^{\triangleleft}}{a} \ar@{-}[rr] \ar@{-}[drr] & & \conjunto{\nu_{i}^{\triangleleft}}{b} \\
				\conjunto{\mu_{i}^{\downarrow}}{} \ar@{-}[rr] \ar@{-}[urr] & & \conjunto{\nu_{i}^{\downarrow}}{} \\
				}
			
				&
				\implies
				&
				
				\xymatrix{
				 & c \ar@{<~}[dl] \ar@{<~}[dr] \ar@{-}[ddl] \ar@{-}[ddr] &                                     \\
				\conjunto{\mu_{i}^{\triangleleft}}{a} \ar@{-}[rr] \ar@{-}[drr]  \ar@{-}[d] & & \conjunto{\nu_{i}^{\triangleleft}}{b} \ar@{-}[d] \\
				\conjunto{\mu_{i}^{\downarrow}}{} \ar@{-}[rr] \ar@{-}[urr] & & \conjunto{\nu_{i}^{\downarrow}}{} \\
				}
				
			\end{array}
			$$
			
			A propriedade do quadrado forçosamente nos leva à situação da direita, onde agora podemos definir uma partição apropriada. Se $\mu_{i}^{\downarrow} \neq \emptyset$ então tome $\mu_{i+1} = \mu_{i}^{\downarrow}$ e $\nu_{i+1} = \mu_{i}^{\triangleleft} \cup \{c\} \cup \nu_{i}^{\triangleleft} \cup \nu_{i}^{\downarrow}$: nesse caso $\mu_{i+1} \downarrow \nu_{i+1}$. Se $\mu_{i}^{\downarrow} = \emptyset$, então se $\nu_{i}^{\downarrow} \neq \emptyset$ tome $\mu_{i+1} = \mu_{i}^{\triangleleft} \cup \{c\} \cup \nu_{i}^{\triangleleft}$ e $\nu_{i+1} = \nu_{i}^{\downarrow}$: de novo $\mu_{i+1} \downarrow \nu_{i+1}$. Se $\mu_{i}^{\downarrow} = \emptyset$ e $\nu_{i}^{\downarrow} = \emptyset$, tome $\mu_{i+1} = \mu_{i}^{\triangleleft} \cup \nu_{i}^{\triangleleft}$ e $\nu_{i+1} = \{c\}$: nesse caso $\mu_{i+1} \triangleleft \nu_{i+1}$. Os outros casos acima, nos quais $a \triangleright c$ e $b \triangleright c$, são tratados de forma análoga.
			
			\item $\mu_{i} \uparrow \nu_{i}$ e existe $a$ em $\mu_{i}$ e $b$ em $\nu_{i}$ tais que $\neg{(a \uparrow c)}$ e $\neg{(b \uparrow c)}$: esse caso é similar ao caso 2.
		
		\end{enumerate}
	
		O passo final do algoritmo acontece quando nenhuma ocorrência de átomos em $\zeta$ já pertence a uma partição.
	
	\end{proof}
	
\end{Theorem}	
	
	A essência da demonstração acima reside na ação combinada da propriedade triangular ($s_{6}$) e na propriedade do quadrado ($s_{7}$). A propriedade triangular reduz o problema a um caso onde apenas duas relações estruturais estão envolvidas, então a propriedade do quadrado é usada para decidir o lado do quadrado que ainda permanece por ser determinado.
	
	O algoritmo fornecido na demonstração acima é não-determinístico, logo ele não pode ser usado de maneira razoável para responder a seguinte pergunta: a cada rede de iteração corresponde uma única estrutura (módulo equivalência)? O seguinte teorema mostra que sim.
	
\begin{Lemma}

	\label{lema:particao}

	Dada uma estrutura $T$, se $(\zeta_{\mu},\zeta_{\nu})$ é uma partição-$\triangleleft$ de $\relweb{T}$ (respectivamente uma partição-$\downarrow$, uma partição-$\uparrow$) então existem duas estruturas $U$ e $V$ tais que $\relweb{U} = \zeta_{\mu}$, $\relweb{V} = \zeta_{\nu}$ e $T = \seqs{U;V}$ (respectivamente, $T = \pars{U,V}$, $T = \aprs{U,V}$).
	
	\begin{proof}
	
		Seja $\relweb{T} = (\occ{T}, \triangleleft, \downarrow, \uparrow)$. Sejam as redes candidatas $\zeta_{\mu} = (\mu, \triangleleft_{\mu}, \downarrow_{\mu}, \uparrow_{\mu})$ e $\zeta_{\nu} = (\nu, \triangleleft_{\nu}, \downarrow_{\nu}, \uparrow_{\nu})$ uma partição-$\triangleleft$ de $\relweb{T}$. Já que $\mu \neq \emptyset \neq \nu$, a estrutura $T$ está em um dos três seguintes casos:
		
		\begin{enumerate}
			
			\item $T = \seqs{T_{1},\hdots,T_{h}}$, onde $h>1$ e, para $1 \leq i \leq h$, temos que $T_{i} \neq \circ$ e $T_{i}$ não é um seq próprio. Então necessariamente temos que $\occ{T_{i}} \subseteq \mu$ ou $\occ{T_{i}} \subseteq \nu$, para todo $i$. Com efeito, suponha o contrário, e suponha que $T_{i} = \pars{T_{i}', T_{i}''}$ para algum $T_{i}'$ e algum $T_{i}''$ tais que $T_{i}' \neq \circ \neq T_{i}''$ (o mesmo argumento vale quando $T_{i} = \aprs{T_{i}', T_{i}''}$, nas mesmas condições). Então é possível achar $a$ em $T_{i}'$ e $b$ em $T_{i}''$, ou $a$ em $T_{i}''$ e $b$ em $T_{i}'$, tais que $a$ está em $\mu$ e $b$ está em $\nu$. Mas então $a \downarrow b$, o que viola a hipótese assumida. Então, para todo $i$, as cocorrências de átomos em $T_{i}$ vêm ou de $\mu$ ou de $\nu$, mas não de ambos. Logo necessariamente temos que há $k$ e $k+1$ em $1, \hdots, h$ tais que todas as ocorrências de átomos de $T_{1},\hdots,T_{k}$ estão em $\mu$ e todas as ocorrências de átomos de $T_{k+1},\hdots,T_{h}$ estão em $\nu$ (de outra forma haveria casos de $b \triangleleft a$ para algum $a$ em $\mu$ e algum $b$ em $\nu$). Então tome $U = \seqs{T_{1};\hdots;T_{k}}$ e $V = \seqs{T_{k+1},\hdots,T_{h}}$.
						
			\item $T = \pars{T',T''}$, onde $T' \neq \circ \neq T''$. Então necessariamente deve haver $a$ em $T'$ e $b$ em $T''$, ou $a$ em $T''$ e $b$ em $T'$, tais qeu $a$ está em $\mu$ e $b$ está em $\nu$. Mas então $a \downarrow b$, o que viola a hipótese. Logo, este caso é de fato impossível.
			
			\item $T = \aprs{T',T''}$, onde $T' \neq \circ \neq T''$. Este caso é análogo ao anterior.
		
		\end{enumerate}
		
		Dessa forma o lema é demonstraçãodo para qualquer partição-$\triangleleft$ de $\relweb{T}$. Um argumento análogo pode ser usado para partições-$\downarrow$ e partições-$\uparrow$.
	
	\end{proof}
	
\end{Lemma}	
	
\begin{Theorem}
	\label{theorem:equivalencia} 

	Duas estruturas são equivalentes se, e somente se, elas possuem a mesma teia de interação.

	\begin{proof}
	
	A demonstração do "somente se" \ é trivial. Vamos mostrar então a demonstração do "se". 
	
	Sejam $S$ e $T$ duas estruturas na forma normal. Temos que demonstrar que se $\relweb{S} = \relweb{T}$ então $S = T$. Vamos aplicar indução estrutural em $S$. Nos casos base $S=\circ$ ou $S=a$ temos trivialmente que $S=T$. Suponha então que existam $P$ e $Q$ tais que $S = \seqs{P;Q}$ e $P \neq \circ \neq Q$. O par $(\relweb{P}, \relweb{Q})$ é então uma partição-$\triangleleft$ de $\relweb{S}$, e portanto de $\relweb{T}$. Pelo lema~\ref{lema:particao}, existem $U$ e $V$ tais que $T = \seqs{U;V}$ e $\relweb{U} = \relweb{P}$ e $\relweb{V} = \relweb{Q}$, e então pode-se aplicar a hipótese de indução. Usam-se argumentos similares quando $S = \pars{P,Q}$ e $S = \aprs{P,Q}$, onde $P \neq \circ \neq Q$.
	
	\end{proof}

\end{Theorem}

\subsection{O Sistema \BV}
\label{section:sistema-bv}

O sistema \BV, ou sistema básico \V\footnote{\emph{Basic system \V}}, é uma extensão conservativa do fragmento multiplicativo da lógica linear acrescido da regra \mix\ (\MLL\ + \mix) e do operador não comutativo e auto-dual seq. Tal sistema foi um dos primeiros propostos dentro do cálculo das estruturas~\cite{guglielmi07}. É para um fragmento deste sistema (o \emph{flat} \BV, ou \FBV) que este trabalho propõe uma estratégia de demonstraçãos.

\begin{Definition}
Uma \emph{regra de inferência} é um esquema da forma 

$$\dernote{\rho}{}{R}{\leaf{T}}$$
onde $\rho$ é o \emph{nome} da regra, $T$ é sua \emph{premissa} e $R$ é sua \emph{conclusão}; nomes de regras são denotados por $\rho$ e $\pi$. Em uma regra de inferência, ou a premissa ou a conclusão podem ser vazias, mas não ambas. Quando premissa e conclusão de uma instância de uma regra de inferência são equivalentes, esta instância é dita \emph{trivial}, ou, caso contrário, é dita \emph{não-trivial}. Um \emph{sistema} (\emph{formal}) é um conjunto de regras de inferência; sistemas formais são denotados por $\mathscr{S}$. Uma \emph{derivação} em um certo sistema formal é uma seqüência de instâncias de regras de inferência do sistema, e pode consistir de uma única estrutura; derivações são denotadas por $\Delta$. A premissa da instância de regra de inferência mais acima na derivação, se presente, é chamada de \emph{premissa} da derivação; se presente, a conclusão da instância de regra mais abaixo na derivação é chamada de \emph{conclusão} da derivação; a premissa e a conclusão de derivações consistindo de uma única estrutura é a própria estrutura. Uma derivação $\Delta$ cuja premissa é $T$, cuja conclusão é $R$ e cujas regras de inferência estão em $\mathscr{S}$ é indicada por $\vcenter{\strder{\Delta}{\mathscr{S}}{R}{\leaf{T}}}$ (o nome $\Delta$ pode ser omitido). O \emph{comprimento} de uma derivação é o número de instâncias de regras de inferência que ela possui. Dois sistemas $\mathscr{S}$ e $\mathscr{S}'$ são \emph{fortemente equivalentes} se para toda derivação $\vcenter{\strder{}{\mathscr{S}}{R}{\leaf{T}}}$ existe uma derivação $\vcenter{\strder{}{\mathscr{S}'}{R}{\leaf{T}}}$, e vice-versa.

\end{Definition}

Existem, no cálculo de seqüentes, duas formas complementares de enxergar uma derivação, que podem ser adaptadas ao cálculo das estruturas:

\begin{enumerate}
	
	\item A visão de baixo para cima: premissas se unem (em árvores) para formar novas conclusões, e a derivação cresce no sentido de sua conclusão. Este pode ser chamado de ponto de vista \emph{dedutivo}.
	
	\item A visão de cima para baixo~\footnote{Em inglês: \emph{bottom-up}}: a conclusão é o ponto de partida e regras de inferência são usadas para atingir as premissas desejadas. Este pode ser chamado de ponto de vista de \emph{construção de demonstraçãos}.
	
\end{enumerate}

Depois de escolher regras de premissa única, existe uma situação de simetria: derivações são seqüências de inferências (simetria \emph{top-down}), par e copar são o mesmo tipo de estrutura com dois nomes diferentes e seq é auto-dual. 

A figura~\ref{fig:sistema-bv} apresenta as regras de inferência para o sistema \BV.

\begin{figure}[!htb]

	\center
	
	\begin{tabular}{|cccc|}
	
		\hline 
		
		& & & \\
	
		\dernote{\circ \downarrow}{}{\circ}{\leaf{}}
		
		&
	
		\dernote{\ruleaidown}{}{S\pars{a,\dual{a}}}{\leaf{S\{\circ\}}}
		
		&
		
		\dernote{\ruleqdown}{}{S\pars{\seqs{R;R'},\seqs{T;T'}}}{\leaf{S\seqs{\pars{R,T};\pars{R',T'}}}}
	
		&
		
		\dernote{\rules}{}{S\pars{\aprs{R,R'},T}}{\leaf{S\aprs{\pars{R,T},R'}}}
		
		\\ 
		
		& & &
		
		\\ \hline
	
	\end{tabular}
	
	\label{fig:sistema-bv}
	\caption{Sistema \BV}

\end{figure}

\begin{Definition}

	Os nomes das regras são:
	
	\begin{description}
		\item[$\circ \downarrow$] \emph{unidade (axioma lógico)};
		\item[$\ruleaidown$] \emph{interação atômica};
		\item[$\ruleqdown$] \emph{seq};
		\item[$\rules$] \emph{switch}.
	\end{description}

\end{Definition}

\begin{Definition}
	Uma \emph{demonstração} é uma derivação cuja regra mais acima seja o axioma lógico $\circ \downarrow$. demonstraçãos são denotadas por $\Pi$. Um sistema formal $\mathscr{S}$ \emph{demonstração} $R$ se existe em $\mathscr{S}$ uma demonstração $\Pi$ cuja conclusão seja $R$, escrita $\vcenter{\strpr{\Pi}{\mathscr{S}}{R}}$ (o nome $\Pi$ pode ser omitido). Dois sistemas são (\emph{fracamente}) \emph{equivalentes} se eles demonstraçãom as mesmas estruturas.
\end{Definition}

\begin{Example}

	demonstração de uma estrutura em \BV:
	
	$$
	\dernote{\ruleqdown}{}{\pars{\seqs{\pars{a,\dual{b}};c},\seqs{\aprs{\dual{a},b};\dual{c}}}}
{\rootnote{\ruleaidown}{}{\seqs{\pars{a,\dual{b},\aprs{\dual{a},b}};\pars{c,\dual{c}}}}
{\root{=}{\seqs{\pars{a,\dual{b},\aprs{\dual{a},b}};\circ}}
{\root{\rules}{\pars{a,\dual{b},\aprs{\dual{a},b}}}
{\root{\ruleaidown}{\pars{a,\aprs{\dual{a},\pars{\dual{b},b}}}}
{\root{=}{\pars{a,\aprs{\dual{a},\circ}}}
{\root{\ruleaidown}{\pars{a,\dual{a}}}
{\root{\ruleodown}{\circ}
{\leaf{}}}}}}}}}$$

\end{Example}

\begin{Definition}
	O sistema \emph{flat} \BV\ (\FBV) é o sistema \BV\  sem a regra $\ruleqdown$. As estruturas permitidas no \FBV\ são apenas estruturas \emph{flat} (sem a estrutura seq). 
\end{Definition}

A figura~\ref{fig:sistema-fbv} apresenta as regras de inferência para o sistema \FBV.

\begin{figure}[!htb]

	\center
	
	\begin{tabular}{|ccc|}
	
		\hline 
		
		& & \\
	
		\dernote{\circ \downarrow}{}{\circ}{\leaf{}}
		
		&
	
		\dernote{\ruleaidown}{}{S\pars{a,\dual{a}}}{\leaf{S\{\circ\}}}
			
		&
		
		\dernote{\rules}{}{S\pars{\aprs{R,R'},T}}{\leaf{S\aprs{\pars{R,T},R'}}}
		
		\\ 
		
		& & 
		
		\\ \hline
	
	\end{tabular}
	
	\label{fig:sistema-fbv}
	\caption{Sistema \FBV}

\end{figure}

\begin{Remark}
	O \FBV\ é um subsistema do sistema \BV, no sentido de que toda estrutura provável em \FBV\ é provável em \BV\ (mas a recíproca não é verdadeira). Mais do que isto, pode-se mostrar que o sistema \BV\ é uma extensão conservativa do sistema \FBV.
\end{Remark}

\begin{Example}

	demonstração de uma estrutura em \FBV:
	
	$$\dernote{\rules}{}{\pars{a,b,\aprs{\dual{a},\dual{b}}}}
{\rootnote{\ruleaidown}{}{\pars{\aprs{\pars{a,\dual{a}},b},\dual{b}}}
{\root{=}{\pars{\aprs{\circ,b},\dual{b}}}
{\root{\ruleaidown}{\pars{b,\dual{b}}}
{\root{\circ \downarrow}{\circ}
{\leaf{}}}}}}$$

\end{Example}

\subsection{O teorema de \emph{splitting}}
\label{section:splitting-principal}

O teorema de \emph{splitting} tem um papel muito importante neste trabalho por dois motivos:

\begin{enumerate}

	\item Ele é a chave para a demonstração de eliminação da regra Cut para o cálculo das estruturas para \MLL\ + \mix + seq (o sistema \BV);
	
	\item Sua idéia central foi a motivadora para a estratégia de demonstraçãos para o sistema \FBV\ com pares de átomos dois a dois distintos proposta neste trabalho.

\end{enumerate}

O segundo item, referente à motivação para a estratégia de demonstraçãos proposta, será retomado com detalhes mais adiante, na seção \ref{section:abordagem}. No momento vamos nos concentrar no primeiro item.

A eliminação da regra Cut é uma propriedade desejável em todo sistema lógico para o qual se deseja uma implementação eficiente. Em cálculo de seqüentes, a regra Cut pode ser escrita como:

$$
\qquad \frac{\Seq{\Gamma_1}{\Delta_1,A}\quad
      \Seq{A,\Gamma_2}{\Delta_2}}
     {\Seq{\Gamma_1,\Gamma_2}{\Delta_1,\Delta_2}}
            \  \hbox{\emph{Cut}}
$$

Vista de baixo para cima, esta regra precisa que uma fórmula $A$ seja ``criada'' do nada, o que exige uma criatividade que o computador não possui. Por isso, quando se foca na implementação, é desejável que a regra Cut possa ser eliminada. Os argumentos clássicos para se demonstrar a eliminação da regra Cut no cálculo de seqüentes se baseiam na propriedade de que, quando as fórmulas principais em uma instância da regra Cut estão ativas em ambas as ramificações, elas determinam quais regras são aplicadas imediatamente acima da regra Cut. Isto é uma conseqüência do fato de que estas fórmulas possuem um conectivo raiz, e regras lógicas são aplicáveis somente a este conectivo e em nenhum outro lugar da fórmula.

Entretanto, no cálculo das estruturas, esta propriedade não é válida. O sistema \BV\ compreende o fragmento multiplicativo da lógica linear (\MLL) mais a regra \mix\ e mais seq. Existe uma versão atômica da regra \emph{Cut} para \BV\ chamada $\ruleaiup$:

$$
\dernote{\ruleaiup}{}{S\{\circ\}}{\leaf{S\aprs{a,\dual{a}}}}
$$

Esta regra é simétrica à regra $\ruleaidown$ e diz que, de baixo para cima, pode-se criar dois átomos de polaridades opostas em \emph{qualquer lugar dentro da estrutura}. Isto torna o argumento para demonstrar a eliminação da regra Cut um tanto mais complicado, pois qualquer coisa pode acontecer acima de uma instância da regra Cut. Por outro lado, uma simplificação considerável pode ser feita porque no sistema \BV\ a regra $\ruleaiup$ cria apenas fórmulas atômicas. A dificuldade restante fica por parte de entender o que acontece, enquanto sobe-se em uma demonstração, ao redor dos átomos produzidos por um Cut atômico. Os dois átomos de um Cut atômico podem ser produzidos dentro de qualquer estrutura, e eles não pertencem a ramos distintos como no cálculo de seqüentes.

Uma abordagem possível para a demonstração da eliminação da regra Cut utiliza o teorema de \emph{splitting}~\cite{guglielmi07}. É importante observar que o sistema \BV\ \emph{não possui a regra Cut como primitiva}. A eliminação da regra Cut a que nos referimos aqui para o sistema \MLL\ + \mix\ + seq significa que:

\emph{Podemos acrescentar a regra Cut $\ruleaiup$ ao sistema \BV\ sem aumentar o conjunto de fórmulas provávies em \MLL\ + \mix + seq. Ou seja, a regra Cut é admissível em \BV.}

Em outras palavras, o sistema \BV\ é completo para \MLL\ + \mix\ + seq sem ter a regra Cut, o que torna sua implementação muito mais prática.

\subsubsection{Teorema de \emph{splitting}}
\label{section:splitting}

A idéia de \emph{splitting} pode ser entendida considerando um sistema de seqüentes  em que não há regras de \emph{weakening} e \emph{contraction}. Considere o exemplo da lógica linear multiplicativa. Se temos uma demonstração do seqüente:

$$ \Seq{}{F\{A \tensor B\}} $$
onde $F\{A \tensor B\}$ é uma fórmula que contém a subfórmula $A \tensor B$, sabemos com certeza que em algum lugar na demonstração existe uma, e apenas uma, instância da regra $\tensor$ que separa $A$ e $B$ juntamente com seu contexto. Temos então a seguinte situação:

\begin{center}

$\strder{\Delta}{}{\Seq{}{F\{A \tensor B\}}}{\leaf{\ddernote{\tensor}{}{\Seq{}{A\tensor B, \Phi, \Psi}}{\leaf{\strpr{\Pi_{1}}{}{\Seq{}{A,\Phi}}}}{\leaf{\strpr{\Pi_{2}}{}{\Seq{}{B,\Psi}}}}}}$ equivale a $\strder{\Delta}{\quad}{\pars{F\aprs{A,B},\Gamma}}{\leaf{\dernote{\rules}{\quad}{\pars{\aprs{A,B},\Phi,\Psi}}{\leaf{\dernote{\rules}{\quad}{\pars{\aprs{\pars{A,\Phi},B},\Psi}}{\leaf{\strder{\Pi_{1}}{\quad}{\aprs{\pars{A,\Phi},\pars{B,\Psi}}}{\leaf{\strpr{\Pi_{2}}{\quad}{\pars{B,\Psi}}}}}}}}}}$

\end{center}

Podemos considerar, como mostrado à esquerda, uma demonstração para o seqüente dado como composta de três partes, $\Delta$, $\Pi_{1}$ e $\Pi_{2}$. No cálculo das estruturas existem muitas demonstraçãos diferentes correspondentes à mesma demonstração em cálculo de seqüentes: elas diferem apenas na possível ordenação das regras. Regras no cálculo das estruturas têm menor granularidade e maior aplicabilidade. Mas, entre todas as demonstraçãos, deve haver uma que se encaixe no esquema da demonstração à direita da figura acima. Este exemplo ilustra precisamente a idéia por trás da técnica de \emph{splitting}.

A derivação $\Delta$ acima implementa uma instância de \emph{context reduction}~\footnote{Como veremos adiante, \emph{context reduction} é um lema necessário para o teorema de \emph{splitting}.} e um \emph{splitting} próprio. De forma geral, podemos enunciar estes princípios da seguinte forma:

\begin{enumerate}

	\item \emph{Context reduction}: Se $S\{R\}$ é provável, então $S\{ \ \}$ pode ser reduzido, de baixo para cima em uma demonstração, à estrutura $\pars{\{ \ \}, U}$, tal que $\pars{R,U}$ é provável. No exemplo acima, $\pars{F\{ \ \},\Gamma}$ é reduzido a $\pars{\{\ \}, \Gamma'}$ para algum $\Gamma'$.
	
	\item (\emph{Shallow}) \emph{splitting}: Se $\pars{\aprs{R,T},P}$ é provável, então $P$ pode ser reduzido, de baixo para cima em uma demonstração, a $\pars{P_{1},P_{2}}$, tal que $\pars{R,P_{1}}$ e $\pars{T,P_{2}}$ são prováveis. No exemplo acima, $\Gamma'$ é reduzido a $\pars{\Phi, \Psi}$.

\end{enumerate}

O conceito de \emph{context reduction} é demonstraçãodo usando o conceito de (\emph{shallow}) \emph{splitting}. Antes de prová-los, vamos introduzir duas proposições simples que serão utilizadas implicitamente ao longo da demonstração.

\begin{Proposition}
O tamanho da premissa de uma derivação em \BV\ não é maior que o tamanho da sua conclusão.
\end{Proposition}

\begin{Proposition}
Em \BV, $\seqs{R;T}$ é provável se, e somente se, $R$ e $T$ são prováveis. Da mesma forma, $\aprs{R,T}$ é provável se, e somente se, $R$ e $T$ são prováveis.
\end{Proposition}

Como o cálculo das estruturas pode ser visto como um sistema de reescrita, podemos introduzir uma terminologia análoga à destes sistemas.

\begin{Definition}
As inferências do sistema \BV\ são todas da forma $\vcenter{\dernote{\rho}{}{S\{U\}}{\leaf{S\{V\}}}}$, onde a estrutura $U$ é chamada de \emph{redex} e $V$ de \emph{contractum} da instância da regra.
\end{Definition}

\begin{Theorem}[Shallow Splitting]
\label{teorema:splitting}

	Para todas as estruturas $R$, $T$ e $P$:
	
	\begin{enumerate}
		
		\item Se $\pars{\seqs{R;T},P}$ é provável em \BV, então existem $P_{1}$, $P_{2}$ e $\vcenter{\strder{}{\BV}{P}{\leaf{\seqs{P_{1};P_{2}}}}}$ tais que $\pars{R,P_{1}}$ e $\pars{T,P_{2}}$ são prováveis em \BV.
		
		\item Se $\pars{\aprs{R,T},P}$ é provável em \BV, então existem $P_{1}$, $P_{2}$ e $\vcenter{\strder{}{\BV}{P}{\leaf{\pars{P_{1},P_{2}}}}}$ tais que $\pars{R,P_{1}}$ e $\pars{T,P_{2}}$ são prováveis em \BV.
		
	\end{enumerate}
	
	\begin{proof}
	
		Todas as derivações da prova são feitas em \BV. Vamos considerar a ordem lexicográfica $\prec$ sobre os números naturais definida como $(m', n') \prec (m,n)$ se, e somente se, ou $m' < m$, ou $m'=m$ e $n'<n$. Considere as seguintes proposições:
		
		$
		\begin{array}{llll}
		S(m,n) = \forall m', n. \forall R,T,P. & ( & (        &  (m',n') \preceq (m,n) \\
																					&   &          & \ \ \ \ \ e \ \ \ \ \ \ m'=|\occ{\pars{\seqs{R;T},P}}| \\
																					&   &          & \ \ \ \ \ e \ \ \ \ \ \ \exists  \vcenter{\strpr{}{}{\pars{\seqs{R;T},P}}} de \ comprimento \ n' ) \\
																					&		& \implies &  \ \exists P_{1}, P_{2}. (	\vcenter{\strder{}{}{P}{\leaf{\seqs{P_{1};P_{2}}}}} \ \ \ \ \ e \ \ \ \ \ \ 
\vcenter{\strpr{}{}{\pars{R,P_{1}}}} \ \ \ \ \ e \ \ \ \ \ \
\vcenter{\strpr{}{}{\pars{T,P_{2}}}}	) ) \\
		\end{array}
		$
		
		$
		\begin{array}{llll}
		C(m,n) = \forall m', n. \forall R,T,P. & ( & (        &  (m',n') \preceq (m,n) \\
																					&   &          & \ \ \ \ \ e \ \ \ \ \ \ m'=|\occ{\pars{\aprs{R,T},P}}| \\
																					&   &          & \ \ \ \ \ e \ \ \ \ \ \ \exists  \vcenter{\strpr{}{}{\pars{\aprs{R,T},P}}} de \ comprimento \ n' ) \\
																					&		& \implies &  \ \exists P_{1}, P_{2}. (	\vcenter{\strder{}{}{P}{\leaf{\pars{P_{1},P_{2}}}}} \ \ \ \ \ e \ \ \ \ \ \ 
\vcenter{\strpr{}{}{\pars{R,P_{1}}}} \ \ \ \ \ e \ \ \ \ \ \
\vcenter{\strpr{}{}{\pars{T,P_{2}}}}	) ) \\
		\end{array}
		$
		
		O enunciado do teorema é equivalente a 
		
		$$ \forall m,n.(S(m,n) \ \ \ \ \ e \ \ \ \ \ C(m,n))$$
		
		Podemos considerar $(m,n)$ como uma medida de $(S(m,n)\ \ \ \ \ e \ \ \ \ \ C(m,n))$, e a prova é uma indução, por $\prec$, sobre esta medida.
		
		\begin{itemize}
			
			\item Casos Base
										
				\begin{enumerate}
					
					\item $R=\circ$
						
						$$ \vcenter{ \strpr{}{}{\pars{\seqs{\circ;T},P}} } \ \ \ \ \ equivale \ a \ \ \ \ \ \vcenter{\strder{}{}{P}{\leaf{\seqs{\circ;P}}}} \ \ \ \ \ e \ \ \ \ \ \vcenter{\strpr{}{}{\pars{\circ,\circ}}} \ \ \ \ \ e \ \ \ \ \ \vcenter{\strpr{}{}{\pars{T,P}}} \ \ \ \ \ equivale \ a \ \ \ \ \ \vcenter{\strpr{}{}{\pars{T,P}}}  $$
						
						$$ \vcenter{ \strpr{}{}{\pars{\aprs{\circ,T},P}} } \ \ \ \ \ equivale \ a \ \ \ \ \ \vcenter{\strder{}{}{P}{\leaf{\pars{\circ,P}}}} \ \ \ \ \ e \ \ \ \ \ \vcenter{\strpr{}{}{\pars{\circ,\circ}}} \ \ \ \ \ e \ \ \ \ \ \vcenter{\strpr{}{}{\pars{T,P}}} \ \ \ \ \ equivale \ a \ \ \ \ \ \vcenter{\strpr{}{}{\pars{T,P}}}  $$
						
					\item $T=\circ$
					
						$$ \vcenter{ \strpr{}{}{\pars{\seqs{R;\circ},P}} } \ \ \ \ \ equivale \ a \ \ \ \ \ \vcenter{\strder{}{}{P}{\leaf{\seqs{P;\circ}}}} \ \ \ \ \ e \ \ \ \ \ \vcenter{\strpr{}{}{\pars{R,P}}} \ \ \ \ \ e \ \ \ \ \ \vcenter{\strpr{}{}{\pars{\circ,\circ}}} \ \ \ \ \ equivale \ a \ \ \ \ \ \vcenter{\strpr{}{}{\pars{R,P}}}  $$
						
						$$ \vcenter{ \strpr{}{}{\pars{\aprs{R,\circ},P}} } \ \ \ \ \ equivale \ a \ \ \ \ \ \vcenter{\strder{}{}{P}{\leaf{\pars{P,\circ}}}} \ \ \ \ \ e \ \ \ \ \ \vcenter{\strpr{}{}{\pars{R,P}}} \ \ \ \ \ e \ \ \ \ \ \vcenter{\strpr{}{}{\pars{\circ,\circ}}} \ \ \ \ \ equivale \ a \ \ \ \ \ \vcenter{\strpr{}{}{\pars{R,P}}}  $$
					
					\item $P=\circ$
					
						$$  \vcenter{ \strpr{}{}{\pars{\seqs{R;T},\circ}} } \ \ \ \ \ equivale \ a \ \ \ \ \ \vcenter{\strder{}{}{\circ}{\leaf{\seqs{\circ;\circ}}}} \ \ \ \ \ e \ \ \ \ \ \vcenter{\strpr{}{}{\pars{R,\circ}}} \ \ \ \ \ e \ \ \ \ \  \vcenter{\strpr{}{}{\pars{T,\circ}}} \ \ \ \ \ equivale \ a \ \ \ \ \ \vcenter{\strpr{}{}{R}} \ \ \ \ \ e \ \ \ \ \ \vcenter{\strpr{}{}{T}} $$
						
						$$  \vcenter{ \strpr{}{}{\pars{\aprs{R,T},\circ}} } \ \ \ \ \ equivale \ a \ \ \ \ \ \vcenter{\strder{}{}{\circ}{\leaf{\pars{\circ,\circ}}}} \ \ \ \ \ e \ \ \ \ \ \vcenter{\strpr{}{}{\pars{R,\circ}}} \ \ \ \ \ e \ \ \ \ \  \vcenter{\strpr{}{}{\pars{T,\circ}}} \ \ \ \ \ equivale \ a \ \ \ \ \ \vcenter{\strpr{}{}{R}} \ \ \ \ \ e \ \ \ \ \ \vcenter{\strpr{}{}{T}} $$
		
				\end{enumerate}

			\item Casos Indutivos
			
				\begin{enumerate}
				
					\item $\forall m', n' . ( (m',n') \prec (m,n) \ \ \ \ \ e \ \ \ \ \ S(m',n') \ \ \ \ \ e \ \ \ \ \ C(m',n')) \implies S(m,n)$
					O tamanho de $\pars{\seqs{R;T},P}$ é $m$ e existe uma prova desta estrutura de tamanho $n$. Vamos considerar a instância de regra de derivação mais abaixo na prova:
					
					$$
						\dernote{\rho}{\quad}{\pars{\seqs{R;T},P}}{\leaf{\strpr{}{}{Q}}}
					$$
					
					Onde assumimos que $\rho$ é não-trivial ($\rho \ \neq \ \ \ \ \ \ equivale \ a \ \ \ \ \ $), pois de outra forma a hipótese de indução se aplicaria. Vamos argumentar baseados na posição do redex de $\rho$ em $\pars{\seqs{R;T},P}$. As possibilidades são:
					
					\begin{enumerate}
					
						\item $\rho  =  \ruleaidown $
						
							\begin{enumerate}
							
								\item O redex está em $R$
								
$$\dernote{\ruleaidown}{\quad}{\pars{\seqs{R;T},P}}{\leaf{\strpr{}{}{\pars{\seqs{R';T},P}}}}  \ \ \ \ \ equivale \ a \ \ \ \ \   \strder{}{}{P}{\leaf{\seqs{P_{1};P_{2}}}} \ \ \ \ \ e \ \ \ \ \ \dernote{\ruleaidown}{}{\pars{R,P_{1}}}{\leaf{\strpr{}{}{\pars{R',P_{1}}}}} \ \ \ \ \ e \ \ \ \ \ \strpr{}{}{\pars{T,P_{2}}} $$																	
							
								\item O redex está em $T$
							$$\dernote{\ruleaidown}{\quad}{\pars{\seqs{R;T},P}}{\leaf{\strpr{}{}{\pars{\seqs{R;T'},P}}}}  \ \ \ \ \ equivale \ a \ \ \ \ \   \strder{}{}{P}{\leaf{\seqs{P_{1};P_{2}}}} \ \ \ \ \ e \ \ \ \ \ \strpr{}{}{\pars{R,P_{1}}} \ \ \ \ \ e \ \ \ \ \							 \dernote{\ruleaidown}{}{\pars{T,P_{2}}}{\leaf{\strpr{}{}{\pars{T',P_{2}}}}}  $$	
							
								\item O redex está em $P$
									
$$\dernote{\ruleaidown}{\quad}{\pars{\seqs{R;T},P}}{\leaf{\strpr{}{}{\pars{\seqs{R;T},P'}}}}  \ \ \ \ \ equivale \ a \ \ \ \ \   \dernote{\ruleaidown}{\quad}{P}{\leaf{\strder{}{}{P'}{\leaf{\seqs{P_{1};P_{2}}}}}} \ \ \ \ \ e \ \ \ \ \ \strpr{}{}{\pars{R,P_{1}}} \ \ \ \ \ e \ \ \ \ \	\strpr{}{}{\pars{T,P_{2}}}						   $$																			
							
							\end{enumerate}																			
						
						\item $\rho  =  \ruleqdown $
						
							\begin{enumerate}
							
									\item O redex está em $R$
								
$$\dernote{\ruleqdown}{\quad}{\pars{\seqs{R;T},P}}{\leaf{\strpr{}{}{\pars{\seqs{R';T},P}}}}  \ \ \ \ \ equivale \ a \ \ \ \ \   \strder{}{}{P}{\leaf{\seqs{P_{1};P_{2}}}} \ \ \ \ \ e \ \ \ \ \ \dernote{\ruleqdown}{}{\pars{R,P_{1}}}{\leaf{\strpr{}{}{\pars{R',P_{1}}}}} \ \ \ \ \ e \ \ \ \ \ \strpr{}{}{\pars{T,P_{2}}} $$																	
							
								\item O redex está em $T$
							$$\dernote{\ruleqdown}{\quad}{\pars{\seqs{R;T},P}}{\leaf{\strpr{}{}{\pars{\seqs{R;T'},P}}}}  \ \ \ \ \ equivale \ a \ \ \ \ \   \strder{}{}{P}{\leaf{\seqs{P_{1};P_{2}}}} \ \ \ \ \ e \ \ \ \ \ \strpr{}{}{\pars{R,P_{1}}} \ \ \ \ \ e \ \ \ \ \							 \dernote{\ruleqdown}{}{\pars{T,P_{2}}}{\leaf{\strpr{}{}{\pars{T',P_{2}}}}}  $$	
							
								\item O redex está em $P$
									
$$\dernote{\ruleqdown}{\quad}{\pars{\seqs{R;T},P}}{\leaf{\strpr{}{}{\pars{\seqs{R;T},P'}}}}  \ \ \ \ \ equivale \ a \ \ \ \ \   \dernote{\ruleqdown}{\quad}{P}{\leaf{\strder{}{}{P'}{\leaf{\seqs{P_{1};P_{2}}}}}} \ \ \ \ \ e \ \ \ \ \ \strpr{}{}{\pars{R,P_{1}}} \ \ \ \ \ e \ \ \ \ \	\strpr{}{}{\pars{T,P_{2}}}						   $$	
									
								\item $R = \seqs{R';R''}$, $P = \pars{\seqs{P';P''},U}$ e
								 $$\vcenter{\dernote{\ruleqdown}{\quad}{\pars{\seqs{R';R'';T},\seqs{P';P''},U}}{\leaf{\strpr{}{}{\pars{\seqs{\pars{R',P'};\pars{\seqs{R'';T},P''}},U}}}}}$$
									
									Queremos mostrar que existem $\Pi_{1}$, $\Pi_{2}$ e $\Pi_{3}$ tais que:
									
									$$\strpr{}{}{\pars{\seqs{R;T},P}} \ \ \ \ \ equivale \ a \ \ \ \ \ \strder{\Pi_{1}}{}{P}{\leaf{\seqs{P_{1};P_{2}}}} \ \ \ \ \ e \ \ \ \ \ \strpr{\Pi_{2}}{}{\pars{R,P_{1}}} \ \ \ \ \ e \ \ \ \ \ \strpr{\Pi_{3}}{}{\pars{T,P_{2}}}$$
									
									Ou seja, queremos mostrar que:
									
									$$ \strpr{}{}{\pars{\seqs{R';R'';T},\seqs{P';P''},U}} \ \ \ \ \ equivale \ a \ \ \ \ \ \strder{\Pi_{1}}{}{\pars{\seqs{P';P''},U}}{\leaf{\seqs{P_{1};P_{2}}}} \ \ \ \ \ e \ \ \ \ \ \strpr{\Pi_{2}}{}{\pars{\seqs{R';R''},P_{1}}} \ \ \ \ \ e \ \ \ \ \ \strpr{\Pi_{3}}{}{\pars{T,P_{2}}} $$	
									
									Uma vez que $|\occ{\pars{\seqs{\pars{R',P'};\pars{\seqs{R'';T},P''}},U}}| = |\occ{\pars{\seqs{R';R'';T},\seqs{P';P''},U}}|$	mas $n'<n$, podemos aplicar a hipótese de indução e obter:
									
									$$ \strder{\Pi_{4}}{}{U}{\leaf{\seqs{U_{1};U_{2}}}} \ \ \ \ \ e \ \ \ \ \ \strpr{\Pi_{5}}{}{\pars{R',P', U_{1}}} \ \ \ \ \ e \ \ \ \ \ \strpr{\Pi_{6}}{}{\pars{\seqs{R'';T},P'',U_{2}}} $$
									
									Uma vez que $|\occ{\pars{\seqs{R'';T},P'',U_{2}}}| < |\occ{\pars{\seqs{R';R'';T},\seqs{P';P''},U}}|$, podemos aplicar a hipótese de indução e obter:
									
									$$ \strder{\Pi_{7}}{}{\pars{P'',U_{2}}}{\leaf{\seqs{P_{1}';P_{2}}}} \ \ \ \ \ e \ \ \ \ \ \strpr{\Pi_{8}}{}{\pars{R'',P_{1}'}} \ \ \ \ \ e \ \ \ \ \ \strpr{\Pi_{9}}{}{\pars{T,U_{2}}} $$
									
									Agora podemos fazer $P_{1} = \seqs{\pars{p',U_{1}'};P_{1}'}$ e construir $\Pi{1}$, $\Pi{2}$ e $\Pi{3}$ da seguinte forma:
									
									\ \newline
									
									\begin{center}
									\center{
									\begin{tabular}{cc}
										
										$\left . 
										\begin{array}{c} \dernote{=}{\quad}{P}{\leaf{\strder{\Pi_{4}}{}{\pars{\seqs{P';P''},U}}{\leaf{\dernote{\ruleqdown}{\quad}{\pars{\seqs{P';P''},\seqs{U_{1};U_{2}}}}{\leaf{\strder{\Pi_{7}}{}{\seqs{\pars{P',U_{1}};\pars{P'',U_{2}}}}{\leaf{\dernote{=}{\quad}{\seqs{\pars{P',U_{1}};P_{1}';P_{2}}}{\leaf{\seqs{P_{1};P_{2}}}}}}}}}}}}
										\end{array}
										\right\} $ & $\Pi_{1}$.
									
									\end{tabular}
									}
									\end{center}
									
									\ \newline
																		
									\begin{center}
									\center{
									\begin{tabular}{cc}
										
										$ \left .
										\begin{array}{c} 	\dernote{=}{\quad}{\pars{R,P_{1}}}{\leaf{\dernote{\ruleqdown}{\quad}{\pars{\seqs{R';R''},\seqs{\pars{P',U_{1}};P_{1}'}}}{\leaf{\strder{\Pi_{8}}{\quad}{\seqs{\pars{R',P',U_{1}};\pars{R'',P_{1}'}}}{\leaf{\strpr{\Pi_{5}}{\quad}{\pars{R',P',U_{1}}}}}}}}}
										\end{array}
										\right\} $ & $\Pi_{2}$.
									
									\end{tabular}
									}			
									\end{center}
									
									\ \newline																																				
									
									\begin{center}
									\center{
									\begin{tabular}{cc}
										
										$ \left .
										\begin{array}{c} 	
										
										\strpr{\Pi_{9}}{\quad}{\pars{T,P_{2}}}
										
										\end{array}
										\right\} $ & $\Pi_{3}$.
									
									\end{tabular}
									}							
									\end{center}
									
									\ \newline												
									
								\item $P = \pars{\seqs{P';P''},U',U''}$ e
								
									$$ \dernote{\ruleqdown}{\quad}{\pars{\seqs{R;T},\seqs{P';P''},U',U''}}{\leaf{\strpr{}{}{\pars{\seqs{\pars{\seqs{R;T},P',U'};P''},U''}}}} $$
									
%
									
									Queremos mostrar que existem $\Pi_{1}$, $\Pi_{2}$ e $\Pi_{3}$ tais que:
																												
									$$ \strpr{}{}{\pars{\seqs{R;T},P}} \ \ \ \ \ equivale \ a \ \ \ \ \ \strder{\Pi_{1}}{}{\pars{\seqs{P';P''},U',U''}}{\leaf{\seqs{P_{1},P_{2}}}} \ \ \ \ \ e \ \ \ \ \ \strpr{\Pi_{2}}{}{\pars{R,P_{1}}} \ \ \ \ \ e \ \ \ \ \ \strpr{\Pi_{3}}{}{\pars{T,P_{2}}} $$														
									
									Uma vez que $|\occ{\pars{\seqs{\pars{\seqs{R;T},P',U'};P''},U''}}| = |\occ{\pars{\seqs{R;T},\seqs{P';P''},U',U''}}|$	mas $n'<n$, podemos aplicar a hipótese de indução e obter:
									
									$$ \strder{\Pi_{4}}{}{U''}{\leaf{\seqs{U_{1};U_{2}}}} \ \ \ \ \ e \ \ \ \ \ \strpr{\Pi_{5}}{}{\pars{\seqs{R;T},P',U',U_{1}}} \ \ \ \ \ e \ \ \ \ \ \strpr{\Pi_{6}}{}{\pars{P'',U_{2}}} $$
									
									Uma vez que $|\occ{\pars{\seqs{R;T},P',U',U_{1}}}| < |\occ{\pars{\seqs{R;T},\seqs{P';P''},U',U''}}|$, podemos aplicar a hipótese de indução e obter:
									
									$$ \strder{\Pi_{7}}{\quad}{\pars{P',U',U_{1}}}{\leaf{\seqs{P_{1}';P_{2}'}}} \ \ \ \ \ equivale \ a \ \ \ \ \ \strpr{\Pi_{8}}{}{\pars{R,P_{1}'}} \ \ \ \ \ e \ \ \ \ \ \strpr{\Pi_{9}}{}{\pars{T,P_{2}'}} $$
									
									Agora podemos fazer $P_{1} = P_{1}'$ e $P_{2} = P_{2}'$ e construir $\Pi_{1}$, $\Pi_{2}$ e $\Pi_{3}$ da seguinte forma:
									
									\ \newline
									
									\center{
									\begin{tabular}{cc}
										
										$ \left .
										\begin{array}{c} 	
										
										\dernote{=}{\quad}{P}{\leaf{\strder{\Pi_{4}}{}{\pars{\seqs{P';P''},U',U''}}{\leaf{\dernote{\ruleqdown}{\quad}{\pars{\seqs{P';P''},U',\seqs{U_{1};U_{2}}}}{\leaf{\strder{\Pi_{6}}{\quad}{\pars{\seqs{\pars{P',U_{1}};\pars{P'',U_{2}}},U'}}{\leaf{\strder{\Pi_{7}}{\quad}{\pars{P',U_{1},U'}}{\leaf{\dernote{=}{\quad}{\seqs{P_{1}';P_{2}'}}{\leaf{\seqs{P_{1};P_{2}}}}}}}}}}}}}}
										
										\end{array}
										\right\} $ & $\Pi_{1}$.
									
									\end{tabular}
									}
									
									\ \newline
																		
									\center{
									\begin{tabular}{cc}
										
										$ \left .
										\begin{array}{c} 	
										
										\dernote{=}{\quad}{\pars{R,P_{1}}}{\leaf{\strpr{\Pi_{8}}{}{\pars{R,P_{1}'}}}}
										
										\end{array}
										\right\} $ & $\Pi_{2}$.
									
									\end{tabular}
									}
									
									\ \newline
									
									\center{
									\begin{tabular}{cc}
										
										$ \left .
										\begin{array}{c} 	
										
										\dernote{=}{\quad}{\pars{T,P_{2}}}{\leaf{\strpr{\Pi_{9}}{}{\pars{T,P_{2}'}}}}
										
										\end{array}
										\right\} $ & $\Pi_{3}$.
									
									\end{tabular}
									}					
									
									\ \newline																																																							
							\end{enumerate}		
						
						\item $\rho = \rules $
						
							\begin{enumerate}
							
								\item O redex está em $R$
							
								$$\dernote{\rules}{\quad}{\pars{\seqs{R;T},P}}{\leaf{\strpr{}{}{\pars{\seqs{R';T},P}}}}  \ \ \ \ \ equivale \ a \ \ \ \ \   \strder{}{}{P}{\leaf{\seqs{P_{1};P_{2}}}} \ \ \ \ \ e \ \ \ \ \ \dernote{\rules}{}{\pars{R,P_{1}}}{\leaf{\strpr{}{}{\pars{R',P_{1}}}}} \ \ \ \ \ e \ \ \ \ \ \strpr{}{}{\pars{T,P_{2}}} $$																	
							
								\item O redex está em $T$
							
								$$\dernote{\rules}{\quad}{\pars{\seqs{R;T},P}}{\leaf{\strpr{}{}{\pars{\seqs{R;T'},P}}}}  \ \ \ \ \ equivale \ a \ \ \ \ \   \strder{}{}{P}{\leaf{\seqs{P_{1};P_{2}}}} \ \ \ \ \ e \ \ \ \ \ \strpr{}{}{\pars{R,P_{1}}} \ \ \ \ \ e \ \ \ \ \							 \dernote{\rules}{}{\pars{T,P_{2}}}{\leaf{\strpr{}{}{\pars{T',P_{2}}}}}  $$	
							
								\item O redex está em $P$
									
$$\dernote{\rules}{\quad}{\pars{\seqs{R;T},P}}{\leaf{\strpr{}{}{\pars{\seqs{R;T},P'}}}}  \ \ \ \ \ equivale \ a \ \ \ \ \   \dernote{\rules}{\quad}{P}{\leaf{\strder{}{}{P'}{\leaf{\seqs{P_{1};P_{2}}}}}} \ \ \ \ \ e \ \ \ \ \ \strpr{}{}{\pars{R,P_{1}}} \ \ \ \ \ e \ \ \ \ \	\strpr{}{}{\pars{T,P_{2}}}						   $$	
																
								\item $P = \pars{\aprs{P';P''},U',U''}$ e
								
									$$ \dernote{\rules}{\quad}{\pars{\seqs{R;T},\aprs{P',P''},U',U''}}{\leaf{\strpr{}{}{\pars{\aprs{\pars{\seqs{R;T},P',U'},P''},U''}}}} $$
									
									Queremos mostrar que existem $\Pi_{1}$, $\Pi_{2}$ e $\Pi_{3}$ tais que:
									
									$$ \strpr{}{\quad}{\pars{\seqs{R;T},P}} \ \ \ \ \ equivale \ a \ \ \ \ \ \strder{\Pi_{1}}{\quad}{\pars{\aprs{P',P''},U',U''}}{\leaf{\seqs{P_{1};P_{2}}}} \ \ \ \ \ e \ \ \ \ \ \strpr{\Pi_{2}}{}{\pars{R,P_{1}}} \ \ \ \ \ e \ \ \ \ \ \strpr{\Pi_{3}}{\quad}{\pars{T,P_{2}}} $$
									
									Uma vez que $|\occ{\pars{\seqs{R;T},\aprs{P',P''},U',U''}}|	= |\occ{\pars{\aprs{\pars{\seqs{R;T},P',U'},P''},U''}}|$ mas $n'<n$, podemos aplicar a hipótese de indução e obter:
									
									$$ \strder{\Pi_{4}}{\quad}{U''}{\leaf{\pars{U_{1},U_{2}}}} \ \ \ \ \ e \ \ \ \ \ \strpr{\Pi_{5}}{}{\pars{\seqs{R;T},P',U',U_{1}}} \ \ \ \ \ e \ \ \ \ \ \strpr{\Pi_{6}}{}{\pars{P'',U_{2}}} $$
									
									Uma vez que $|\occ{\pars{\seqs{R;T},P',U',U_{1}}}| < |\occ{\pars{\seqs{R;T},\aprs{P',P''},U',U''}}|$, podemos aplicar a hipótese de indução e obter:
								
									$$ \strder{\Pi_{7}}{\quad}{\pars{P',U',U_{1}}}{\leaf{\seqs{P_{1}';P_{2}'}}} \ \ \ \ \ e \ \ \ \ \ \strpr{\Pi{8}}{\quad}{\pars{R,P_{1}'}} \ \ \ \ \ e \ \ \ \ \  \strpr{\Pi{9}}{\quad}{\pars{T,P_{2}'}} $$
									
									Agora podemos fazer $P_{1} = P_{1}'$ e $P_{2} = P_{2}'$ e construir $\Pi_{1}$, $\Pi_{2}$ e $\Pi_{3}$ da seguinte forma:
									
									\ \newline
									
									\center{
									\begin{tabular}{cc}
										
										$\left .
										\begin{array}{c}
										
											$ \dernote{=}{\quad}{P}{\leaf{\strder{\Pi_{7}}{\quad}{\pars{\aprs{P',P''},U',U''}}{\leaf{\dernote{\rules}{\quad}{\pars{\aprs{P',P''},U',U_{1},U_{2}}}{\leaf{\dernote{\rules}{\quad}{\pars{\aprs{\pars{P',U',U_{1}},P''},U_{2}}}{\leaf{\strder{\Pi_{6}}{\quad}{\aprs{\pars{P'',U_{2}},\pars{P',U',P_{1}}}}{\leaf{\dernote{=}{\quad}{\strder{\Pi_{7}}{\quad}{\pars{P',U',P_{1}}}{\leaf{\seqs{P_{1}';P_{2}'}}}}{\leaf{\seqs{P_{1};P_{2}}}}}}}}}}}}}} $
										
										\end{array}
										\right\}
										$ & $\Pi_{1}$
									
									\end{tabular}
									}
									
									\ \newline
									
									\center{
									\begin{tabular}{cc}									
										
										$\left .
										\begin{array}{c}
										
											$\dernote{=}{\quad}{\pars{R,P_{1}}}{\leaf{\strpr{\Pi_{8}}{\quad}{\pars{R,P_{1}'}}}}$
										 										
										\end{array}
										\right\}
										
										$ & $\Pi_{2}$
									
									\end{tabular}							
									}
									
									\ \newline
									
									\center{
									\begin{tabular}{cc}									
										
										$\left .
										\begin{array}{c}
										
											$\dernote{=}{\quad}{\pars{T,P_{2}}}{\leaf{\strpr{\Pi_{9}}{\quad}{\pars{T,P_{2}'}}}}$
										 										
										\end{array}
										\right\}
										
										$ & $\Pi_{3}$
									
									\end{tabular}		
									}
									
									\ \newline
															
							\end{enumerate}		
					
					\end{enumerate}

					\item $\forall m', n' . ( (m',n') \prec (m,n) \ \ \ \ \ e \ \ \ \ \ S(m',n') \ \ \ \ \ e \ \ \ \ \ C(m',n')) \implies C(m,n)$
					O tamanho de $\pars{\aprs{R,T},P}$ é $m$ e existe uma prova desta estrutura de tamanho $n$. Vamos considerar a instância de regra de derivação mais abaixo na prova:
					
					$$
						\dernote{\rho}{\quad}{\pars{\aprs{R,T},P}}{\leaf{\strpr{}{}{Q}}}
					$$
					
					Onde assumimos que $\rho$ é não-trivial ($\rho \ \neq \ \ \ \ \ \ equivale \ a \ \ \ \ \ $), pois de outra forma a hipótese de indução se aplicaria. Vamos argumentar baseados na posição do redex de $\rho$ em $\pars{\aprs{R,T},P}$. As possibilidades são:
					
					\begin{enumerate}
					
							\item $\rho  =  \ruleaidown $
						
							\begin{enumerate}
							
								\item O redex está em $R$
								
$$\dernote{\ruleaidown}{\quad}{\pars{\aprs{R,T},P}}{\leaf{\strpr{}{}{\pars{\aprs{R',T},P}}}}  \ \ \ \ \ equivale \ a \ \ \ \ \   \strder{}{}{P}{\leaf{\pars{P_{1},P_{2}}}} \ \ \ \ \ e \ \ \ \ \ \dernote{\ruleaidown}{}{\pars{R,P_{1}}}{\leaf{\strpr{}{}{\pars{R',P_{1}}}}} \ \ \ \ \ e \ \ \ \ \ \strpr{}{}{\pars{T,P_{2}}} $$																	
							
								\item O redex está em $T$
							$$\dernote{\ruleaidown}{\quad}{\pars{\aprs{R,T},P}}{\leaf{\strpr{}{}{\pars{\aprs{R,T'},P}}}}  \ \ \ \ \ equivale \ a \ \ \ \ \   \strder{}{}{P}{\leaf{\pars{P_{1},P_{2}}}} \ \ \ \ \ e \ \ \ \ \ \strpr{}{}{\pars{R,P_{1}}} \ \ \ \ \ e \ \ \ \ \							 \dernote{\ruleaidown}{}{\pars{T,P_{2}}}{\leaf{\strpr{}{}{\pars{T',P_{2}}}}}  $$	
							
								\item O redex está em $P$
									
$$\dernote{\ruleaidown}{\quad}{\pars{\aprs{R,T},P}}{\leaf{\strpr{}{}{\pars{\aprs{R,T},P'}}}}  \ \ \ \ \ equivale \ a \ \ \ \ \   \dernote{\ruleaidown}{\quad}{P}{\leaf{\strder{}{}{P'}{\leaf{\pars{P_{1},P_{2}}}}}} \ \ \ \ \ e \ \ \ \ \ \strpr{}{}{\pars{R,P_{1}}} \ \ \ \ \ e \ \ \ \ \	\strpr{}{}{\pars{T,P_{2}}}						   $$														
							
							\end{enumerate}		
							
						\item $\rho = \ruleqdown$
						
							\begin{enumerate}
													
								\item O redex está em $R$
								
$$\dernote{\ruleqdown}{\quad}{\pars{\aprs{R,T},P}}{\leaf{\strpr{}{}{\pars{\aprs{R',T},P}}}}  \ \ \ \ \ equivale \ a \ \ \ \ \   \strder{}{}{P}{\leaf{\pars{P_{1},P_{2}}}} \ \ \ \ \ e \ \ \ \ \ \dernote{\ruleqdown}{}{\pars{R,P_{1}}}{\leaf{\strpr{}{}{\pars{R',P_{1}}}}} \ \ \ \ \ e \ \ \ \ \ \strpr{}{}{\pars{T,P_{2}}} $$																	
							
								\item O redex está em $T$
							$$\dernote{\ruleqdown}{\quad}{\pars{\aprs{R,T},P}}{\leaf{\strpr{}{}{\pars{\aprs{R,T'},P}}}}  \ \ \ \ \ equivale \ a \ \ \ \ \   \strder{}{}{P}{\leaf{\pars{P_{1},P_{2}}}} \ \ \ \ \ e \ \ \ \ \ \strpr{}{}{\pars{R,P_{1}}} \ \ \ \ \ e \ \ \ \ \							 \dernote{\ruleqdown}{}{\pars{T,P_{2}}}{\leaf{\strpr{}{}{\pars{T',P_{2}}}}}  $$	
							
								\item O redex está em $P$
									
$$\dernote{\ruleqdown}{\quad}{\pars{\aprs{R,T},P}}{\leaf{\strpr{}{}{\pars{\aprs{R,T},P'}}}}  \ \ \ \ \ equivale \ a \ \ \ \ \   \dernote{\ruleqdown}{\quad}{P}{\leaf{\strder{}{}{P'}{\leaf{\pars{P_{1},P_{2}}}}}} \ \ \ \ \ e \ \ \ \ \ \strpr{}{}{\pars{R,P_{1}}} \ \ \ \ \ e \ \ \ \ \	\strpr{}{}{\pars{T,P_{2}}}						   $$	
																		
								\item	$P = \pars{\seqs{P';P''},U',U''}$ e 
								
									$$ \dernote{\ruleqdown}{\quad}{\pars{\aprs{R,T},\seqs{P';P''},U',U''}}{\leaf{\strpr{}{}{\pars{\seqs{\pars{\aprs{R,T},P',U'};P''},U''}}}} $$
									
%
									
									Queremos mostrar que existem $\Pi_{1}$, $\Pi_{2}$ e $\Pi_{3}$ tais que:
																												
									$$ \strpr{}{}{\pars{\aprs{R,T},P}} \ \ \ \ \ equivale \ a \ \ \ \ \ \strder{\Pi_{1}}{}{\pars{\seqs{P';P''},U',U''}}{\leaf{\pars{P_{1};P_{2}}}} \ \ \ \ \ e \ \ \ \ \ \strpr{\Pi_{2}}{}{\pars{R,P_{1}}} \ \ \ \ \ e \ \ \ \ \ \strpr{\Pi_{3}}{}{\pars{T,P_{2}}} $$														
									
									Uma vez que $|\occ{\pars{\seqs{\pars{\aprs{R,T},P',U'};P''},U''}}| = |\occ{\pars{\aprs{R,T},\seqs{P';P''},U',U''}}|$	mas $n'<n$, podemos aplicar a hipótese de indução e obter:
									
									$$ \strder{\Pi_{4}}{}{U''}{\leaf{\seqs{U_{1};U_{2}}}} \ \ \ \ \ e \ \ \ \ \ \strpr{\Pi_{5}}{}{\pars{\aprs{R,T},P',U',U_{1}}} \ \ \ \ \ e \ \ \ \ \ \strpr{\Pi_{6}}{}{\pars{P'',U_{2}}} $$
									
									Uma vez que $|\occ{\pars{\aprs{R,T},P',U',U_{1}}}| < |\occ{\pars{\seqs{\pars{\aprs{R,T},P',U'};P''},U''}}|$, podemos aplicar a hipótese de indução e obter:
									
									$$ \strder{\Pi_{7}}{\quad}{\pars{P',U',U_{1}}}{\leaf{\pars{P_{1}',P_{2}'}}} \ \ \ \ \ equivale \ a \ \ \ \ \ \strpr{\Pi_{8}}{}{\pars{R,P_{1}'}} \ \ \ \ \ e \ \ \ \ \ \strpr{\Pi_{9}}{}{\pars{T,P_{2}'}} $$

									Agora podemos fazer	$P_{1} = P_{1}'$ e $P_{2} = P_{2}'$ e construir $\Pi_{1}$, $\Pi_{2}$ e $\Pi_{3}$ da seguinte forma:							
							
									\ \newline
							
									\center{
									\begin{tabular}{cc}
										
										$ \left .
										\begin{array}{c} 	
										
\dernote{=}{\quad}{P}{\leaf{\strder{\Pi_{4}}{}{\pars{\seqs{P';P''},U',U''}}{\leaf{\dernote{\ruleqdown}{\quad}{\pars{\seqs{P';P''},U',\seqs{U_{1};U_{2}}}}{\leaf{\strder{\Pi_{6}}{\quad}{\pars{\seqs{\pars{P',U_{1}};\pars{P'',U_{2}}},U'}}{\leaf{\strder{\Pi_{7}}{\quad}{\pars{P',U_{1},U'}}{\leaf{\dernote{=}{\quad}{\pars{P_{1}',P_{2}'}}{\leaf{\pars{P_{1},P_{2}}}}}}}}}}}}}}
										
										\end{array}
										\right\} $ & $\Pi_{1}$.
									
									\end{tabular}
									}
									
									\ \newline
									
									\center{
									\begin{tabular}{cc}
										
										$ \left .
										\begin{array}{c} 	
										
										\dernote{=}{\quad}{\pars{R,P_{1}}}{\leaf{\strpr{\Pi_{8}}{}{\pars{R,P_{1}'}}}}
										
										\end{array}
										\right\} $ & $\Pi_{2}$.
									
									\end{tabular}
									}
									
									\ \newline
									
									\center{
									\begin{tabular}{cc}
										
										$ \left .
										\begin{array}{c} 	
										
										\dernote{=}{\quad}{\pars{T,P_{2}}}{\leaf{\strpr{\Pi_{9}}{}{\pars{T,P_{2}'}}}}
										
										\end{array}
										\right\} $ & $\Pi_{3}$.
									
									\end{tabular}
									}					
									
									\ \newline									
								
							\end{enumerate}
							
						\item $\rho = \rules$													
						
							\begin{enumerate}
							
								\item O redex está em $R$
							
								$$\dernote{\rules}{\quad}{\pars{\aprs{R,T},P}}{\leaf{\strpr{}{}{\pars{\aprs{R',T},P}}}}  \ \ \ \ \ equivale \ a \ \ \ \ \   \strder{}{}{P}{\leaf{\pars{P_{1},P_{2}}}} \ \ \ \ \ e \ \ \ \ \ \dernote{\rules}{}{\pars{R,P_{1}}}{\leaf{\strpr{}{}{\pars{R',P_{1}}}}} \ \ \ \ \ e \ \ \ \ \ \strpr{}{}{\pars{T,P_{2}}} $$																	
							
								\item O redex está em $T$
							
								$$\dernote{\rules}{\quad}{\pars{\aprs{R,T},P}}{\leaf{\strpr{}{}{\pars{\aprs{R,T'},P}}}}  \ \ \ \ \ equivale \ a \ \ \ \ \   \strder{}{}{P}{\leaf{\pars{P_{1},P_{2}}}} \ \ \ \ \ e \ \ \ \ \ \strpr{}{}{\pars{R,P_{1}}} \ \ \ \ \ e \ \ \ \ \							 \dernote{\rules}{}{\pars{T,P_{2}}}{\leaf{\strpr{}{}{\pars{T',P_{2}}}}}  $$	
							
								\item O redex está em $P$
									
$$\dernote{\rules}{\quad}{\pars{\aprs{R,T},P}}{\leaf{\strpr{}{}{\pars{\aprs{R,T},P'}}}}  \ \ \ \ \ equivale \ a \ \ \ \ \   \dernote{\rules}{\quad}{P}{\leaf{\strder{}{}{P'}{\leaf{\pars{P_{1},P_{2}}}}}} \ \ \ \ \ e \ \ \ \ \ \strpr{}{}{\pars{R,P_{1}}} \ \ \ \ \ e \ \ \ \ \	\strpr{}{}{\pars{T,P_{2}}}						   $$	
																
								\item $R = \aprs{R',R''}, T = \aprs{T',T''}, P = \pars{P',P''}$ e
																
									$$ \dernote{\rules}{\quad}{\pars{\aprs{R',R'',T',T''},P',P''}}{\leaf{\pars{\aprs{\pars{\aprs{R',T'},P'},R'',T''},P''}}} $$
																			
									Queremos mostrar que existem $\Pi_{1}$, $\Pi_{2}$ e $\Pi_{3}$ tais que:
									
									$$ \strpr{}{\quad}{\pars{\aprs{R,T},P}} \ \ \ \ \ equivale \ a \ \ \ \ \ \strder{\Pi_{1}}{\quad}{\pars{P',P''}}{\leaf{\pars{P_{1};P_{2}}}} \ \ \ \ \ e \ \ \ \ \ \strpr{\Pi_{2}}{}{\pars{\aprs{R',R''},P_{1}}} \ \ \ \ \ e \ \ \ \ \ \strpr{\Pi_{3}}{\quad}{\pars{\aprs{T',T''},P_{2}}} $$
									
									Uma vez que $|\occ{\pars{\aprs{\pars{\aprs{R',T'},P'},R'',T''},P''}}| = |\occ{\pars{\aprs{R',R'',T',T''},P',P''}}|$	mas $n'<n$, podemos aplicar a hipótese de indução e obter:
									
									$$ \strder{\Pi_{4}}{\quad}{P''}{\leaf{\pars{P_{1}',P_{2}'}}} \ \ \ \ \ e \ \ \ \ \ \strpr{\Pi_{5}}{}{\pars{\aprs{R';T'},P',P_{1}'}} \ \ \ \ \ e \ \ \ \ \ \strpr{\Pi_{6}}{}{\pars{\aprs{R'',T''},P_{2}'}} $$
									
									Uma vez que $|\occ{\pars{\aprs{R';T'},P',P_{1}'}}| < |\occ{\pars{\aprs{R',R'',T',T''},P',P''}}|$, podemos aplicar a hipótese de indução e obter:
								
									$$ \strder{\Pi_{7}}{\quad}{\pars{P',P_{1}'}}{\leaf{\pars{P_{1}'',P_{2}''}}} \ \ \ \ \ e \ \ \ \ \ \strpr{\Pi{8}}{\quad}{\pars{R',P_{1}''}} \ \ \ \ \ e \ \ \ \ \  \strpr{\Pi{9}}{\quad}{\pars{T',P_{2}''}} $$
									
									Uma vez que $|\occ{\pars{\aprs{R'',T''},P_{2}'}}| < |\occ{\pars{\aprs{R',R'',T',T''},P',P''}}|$, podemos aplicar a hipótese de indução e obter:
								
									$$ \strder{\Pi_{10}}{\quad}{P_{2}''}{\leaf{\pars{P_{1}''',P_{2}'''}}} \ \ \ \ \ e \ \ \ \ \ \strpr{\Pi{11}}{\quad}{\pars{R'',P_{1}'''}} \ \ \ \ \ e \ \ \ \ \  \strpr{\Pi{12}}{\quad}{\pars{T'',P_{2}'''}} $$
									
									Agora podemos fazer $P_{1} = \pars{P_{1}'',P_{1}'''}$ e $P_{2} = \pars{P_{2}'',P_{2}'''}$ e construir $\Pi_{1}$, $\Pi_{2}$ e $\Pi_{3}$ da seguinte forma:
									
								  \ \newline
								
									\begin{center}
									\begin{tabular}{cc}
										
										$\left .
										\begin{array}{c}
										
											$
\dernote{=}{\quad}{P}{\leaf{\strder{\Pi_{4}}{\quad}{\pars{P',P''}}{\leaf{\strder{\Pi_{7}}{\quad}{\pars{P',P_{1}',P_{2}'}}{\leaf{\strder{\Pi_{10}}{\quad}{\pars{P_{1}'',P_{2}'',P_{2}'}}{\leaf{\dernote{=}{\quad}{\pars{P_{1}'',P_{2}'',P_{1}''',P_{2}'''}}{\leaf{\pars{P_{1},P_{2}}}}}}}}}}}} $										
										\end{array}
										\right\}
										$ & $\Pi_{1}$
									
									\end{tabular}
																	
									\ \newline
									
									\begin{tabular}{cc}									
										
										$\left .
										\begin{array}{c}
										
$\dernote{=}{\quad}{\pars{R,P_{1}}}{\leaf{\dernote{\rules}{\quad}{\pars{\aprs{R',R''},P_{1}'',P_{1}'''}}{\leaf{\strder{\Pi_{8}}{\quad}{\pars{\aprs{\pars{R',P_{1}''},R''},P_{1}'''}}{\leaf{\strpr{\Pi_{11}}{\quad}{\pars{R'',P_{1}'''}}}}}}}}$
										 										
										\end{array}
										\right\}
										
										$ & $\Pi_{2}$
									
									\end{tabular}							
									
									\ \newline
																		
									\begin{tabular}{cc}									
										
										$\left .
										\begin{array}{c}
										
											$\dernote{=}{\quad}{\pars{T,P_{2}}}{\leaf{\dernote{\rules}{\quad}{\pars{\aprs{T',T''},P_{2}'',P_{2}'''}}{\leaf{\strder{\Pi_{9}}{\quad}{\pars{\aprs{\pars{T',P_{2}''},T''},P_{2}'''}}{\leaf{\strpr{\Pi_{12}}{\quad}{\pars{T'',P_{2}'''}}}}}}}}$
										 										
										\end{array}
										\right\}
										
										$ & $\Pi_{3}$
									
									\end{tabular}		
									\end{center}
									
								\ \newline																								
								
								\item $P = \pars{\aprs{P',P''},U',U''}$ e
																$$\dernote{\rules}{\quad}{\pars{\aprs{R,T},\aprs{P',P''},U',U''}}{\leaf{\pars{\aprs{\pars{\aprs{R,T},P',U'},P''},U''}}}$$
																
									Queremos mostrar que existem $\Pi_{1}$, $\Pi_{2}$ e $\Pi_{3}$ tais que:
									
									$$ \strpr{}{\quad}{\pars{\aprs{R,T},P}} \ \ \ \ \ equivale \ a \ \ \ \ \ \strder{\Pi_{1}}{\quad}{\pars{\aprs{P',P''},U',U''}}{\leaf{\pars{P_{1},P_{2}}}} \ \ \ \ \ e \ \ \ \ \ \strpr{\Pi_{2}}{}{\pars{R,P_{1}}} \ \ \ \ \ e \ \ \ \ \ \strpr{\Pi_{3}}{\quad}{\pars{T,P_{2}}} $$
									
									Uma vez que $|\occ{\pars{\aprs{\pars{\aprs{R,T},P',U'},P''},U''}}| = |\occ{\pars{\aprs{R,T},\aprs{P',P''},U',U''}}|$	mas $n'<n$, podemos aplicar a hipótese de indução e obter:
									
									$$ \strder{\Pi_{4}}{\quad}{U''}{\leaf{\pars{U_{1},U_{2}}}} \ \ \ \ \ e \ \ \ \ \ \strpr{\Pi_{5}}{}{\pars{\aprs{R,T},P',U',U_{1}}} \ \ \ \ \ e \ \ \ \ \ \strpr{\Pi_{6}}{}{\pars{P'',U_{2}}} $$
									
									Uma vez que $|\occ{\pars{\aprs{R,T},P',U',U_{1}}}| < |\occ{\pars{\aprs{\pars{\aprs{R,T},P',U'},P''},U''}}|$, podemos aplicar a hipótese de indução e obter:
								
									$$ \strder{\Pi_{7}}{\quad}{\pars{P',U',U_{1}}}{\leaf{\pars{P_{1}',P_{2}'}}} \ \ \ \ \ e \ \ \ \ \ \strpr{\Pi{8}}{\quad}{\pars{R,P_{1}'}} \ \ \ \ \ e \ \ \ \ \  \strpr{\Pi{9}}{\quad}{\pars{T,P_{2}'}} $$
									
									Agora podemos fazer $P_{1} = P_{1}'$ e $P_{2} = P_{2}'$ e construir $\Pi_{1}$, $\Pi_{2}$ e $\Pi_{3}$ da seguinte forma:
									
									\ \newline
									
									\begin{center}
									\begin{tabular}{cc}
										
										$\left .
										\begin{array}{c}
										
											$\dernote{=}{\quad}{P}{\leaf{\strder{\Pi{4}}{\quad}{\pars{\aprs{P',P''},U',U''}}{\leaf{\dernote{\rules}{\quad}{\pars{\aprs{P',P''},U',U_{1},U_{2}}}{\leaf{\dernote{\rules}{\quad}{\pars{\aprs{\pars{P',U',U_{1}},P''},U_{2}}}{\leaf{\strder{\Pi_{6}}{\quad}{\aprs{\pars{P',U',U_{1}},\pars{P'',U_{2}}}}{\leaf{\strder{\Pi_{7}}{\quad}{\pars{P',U',U_{1}}}{\leaf{\dernote{=}{\quad}{\pars{P_{1}',P_{2}'}}{\leaf{\pars{P_{1},P_{2}}}}}}}}}}}}}}}}$
								
										\end{array}
										\right\}
										$ & $\Pi_{1}$
									
									\end{tabular}
																	
									\ \newline
									
									\begin{tabular}{cc}									
										
										$\left .
										\begin{array}{c}
										
											$\dernote{=}{\quad}{\pars{R,P_{1}}}{\leaf{\strpr{\Pi_{8}}{}{\pars{R,P_{1}'}}}}$
										 										
										\end{array}
										\right\}
										
										$ & $\Pi_{2}$
									
									\end{tabular}							
														
									\ \newline														
																		
									\begin{tabular}{cc}									
										
										$\left .
										\begin{array}{c}
										
											$\dernote{=}{\quad}{\pars{T,P_{2}}}{\leaf{\strpr{\Pi_{9}}{}{\pars{T,P_{2}'}}}}$
										 										
										\end{array}
										\right\}
										
										$ & $\Pi_{3}$
									
									\end{tabular}			
									\end{center}																							
								
									\ \newline
															
							\end{enumerate}										
					
					\end{enumerate}					
				
				\end{enumerate}
			
		\end{itemize}
		
	\end{proof}
	
\end{Theorem}

\begin{Theorem}[Context Reduction]
\label{teorema:context-reduction}

	Para todas as estruturas R e para todos os contextos $S\{\ \}$ tais que $S\{R\}$ é provável em \BV, existe uma estrutura $U$ tal que, para todas as estruturas $X$, existem derivações: 
	
	\begin{center}
	$\strder{}{\BV}{S\{X\}}{\leaf{\pars{X,U}}} \ \ \ \ \ e \ \ \ \ \ \strpr{}{\BV}{\pars{R,U}}$
	\end{center}
	
	\begin{proof} 
	
		Todas as derivações da prova são feitas em \BV. Por indução no tamanho de $S\{\circ\}$:
		
		\begin{itemize}
			
			\item Caso Base
			
				$$|S\{\circ\}| = 0 \implies S\{X\} = X \implies \strder{}{\quad}{S\{X\}}{\leaf{\pars{X,U}}} \ \ \ \ \ equivale \ a \ \ \ \ \ \strder{}{\quad}{X}{\leaf{X}} \implies U = \circ$$
				
				Logo:
				
				$$\strder{}{\quad}{S\{X\}}{\leaf{\pars{X,U}}} \ \ \ \ \ e \ \ \ \ \ \strpr{}{\quad}{\pars{R,U}} \ \ \ \ \ equivale \ a \ \ \ \ \ \strder{}{\quad}{X}{\leaf{X}} \ \ \ \ \ e \ \ \ \ \ \strpr{}{\quad}{U} $$							
			
			\item Casos Indutivos
			
				\begin{enumerate}
				
					\item $S = \seqs{S'\{\};P}$, $P \neq \circ$
						
						$$\strpr{}{}{\seqs{S'\{R\};P}} \ \ \ \ \ equivale \ a \ \ \ \ \ \strpr{\Pi_{1}}{}{S'\{R\}} \ \ \ \ \ e \ \ \ \ \ \strpr{\Pi_{2}}{}{P} $$
						
						Logo podemos fazer a seguinte derivação:
						
						$$\strder{\Pi_{2}}{}{\seqs{S'\{X\};P}}{\leaf{S'\{X\}}}$$
						
						Uma vez que $|\occ{S'\{X\}}| < |\occ{\seqs{S'\{X\};P}}|$, podemos aplicar a hipótese de indução e obter:
						
						$$\strder{}{}{S'\{X\}}{\leaf{\pars{X,U}}} \ \ \ \ \ e \ \ \ \ \ \strpr{}{}{\pars{R,U}}$$												
					
					\item $S\{\} = \pars{S'\{\},P}$, $P \neq \circ$ e $S'\{\}$ não é uma estrutura par própria.
						
						Subcasos: 
						\begin{itemize}
							
							\item $S'\{\circ\} = \circ$
							
								Então nós temos:
								
								$$S\{R\} \ \ \ \ \ equivale \ a \ \ \ \ \ \pars{S'\{R\},P} \ \ \ \ \ equivale \ a \ \ \ \ \ \pars{R,P} $$
								
								Fazendo $U=P$ temos então:
								
								$$\strder{}{}{\pars{X,P}}{\leaf{\pars{X,P}}} \ \ \ \ \ e \ \ \ \ \ \strpr{}{}{\pars{R,P}}$$
												
							\item $S'\{\} = \seqs{S''\{\};P'}$, $P' \neq \circ$
							
								Temos:	
							
								$$\strpr{}{}{S\{R\}} \ \ \ \ \ equivale \ a \ \ \ \ \ \strpr{}{}{\pars{S'\{R\},P}} \ \ \ \ \ equivale \ a \ \ \ \ \ \strpr{}{}{\pars{\seqs{S''\{R\};P'},P}} $$
								
								Pelo teorema de Shallow Splitting, temos:
								
								$$ \strpr{}{}{\pars{\seqs{S''\{R\},P'},P}} \implies \strder{\Pi_{1}}{\quad}{P}{\leaf{\seqs{P_{1};P_{2}}}} \ \ \ \ \ e \ \ \ \ \ \strpr{\Pi_{2}}{\quad}{\pars{S''\{R\},P_{1}}} \ \ \ \ \ e \ \ \ \ \ \strpr{\Pi_{3}}{}{\pars{P',P_{2}}} $$
							
								Logo:
								
								$$ \strder{\Pi{1}}{\quad}{\pars{\seqs{S''\{R\};P'},P}}{\leaf{\dernote{\ruleqdown}{\quad}{\pars{\seqs{S''\{R\};P'},\seqs{P_{1};P_{2}}}}{\leaf{\strder{\Pi_{3}}{\quad}{\seqs{\pars{S''\{R\},P_{1}};\pars{P',P_{2}}}}{\leaf{\pars{S''\{R\},P_{1}}}}}}}} $$
								
								Uma vez que $|\occ{\pars{S''\{\},P_{1}}}| < |\occ{S\{\}}|$, podemos aplicar a hipótese de indução e obter:
								
								$$ \strder{}{}{\pars{S''\{X\},P_{1}}}{\leaf{\pars{X,U}}} \ \ \ \ \ e \ \ \ \ \ \strpr{}{}{\pars{R,U}} $$									
								Podemos proceder analogamente quando $S'\{\} = \seqs{P';S''\{\}}$.
															
							\item $S'\{\} = \aprs{S''\{\},P'}, P\neq \circ$
							
								Temos que:
								
								$$\strpr{}{}{S\{R\}} \ \ \ \ \ equivale \ a \ \ \ \ \ \pars{S'\{R\},P} \ \ \ \ \ equivale \ a \ \ \ \ \ \pars{\aprs{S''\{R\},P'},P} $$
								
								Pelo teroema de Shallow Splitting, temos:
								
								$$ \strpr{}{}{\pars{\aprs{S''\{R\},P'},P}} \implies \strder{\Pi_{1}}{\quad}{P}{\leaf{\pars{P_{1},P_{2}}}} \ \ \ \ \ e \ \ \ \ \ \strpr{\Pi_{2}}{\quad}{\pars{S''\{R\},P_{1}}} \ \ \ \ \ e \ \ \ \ \ \strpr{\Pi_{3}}{\quad}{\pars{P',P_{2}}} $$
							
								Logo:
								
								$$ \strder{\Pi_{1}}{\quad}{\pars{\aprs{S''\{R\},P'},P}}{\leaf{\dernote{\rules}{\quad}{\pars{\aprs{S''\{R\},P'},P_{1},P_{2}}}{\leaf{\strder{\Pi_{3}}{\quad}{\pars{\aprs{\pars{P',P_{2}},S''\{R\}},P_{1}}}{\leaf{\pars{S''\{R\},P_{1}}}}}}}} $$
								
								Uma vez que $|\occ{\pars{S''\{\},P_{1}}}| < |\occ{S\{\}}|$, podemos aplicar a hipótese de indução e obter:
								
								$$ \strder{}{\quad}{\pars{S''\{X\},P_{1}}}{\leaf{\pars{X,U}}} \ \ \ \ \ e \ \ \ \ \ \strpr{}{}{\pars{R,U}} $$
								
								Podemos proceder analogamente quando $S'\{\} = \aprs{P';S''\{\}}$.							
							
						\end{itemize}
					
					\item $S\{\} = \aprs{S'\{\},P}$, $P \neq \circ$
					
							Temos que:
					
							$$ \strpr{}{}{\aprs{S'\{R\},P}} \ \ \ \ \ equivale \ a \ \ \ \ \ \strpr{\Pi_{1}}{}{S'\{R\}} \ \ \ \ \ e \ \ \ \ \ \strpr{\Pi_{2}}{}{P} $$
							
							Logo:
							
							$$ \strder{\Pi_{2}}{\quad}{\aprs{S'\{X\},P}}{\leaf{S'\{X\}}} $$
							
							Uma vez que $|\occ{S'\{\}}| < |\occ{S\{\}}|$, podemos aplicar a hipótese de indução e obter:
							
							$$ \strder{}{}{\aprs{S'\{X\},P}}{\leaf{\strder{}{}{S'\{X\}}{\leaf{\pars{X,U}}}}} \ \ \ \ \ e \ \ \ \ \ \strpr{}{}{\pars{R,U}}$$								
				
				\end{enumerate}
			
		\end{itemize}
		
	\end{proof}

\end{Theorem}

\begin{Corollary}[Splitting]
\label{corolario:splitting}

	Para todas as estruturas R, T e para todos os contextos S\{\}:
	
	\begin{enumerate}
		
		\item Se $S\seqs{R;T}$ é provável em \BV, então existem estruturas $S_{1}$ e $S_{2}$ tais que, para todas estrutura $X$, existe uma derivação:
		
		$$ \strder{}{\BV}{S\{X\}}{\leaf{\pars{X,\seqs{S_{1};S_{2}}}}} \ \ \ \ \ e \ \ \ \ \ \strpr{}{\BV}{\pars{R,S_{1}}} \ \ \ \ \ e \ \ \ \ \ \strpr{}{\BV}{T,S_{2}} $$
		
		\item Se $S\aprs{R,T}$ é provável em \BV, então existem estruturas $S_{1}$ e $S_{2}$ tais que, para todas estrutura $X$, existe uma derivação:
		
		$$ \strder{}{\BV}{S\{X\}}{\leaf{\pars{X,S_{1},S_{2}}}} \ \ \ \ \ e \ \ \ \ \ \strpr{}{\BV}{\pars{R,S_{1}}} \ \ \ \ \ e \ \ \ \ \ \strpr{}{\BV}{T,S_{2}} $$
		
	\end{enumerate}

	\begin{proof}
	
		Todas as derivações da prova são feitas em \BV. Primeiramente aplicamos o teorema de \emph{Context Reduction} e então aplicamos o teorema de \emph{Shallow Splitting}.
		
		\begin{enumerate}
		
			\item $$\begin{array}{c}		
					\strpr{}{}{S\seqs{R;T}} \\ \\
					\textbf{Context \ Reduction} \\
					 \Downarrow \\ \\
					 \strder{\Pi_{1}}{}{S\{X\}}{\leaf{\pars{U,X}}} \ \ \ \ \ e \ \ \ \ \ \strpr{\Pi_{2}}{}{\pars{\seqs{R;T},U}} \\ \\
					\textbf{Shallow \ Splitting} \\
					 \Downarrow\\ \\
					 \strder{\Pi_{1}}{\quad}{S\{X\}}{\leaf{\pars{U,X}}} \ \ \ \ \ e \ \ \ \ \  \strder{\Pi_{3}}{\quad}{U}{\leaf{\seqs{S_{1};S_{2}}}} \ \ \ \ \ e \ \ \ \ \ \strpr{\Pi_{4}}{\quad}{\pars{R,S_{1}}} \ \ \ \ \ e \ \ \ \ \ \strpr{\Pi_{5}}{\quad}{\pars{T,S_{2}}} \\				
					\end{array}$$										
		
				Então nós combinamos $\Pi_{1}$ e $\Pi_{3}$ para obter $\Pi_{6}$:
				
				$$ \strder{\Pi_{1}}{\quad}{S\{X\}}{\leaf{\strder{\Pi_{3}}{\quad}{\pars{U,X}}{\leaf{\pars{X,\seqs{S_{1};S_{2}}}}}}} \ \ \ \ \ equivale \ a \ \ \ \ \ \strder{\Pi_{6}}{\quad}{S\{X\}}{\leaf{\pars{X,\seqs{S_{1};S_{2}}}}} $$
				
				E com $\Pi_{4}$, $\Pi_{5}$ e $\Pi_{6}$ temos o caso $1$ provado.
		
			\item $$\begin{array}{c}		
					\strpr{}{}{S\aprs{R,T}} \\ \\
					\textbf{Context \ Reduction} \\
					 \Downarrow \\ \\
					 \strder{\Pi_{1}}{}{S\{X\}}{\leaf{\pars{X,U}}} \ \ \ \ \ e \ \ \ \ \ \strpr{\Pi_{2}}{}{\pars{\aprs{R,T},U}} \\ \\
					\textbf{Shallow \ Splitting} \\
					 \Downarrow\\ \\
					 \strder{\Pi_{1}}{\quad}{S\{X\}}{\leaf{\pars{X,U}}} \ \ \ \ \ e \ \ \ \ \  \strder{\Pi_{3}}{\quad}{U}{\leaf{\pars{S_{1},S_{2}}}} \ \ \ \ \ e \ \ \ \ \ \strpr{\Pi_{4}}{\quad}{\pars{R,S_{1}}} \ \ \ \ \ e \ \ \ \ \ \strpr{\Pi_{5}}{\quad}{\pars{T,S_{2}}} \\				
					\end{array}$$										
		
				Então nós combinamos $\Pi_{1}$ e $\Pi_{3}$ para obter $\Pi_{6}$:
				
				$$ \strder{\Pi_{1}}{\quad}{S\{X\}}{\leaf{\strder{\Pi_{3}}{\quad}{\pars{X,U}}{\leaf{\pars{X,S_{1},S_{2}}}}}} \ \ \ \ \ equivale \ a \ \ \ \ \ \strder{\Pi_{6}}{\quad}{S\{X\}}{\leaf{\pars{X,S_{1},S_{2}}}} $$
				
				E com $\Pi_{4}$, $\Pi_{5}$ e $\Pi_{6}$ temos o caso $2$ provado.
		
		\end{enumerate}
	
	\end{proof}

\end{Corollary}

\subsubsection{Eliminação da regra Cut}
\label{section:eliminacao-corte}

Antes de demonstrar que a regra de Cut $\ruleaiup$ é admissível, é preciso introduzir uma proposição importante:

\begin{Proposition}
\label{prop:derqs}
Para todo contexto $S\{ \ \}$ e estruturas $R$, $T$ existe uma derivação $\strder{}{\{ \ruleqdown,\rules \}}{\pars{S\{ R \},T}}{\leaf{S\pars{R,T}}}$.
\end{Proposition}

\begin{Theorem}

A regra $\ruleaiup$ é admissível em \BV.

	\begin{proof}
		Considere a demonstração 
		
		$$ \dernote{\ruleaiup}{}{S\{\circ \}}{\leaf{\strpr{}{\BV}{S\aprs{a,\dual{a}}}}} $$
		
		Pelo corolário~\ref{corolario:splitting}, existem $S_{1}$ e $S_{2}$ tais que há derivações:
		
		\begin{center}
		
			$ \strder{}{\BV}{S\{\circ \}}{\leaf{\pars{S_{1},S_{2}}}} $, \ \ \ \ \  $\strpr{\Pi_{1}}{\BV}{\pars{a,S_{1}}}$ \ \ \ \ \  e \ \ \ \ \ $\strpr{\Pi_{2}}{\BV}{\pars{\dual{a},S_{2}}}$

		\end{center}
		
		Considere a demonstração $\Pi_{1}$. Deve haver um contexto $S_{1}'\{ \ \}$ tal que $S_{1} = S_{1}'\{ \dual{a} \}$ e 
		
		$$ \strder{\Pi_{1}}{\BV}{\pars{a,S_{1}'\{ \dual{a} \}}}{\leaf{\dernote{\ruleaidown}{}{S''\pars{a,\dual{a}}}{\leaf{\strpr{}{\BV}{S''\{\circ\}}}}}} $$
para algum $S''\{ \ \}$, onde nós destacamos a instância de $\ruleaidown$ na qual a ocorrência de $a$ interage com a ocorrência de $\dual{a}$ que vem de $S_{1}'\{ \dual{a}\}$. Podemos substituir cada ocorrência de $a$ e $\dual{a}$ em $\Pi_{1}$ por $\circ$, e obtemos em \BV\ uma demonstração de $S_{1}'\{ \circ \}$. Analogamente, podemos transformar $\Pi_{2}$ em uma demonstração em \BV\ de $S_{2}'\{ \circ \}$ tal que $S_{2} = S_{2}'\{a\}$. Podemos construir então a seguinte demonstração:
		
		$$ \strder{}{\BV}{S\{\circ\}}{\leaf{\strder{}{\BV}{\pars{S_{1}'\{ \dual{a}\},S_{2}'\{a\}}}{\leaf{\dernote{\ruleaidown}{}{S_{1}'\{ S_{2}'\pars{a,\dual{a}} \}}{\leaf{\strder{}{\BV}{S_{1}'\{ S_{2}'\{ \circ \} \}}{\leaf{\strpr{}{\BV}{S_{1}'\{ \circ \}}}}}}}}}} $$
onde usamos duas vezes a proposição \ref{prop:derqs}. Podemos repetir indutivamente este argumento, começando do topo, para qualquer demonstração em \BV\ $\cup$ $\{ \ruleaiup \}$, eliminando todas as instâncias de $\ruleaiup$ uma a uma.

	\end{proof}

\end{Theorem}

\section{Reduzindo o não determinismo no cálculo das estruturas}
\label{section:reduzindo-nao-determinismo}

Uma das possíveis abordagens para reduzir o não-determinismo do cálculo das estruturas é limitar as possibilidades de aplicação de regras durante a construção de uma demonstração. Em ~\cite{ozan06} esta abordagem é tomada, e um sistema equivalente ao sistema \BV\ é proposto, mas com regras reprojetadas de forma a diminuir o não-determinismo de sua aplicação. A solução proposta realmente diminui o não-determinismo, mas é uma abordagem basicamente operacional. Esta seção apresenta a abordagem proposta e os principais resultados obtidos.

A idéia central é reprojetar as regras \emph{switch} e \emph{seq} de tal forma que sua aplicação contribua para encurtar o tamanho da demonstração. Freqüentemente, regras de inferência podem ser aplicadas de muitas formas diferentes. Entretanto, apenas algumas dessas aplicações podem conduzir a uma demonstração. Por exemplo, para a estrutura $\pars{\aprs{\dual{a},\dual{b}},a,b}$ existem doze formas distintas de se aplicar, de baixo para uma instância de \emph{switch}, duas delas sendo: 

\begin{center}
$\dernote{s}{}{\pars{\aprs{\dual{a},\dual{b}},a,b}}{\leaf{\aprs{\pars{\dual{a},a,b},\dual{b}}}}$ \ \ \ \ \ e \ \ \ \ \ $\dernote{s}{}{\pars{\aprs{\dual{a},\dual{b}},a,b}}{\leaf{\pars{\aprs{\pars{\dual{a},a},\dual{b}},b}}}$
\end{center}

A primeira instância não leva a uma demonstração, enquanto a segunda pode levar. De fato, das doze aplicações possíveis, apenas duas podem levar a uma demonstração.

\begin{Definition}
Dada uma estrutura $S$, a notação \at{S} indica o conjunto de todos os átomos que aparecem em $S$.
\end{Definition}

\begin{Proposition}
Se uma estrutura $R$ tem uma demonstração em \BV\ então, para cada átomo $a$ que aparece em $R$, há um átomo $\dual{a}$ em $R$ tal que $a \downarrow_{R} \dual{a}$.
\end{Proposition}

\begin{Definition}
Seja a regra \emph{interaction switch}:

$$ \dernote{\ruleis}{}{S\pars{\aprs{R,T},W}}{\leaf{S\aprs{\pars{R,W},T}}} $$

onde $\at{\overline{W}} \cap \at{R} \neq \emptyset$

\end{Definition}

O objetivo do \emph{interaction switch} é evitar que apliquemos o \emph{switch} desnecessariamente, uma vez que só faz sentido colocar as estruturas $R$ e $W$ em par se elas tiverem átomos opostos, de forma a facilitar o futuro uso de uma regra $\ruleaidown$.

\begin{Definition}
Seja o sistema \BV\ com \emph{interaction switch}, ou sistema \BVs, o sistema $\{ \ruleodown, \ruleaidown, \ruleis, \ruleqdown \}$. Seja o sistema BV com \emph{lazy interaction switch}, ou sistema \BVsl, o sistema resultante da substituição da regra \ruleis\ em \BVs\  pela sua instância, chamada de \emph{lazy interaction switch}, ou \rulelis, onde a estrutura $W$ não é um par próprio.
\end{Definition}

É possível demonstrar que:

\begin{Theorem}
Os sistemas \BV, \BVsl\ e \BVs\ são fortemente equivalentes.
\end{Theorem}

A demonstração encontra-se em \cite{ozan06}.

À primeira vista, as regras \emph{switch} e \emph{seq} parecem ter naturezas diferentes devido aos operadores diferentes sobre os quais elas trabalham. Entretanto, uma inspeção mais cuidadosa mostra que ambas as regras lidam com o contexto das estruturas sobre as quais se aplicam: enquanto a regra \emph{switch} reduz as interações nas estruturas que envolvem uma estrutura copar (numa visão de baixo para cima), a regra \emph{seq} faz a mesma coisa com as estruturas envolvendo uma estrutura seq. Nesse sentido, podem-se estender os conceitos de redefinição da regra \emph{switch}  para a regra \emph{seq}.

\begin{Definition}
Seja o sistema \emph{lazy seq} \V, ou \QVI, o sistema que consiste das seguintes regras 

\begin{center}
$\dernote{\ruleqidown}{}{S\pars{\seqs{R;U},\seqs{T;V}}}{\leaf{S\seqs{\pars{R,T};\pars{U,V}}}}$ \ \ \ \ \
$\dernote{\ruleqiidown}{}{S\pars{R,T}}{\leaf{S\seqs{R;T}}}$ \ \ \ \ \
$\dernote{\rulelqiiidown}{}{S\pars{W,\seqs{R;T}}}{\leaf{S\seqs{\pars{R,W};T}}}$ \ \ \ \ \
$\dernote{\rulelqiiidown}{}{S\pars{W,\seqs{R;T}}}{\leaf{S\seqs{R;\pars{T,W}}}}$
\end{center}

onde $W$ não é uma estrutura par própria, e nenhuma das estruturas $R$, $T$, $U$, $V$ ou $W$ é a unidade $\circ$.

\end{Definition}

\begin{Proposition}
Os sistemas \QVI\ e \{$\ruleqdown$\} são equivalentes.
\end{Proposition}

\begin{Proposition}
Seja o sistema $\mathscr{S} \in$ \{\BV, \BVs, \BVsl \}. O sistema resultante da substituição da regra $\ruleqdown$ em $mathscr{S}$ pelo sistema \QVI\ é equivalente ao sistema \BV.
\end{Proposition}

\begin{Definition}
As seguintes regras são chamadas de \emph{interaction seq rule 1}, \emph{lazy interaction seq rule 3} e \emph{lazy interaction seq rule 4}, respectivamente:

\begin{center}
$\dernote{\ruleiqidown}{}{S\pars{\seqs{R;U},\seqs{T;V}}}{\leaf{S\seqs{\pars{R,T};\pars{U,V}}}}$ \ \ \ \ \
$\dernote{\ruleliqiiidown}{}{S\pars{W,\seqs{R;T}}}{\leaf{S\seqs{\pars{R,W};T}}}$ \ \ \ \ \
$\dernote{\ruleliqivdown}{}{S\pars{W,\seqs{R;T}}}{\leaf{S\seqs{R;\pars{T,W}}}}$
\end{center}
onde em \ruleiqidown temos que $\at{\dual{R}} \cap \at{T} \neq \emptyset$ e $\at{\dual{U}} \cap \at{V} \neq \emptyset$; em \ruleliqiiidown e em \ruleliqivdown temos que $\at{\dual{R}} \cap \at{W} \neq \emptyset$ e $W$ não é uma estrutura par própria. O sistema resultante da substituição da regra seq no sistema \BVsl\ pelas regras \ruleiqidown, \ruleqiidown, \ruleliqiiidown e \ruleliqivdown é chamado de \emph{interaction} \BV, ou \BVi.

\end{Definition}

\begin{Theorem}
Os sistemas \BV\ e \BVi\ são equivalentes.
\end{Theorem}

Veja \cite{ozan06} para a demonstração do teorema.

Na aplicação de regras switch e seq de baixo para cima durante a construção de uma demonstração, além de promover a interação entre alguns átomos, a interação entre alguns átomos pode ser quebrada. Entretanto, se a estrutura a ser demonstraçãoda consiste de pares de átomos distintos, quebrar a interação entre átomos duais (numa visão de baixo para cima), conduz a uma estrutura não-provável. As definições seguintes introduzem uma nova restrição sobre as regras de inferência, que explora esta observação e permite que apenas instâncias ``cautelosas'' das regras de inferência, ou seja, que não quebram essas interações entre átomos duais.

\begin{Definition}
Seja \emph{pruned switch} a regra $\ruleps$ abaixo, onde $\at{T} \cap \at{W} = \emptyset$, e seja \emph{pruned seq} a regra $\rulepqdown$ abaixo, onde $\at{\dual{T}} \cap \at{U} = \emptyset$ e $\at{\dual{R}} \cap \at{V} = \emptyset$:

$$ \dernote{\ruleps}{}{S\pars{\aprs{R,T},W}}{\leaf{S\aprs{\pars{R,W},T}}} \ \ \ \ \ \dernote{\rulepqdown}{}{S\pars{\seqs{R;U},\seqs{T;V}}}{\leaf{\seqs{\pars{R,T};\pars{U,V}}}}$$

Seja \emph{pruned} \BV, ou sistema \BVp, o sistema $\{\ruleodown, \ruleaidown, \ruleps, \rulepqdown \}$.

\end{Definition}

\begin{Proposition}
\label{prop:bvi}
Seja $P$ uma estrutura em \BV\ que consista de pares de átomos distintos e $\pi$ uma demonstração de $P$ em \BVi. Em $\pi$ todas as intâncias da regra $\rules$ são instâncias da regra $\ruleps$; e todas as instâncias das regras $\ruleiqidown$, $\ruleqiidown$, $\ruleliqiiidown$ e $\ruleliqivdown$ são instâncias da regra $\rulepqdown$.
\end{Proposition}

A demonstração encontra-se em \cite{ozan06}.

Em outras palavras, a proposição \ref{prop:bvi} diz que, em uma demonstração, todas as instâncias de switch e seq devem obedecer às restrições do sistema \BVp. Em termos práticos, isso significa que, ao implementar o sistema \BVi, deve-se checar, quando da aplicação de uma regra, se ela satisfaz \BVp. Se não satisfizer, ela não deve ser aplicada, pois não levará a uma demonstração. 

\subsection{Análise do estado da arte}
\label{section:analise-estado-arte}

A solução apresentada na seção \ref{section:reduzindo-nao-determinismo} realmente diminui o não determinismo do cálculo das estruturas. A abordagem consiste em checar, a cada passo na construção de uma demonstração, se a aplicação da regra pode ser realizada sem atingir um estágio do qual não se pode sair. Trata-se de uma solução válida, mas que apresenta as seguintes restrições:

\begin{enumerate}

	\item É uma solução \emph{operacional}. Durante a implementação do sistema, uma série de checagens é necessária para verificar se as regras podem ser aplicadas ou não.
	
	\item Não emprega o conceito de \emph{demonstração uniforme}, no sentido de que não apresenta uma estratégia de demonstração \emph{teórica}.

\end{enumerate}

Uma questão que permanece em aberto é: \emph{seria possível resolver o problema do não-determinismo sem recorrer a uma solução operacional?} É esta a principal contribuição deste trabalho: uma solução não operacional para a implementação de um fragmento do sistema \FBV. A solução é apresentada no Capítulo \ref{cap:aspectos-computacionais-cos}.

\chapter{Aspectos computacionais do cálculo das estruturas}
\label{cap:aspectos-computacionais-cos}
\section{Objetivo}
\label{section:objetivo}

É interessante ter a liberdade e expressividade do cálculo das estruturas. Entretanto, quando se está tentando construir uma demonstração, mais liberdade significa um espaço de busca maior. É desejável restringir a busca caso estejamos procurando por implementações eficientes dos sistemas dedutivos. Teoremas de \emph{splitting} representam um papel chave para reduzir o não-determinismo, uma vez que eles proporcionam um balanço entre tamanho de demonstrações e largura do espaço de busca, e essa proporção pode ser variada conforme se desejar.

Quanto ao cálculo das estruturas, estamos hoje na mesma situação em que se encontrava o cálculo de seqüentes antes que Dale Miller introduzisse a noção de demonstrabilidade uniforme, que é uma maneira de casar demonstrabilidade e a demonstrações dirigidas a objetivo\footnote{Em inglês: \emph{goal directed proofs}}, necessárias em implementações.

Nesse momento surge a pergunta: \emph{qual seria uma boa noção de uma demonstração dirigida a objetivo no caso mais geral do cálculo das estruturas}? Esta pergunta poderia, a princípio, ser respondida de várias maneiras, e as noções a serem desenvolvidas podem ou não ser dependentes de lógicas particulares. A calibração desta dependência pode tornar o projeto mais específico ou mais abrangente.

Nesse sentido, o objetivo principal do trabalho é:

\emph{Discutir caminhos possíveis para uma estratégia de busca por demonstrações em cálculo das estruturas adequada para a implementação computacional. Esta estratégia deve ser, de preferência, teórica e não puramente operacional.}

Ou seja, desejamos aprofundar o conhecimento existente sobre o cálculo das estruturas no sentido de contribuir para, no futuro, conseguir um resultado similar ao obtido através conceito de demonstrações uniformes para o cálculo de seqüentes.

\section{Abordagem}
\label{section:abordagem}

Para começar, é importante notar que o desenvolvimento de um conceito de demonstrações uniformes para cálculo das estruturas esbarra nos seguintes problemas:

\begin{enumerate}

	\item O comportamento operacional do cálculo das estruturas está longe de ser claro.
	\item Existe um alto grau de não determinismo no cálculo das estruturas.

\end{enumerate}

Para resolver estes problemas, estudamos uma gama abordagens possíveis, que serão expostas a seguir.

\subsection{O problema do comportamento operacional}
\label{section:comportamento-operacional}

Uma primeira idéia que estudamos está relacionada à suposição de que a noção de demonstrações uniformes em cálculo das estruturas se basearia principalmente na relação entre implicação ($\cimp$) e validade ($\vdash$). Esta noção pode ser melhor entendida com o exemplo de cláusulas de Horn e Prolog. Em Prolog, uma fórmula como 
$$B_{1} \wedge \hdots \wedge B_{h} \implies H$$
especifica o comportamento de um seqüente da forma
$$\Seq{B_{1} , \hdots , B_{h}}{H}$$
quando este seqüente é demonstrado seguindo uma demonstração uniforme. O que é importante ressaltar é que a fórmula já indica a forma da demonstração uniforme, devido à conexão clara entre implicação e validade para este caso. A construção de demonstrações para cláusulas de Horn faz uso constante desta conexão. Esta analogia pode ser estendida para cláusulas de Harrop e para a lógica linear clássica, mas não para a lógica clássica completa.
		
		O problema com cálculo das estruturas é que não há analogia entre implicação e validade. Além disso:
		
		\begin{enumerate}
		
			\item a princípio, o cálculo das estruturas não apresenta implicação explícita;
			
			\item usualmente, a negação não é definida explicitamente;
			
			\item o formalismo normalmente é usado para lógicas clássicas;
			
			\item existe a \emph{deep inference}.
		
		\end{enumerate}
		
		Os problemas $1$, $2$ e $3$ podem ser resolvidos movendo o foco para linguagens adequadas. O problema $4$, entretanto, é crucial. Vamos analisar cada problema com um pouco mais de cuidado:
		
		\begin{itemize}
		
			\item Problemas $1$ e $2$.
			
				Dada a estrutura:
				$$ \pars{a,b,\dual{c},\dual{d}}$$
				
				ela representa $\aprs{c,d} \implies \pars{a,b}$ ou  $ \aprs{\dual{a},\dual{b}} \implies \pars{\dual{c},\dual{d}} $?
				
				O Problema $1$ pode ser resolvido ao se adotar um símbolo explícito para a implicação, enquanto o Problema $2$ ao se estabelecer que existe um símbolo de negação na linguagem, tal que $\neg{a}$ é negativo enquanto $a$ é positivo. Com isso, há uma conseqüência importante: temos que lidar com ``contextos negativos''. Isso quer dizer que dentro de negações e dentro de premissas de implicações teremos que usar as regras de inferências duais às regras normais (positivas). Isto dobraria o número de regras dos sistemas dedutivos.
				
				Note que o Problema $2$ é separado do Problema $1$. De fato, analisando o Problema $2$, poderíamos estipular que a linguagem fica restrita de forma que o símbolo de negação só é permitido sobre átomos, implicando que só trabalhemos com fórmulas na forma normal. Trata-se de uma solução bastante operacional.
				
				\item Problema $3$.
				
					Claramente, deveríamos esperar que lógicas não-clássicas tenham maior propensão a apresentar demonstrações uniformes em cálculo das estruturas do que lógicas clássicas. Depois de um artigo sobre lógica intuicionista em cálculo das estruturas~\cite{tiu06}, outros para outras lógicas não-clássicas estão sendo lançados. Mas este não é um problema que representa um grande obstáculo.
					
				\item Problema $4$.
				
					Este é um problema grave. Por exemplo, como visto na seção ~\ref{section:horn-harrop}, dada uma fórmula de Harrop, pode-se vê-la como um programa: sabemos como os conectivos se comportam e, crucialmente, sabemos que o programa se desenvolve de fora para dentro da fórmula. Em outras palavras, o aninhamento de fórmulas tem um papel direto no que se refere à ordem de execução. Claramente, se usamos \emph{deep inference}, este não é mais o caso. 
					
					Considere que algumas interpretações operacionais de fórmulas dependam da estrutura de ramificação do cálculo de seqüentes. Por exemplo, módulos são mantidos em separados em ramos diferentes de uma árvore de cálculo de seqüentes. Se adotássemos um sistema de cálculo das estruturas que contivesse a regra \emph{medial}\footnote{A regra \emph{medial} não está presente no sistema \BV. É uma regra da lógica clássica, proposta por Alwen e Kai\cite{kai06}: $$\dernote{\rulem}{}{S\aprs{\pars{A,B},\pars{C,D}}}{\leaf{S\pars{\aprs{A,C},\aprs{B,D}}}}$$ cuja presença permite que a aplicação da regra \emph{contraction} possa ser restrita a átomos - um resultado notável, pois é a primeira vez que se vê um sistema com tal regra definida localmente.}, teríamos que módulos diferentes (ou seja, premissas de diferentes implicações em uma fórmula) seriam capazes de interagir com a demonstração, o que é indesejável\footnote{Essa situação corresponde ao problema de \emph{scope extrusion}, discutido na Seção~\ref{section:demonstracoes-uniformes}}. E isto contrasta com a noção usual de demonstrações uniformes.
		
		\end{itemize}
		
\subsection{O problema do não-determinismo}
\label{section:nao-determinismo}

	Como já foi visto, demonstrações uniformes propiciam menor grau de não-determinismo, algo realmente desejável em cálculo das estruturas. Uma abordagem para o problema poderia ser partir de demonstrações uniformes em cálculo de seqüentes e tentar entender como elas poderiam nos ajudar a reduzir o não-determinismo em cálculo das estruturas.
	
	Existe uma certa dualidade de jogos em \emph{deep inference}. No cálculo de seqüentes, regras à direita e à esquerda se comportam como jogador e oponente, respectivamente, e o jogo prossegue com cada um se movendo em turnos. Claramente, apenas conseguiríamos este tipo de comportamento em cálculo das estruturas se adotássemos uma das seguintes ``soluções'': implicação explícita, polaridades ou contextos positivos e negativos. Esta é uma idéia que poderia ser explorada.
	
	Entretanto, já temos uma fonte de jogos em cálculo das estruturas, num sentido um tanto mais inesperado que o citado acima. Este seria o teorema de \emph{splitting}. Relembrando, o teorema de \emph{splitting} pode ser enunciado como: 
	$$\strpr{}{\BV}{S\aprs{R,T}} \ \ \ equivale \ \ \ a \ \ \ \strder{}{\BV}{S\{ X \}}{\leaf{\pars{X, K_{R},K_{T}}}} \ \ \ e \ \ \ \strpr{}{\BV}{\pars{R,K_{R}}} \ \ \ e \ \ \ \strpr{}{\BV}{\pars{T,K_{T}}} $$
onde $X$ é uma estrutura qualquer.
	
	Utilizando a mesma metáfora acima, poderíamos dizer que o jogador escolhe $S\aprs{R,T}$, ou seja,  o contexto e a conjunção a ser dividida, e o oponente terá que achar os \emph{killers} $K_{R}$ e $K_{T}$. O jogo então prossegue sobre as novas estruturas a serem demonstradas $\pars{R,K_{R}}$ e $\pars{T,K_{T}}$.
	
	Então, uma possível abordagem para buscar demonstrações uniformes em cálculo das estruturas seria entender como a noção de jogos se relaciona com demonstrações uniformes em cálculo de seqüentes e trazer esta noção para o cálculo das estruturas.
	
	O teorema de \emph{splitting} garante demonstrações para qualquer $S\aprs{R,T}$, entretanto, a maneira como $S$, $R$ e $T$ são escolhidos não é elucidada, ela é completamente arbitrária. Então, a questão que surge é: como escolhê-los?
		
	Quanto a uma outra fonte de não-determinismo, a aplicação de regras, o trabalho~\cite{ozan06} trata do assunto, mas de uma maneira completamente operacional. Tal trabalho já foi discutido na Seção~\ref{section:reduzindo-nao-determinismo}.
		
\section{A tentativa de solução proposta}
\label{section:solucao}	

	De acordo com os estudos realizados, dentre os caminhos estudados o que apresenta maior probabilidade de produzir frutos segue mais ou menos a linha do teorema de \emph{splitting}. Dessa forma, baseamo-nos em resultados teóricos que apresentaremos a seguir para conjecturar que eles levarão a uma estratégia de demonstração uniforme para cálculo das estruturas. Entretanto, no atual estágio de desenvolvimento da pesquisa, podemos dizer que este trabalho apresenta uma \textit{tentativa de estratégia}.
	
	Dizemos \emph{tentativa de estratégia} porque, apesar de se basear em resultados teóricos apresentados formalmente e de ter funcionado perfeitamente em todos os testes que executamos (o Capítulo \ref{cap:implementacao} apresenta a implementação computacional da tentativa de estratégia), a demonstração de sua correção ainda está em curso (apesar de a maior parte do trabalho neste sentido já estar pronta).
	
	Antes de aprofundar em resultados teóricos mais precisos, abaixo damos uma idéia geral de como a tentativa de estratégia lida com a busca por demonstrações uniformes em \FBV\ com pares de átomos dois a dois distintos:
	
	\begin{enumerate}
	
		\item Em vez de escolhermos um contexto $S\aprs{R,T}$ a ser divido, escolheríamos um átomo $a$ em $S\{a\}$ a ser eliminado. No caso trivial, $a$ não estaria em uma estrutura copar, nos casos mais complicados, teríamos $S\aprs{R,T}$, onde $R = S'\{a\}$. A escolha de quem seria o átomo $a$ a ser eliminado seria feita baseada no conceito de \emph{número de incoerência}, que será introduzido mais adiante.
		
		\item Uma vez escolhido o átomo a ser eliminado, determinaríamos o seu \emph{killer}, que seria sempre $\dual{a}$. Por isso a estratégia seria aplicável somente a estruturas com pares de átomos distintos dois a dois, ou seja, só existe um átomo com o mesmo ``nome''.
		
		\item Escolhido o átomo e o seu \emph{killer}, procederíamos aplicando regras \emph{switch} até que ambos se encontrassem em uma estrutura par. Então poderíamos aplicar a regra $\ruleaidown$ para eliminá-los.
		
		\item O processo seguiria até que a estrutura fosse demonstrada, ou até que atingíssemos um estágio sobre o qual não fosse possível avançar, indicando que a estrutura não é demonstrável.
	
	\end{enumerate}
	
	É interessante notar que, se a estratégia for demonstrada correta, ela apresentará as seguintes características importantes:
	
	\begin{enumerate}
	
		\item Ela funcionaria para o sistema \FBV, restrita a estruturas com pares de átomos distintos dois a dois, ou seja, só poderia haver uma única ocorrência de cada átomo.
		
		\item Ela não exigiria \emph{backtracking} na construção de uma demonstração: uma vez tomado um passo na derivação, não seria preciso reconsiderá-lo.
		
		\item Se um estágio fosse atingido sobre o qual não seria mais possível avançar, é porque a estrutura inicial não era demonstrável.
	
	\end{enumerate}
	
	Estamos prontos para enunciar os resultados principais desta dissertação. Nesse ponto, é importante ressaltar que tais resultados são \emph{inéditos} e constituem em um avanço na busca de um conceito de demonstrações uniformes em cálculo das estruturas. Ainda muito deve ser feito, uma vez que nossos resultados estão restritos a um fragmento pequeno do sistema \BV. Mas é o \emph{primeiro} resultado teórico obtido na área. A seção \ref{section:definicoes} apresenta conceitos e definições necessárias, e a tentativa de  estratégia é formalmente apresentada e explicada em detalhes na seção \ref{section:resultados-principais}.
	
\subsection{Resultados fundamentais e definições}
\label{section:definicoes}

Vamos começar com um teorema que nos dá condições necessárias para que uma estrutura seja demonstrável em \FBV. Em seguida, vamos introduzir o conceito fundamental para a tentativa de estratétiga proposta: \emph{número de incoerência}.

\subsubsection{Condições necessárias para demonstrabilidade}
\label{section:teorema-condicoes}

Vamos começar com um lema necessário:

\begin{Lemma}[Propriedade X]
\label{lema:X}

	Seja uma estrutura $S$ da forma $S\pars{P\aprs{X\{x\}, Y\{y\}},Q\aprs{Z\{z\}, W\{w\}}}$. Não existem  derivações $\Pi_{1}$ e $\Pi_{2}$ em \FBV\ tais que:
	
	\begin{center}
		\begin{tabular}{cc}	
			$
			\strder{\Pi_{1}}{\FBV}{S\pars{P\aprs{X\{x\}, Y\{y\}},Q\aprs{Z\{z\}, W\{w\}}}}{\leaf{S'\aprs{P'\pars{X'\{x\}, W'\{w\}},Q'\pars{Y'\{y\}, Z'\{z\}}}}}
			$
	
			&		
			
			$
			\strder{\Pi_{2}}{\FBV}{S\pars{P\aprs{X\{x\}, Y\{y\}},Q\aprs{Z\{z\}, W\{w\}}}}{\leaf{S'\aprs{P'\pars{X'\{x\}, Z'\{z\}},Q'\pars{Y'\{y\}, W'\{w\}}}}}
			$	
		\end{tabular}
	\end{center}
		
	Equivalentemente, podemos enunciar o lema como:
	
	Uma subrede de interação da forma $\textsf{w}_{0}$ não pode ser transformada em uma subrede de interação da forma $\textsf{w}_{1}$ ou em uma subrede de interação da forma $\textsf{w}_{2}$, via regras de infererência de \FBV\ durante a construção de uma demonstração (de baixo para cima):

	$$
		\begin{tabular}{cccccc}

			& $\textsf{w}_{0}$ &  & $\textsf{w}_{1}$ & & $\textsf{w}_{2}$ \\
			
			Partindo de
			
			&		
			
			$
				\xymatrix{
					x \ar@{-}[r] \ar@{-}[d] \ar@{~}[dr]  & w \ar@{-}[d] \\
					z \ar@{-}[r] \ar@{~}[ur] &  y \\
				}	
			$
			
			&
						
			não podemos chegar a			
						
			&
			
			$
				\xymatrix{
					x \ar@{-}[r] \ar@{~}[d] \ar@{~}[dr]  & w \ar@{~}[d] \\
					z \ar@{-}[r] \ar@{~}[ur] &  y \\
				}	
			$
			
			&
			
			nem a 
			
			&
			
			$
				\xymatrix{
					x \ar@{~}[r] \ar@{-}[d] \ar@{~}[dr]  & w \ar@{-}[d] \\
					z \ar@{~}[r] \ar@{~}[ur] &  y \\
				}	
			$

			\\				
		\end{tabular}
	$$			

\begin{proof}

	Sabemos que em \FBV\ o número de aplicações de regras de inferência sobre uma estrutura $S_{1}$, demonstrável ou não, é finito. Seja $n(S_{1}) = |\rho|$ onde $\rho \in \{\ruleodown, \ruleaidown, \rules\}$ são todas as instâncias de regras de inferência aplicáveis em \FBV\ tais que:
	
	$$
		\dernote{\rho}{}{S_{1}}
		{\leaf{S_{2}}}
	$$
	
	Temos que, necessariamente, $n(S_{2}) < n(S_{1})$
	
	Portanto, vamos utilizar este fato para demonstrar o lema por indução no número de possibilidades de aplicações de regras $n(S)$. 
	
	\begin{description}
		
		\item[Casos base:]
		
				\ \\			
				\begin{enumerate}
					
					\item $P\{\circ \} = \circ$
					
						Uma possibilidade é:
					
						$$
							\dernote{\rules}{}{S\pars{\aprs{X\{ x \}, Y\{ y \}},Q\aprs{Z\{ z \}, W\{ w \}}}}
							{\leaf{S\aprs{\pars{X\{ x \},Q\aprs{Z\{ z \},W\{ w \}}},Y\{ y \}}}}
						$$
												
						Mas, com isso, $y \uparrow z$ (o que vai contra a criação da rede $\textsf{w}_{1}$) e $y \uparrow w$ (o que vai contra a criação a rede $\textsf{w}_{2}$). Dado que uma vez que criamos uma relação $\uparrow$ de baixo para cima em uma demonstração ela não pode mais ser desfeita, então não poderemos criar as desejadas $\textsf{w}_{1}$ ou $\textsf{w}_{2}$.
						
						A outra possibilidade é:
						
						$$
							\dernote{\rules}{}{S\pars{\aprs{X\{ x \}, Y\{ y \}},Q\aprs{Z\{ z \}, W\{ w \}}}}
							{\leaf{S\aprs{\pars{Y\{ y \},Q\aprs{Z\{ z \},W\{ w \}}},X\{ x \}}}}
						$$
						
						Mas, com isso, $x \uparrow w$ (o que vai contra a criação da rede $\textsf{w}_{1}$) e $x \uparrow z$ (o que vai contra a criação da rede $\textsf{w}_{2}$). Dado que uma vez que criamos uma relação $\uparrow$ de baixo para cima em uma demonstração ela não pode mais ser desfeita, então não poderemos criar as desejadas $\textsf{w}_{1}$ ou $\textsf{w}_{2}$.
						
					\item $Q\{\circ \} = \circ$
						
						Análogo ao caso anterior.		
						
					\item Aplica-se uma instância de $\ruleaidown$ sobre um ocorrência de átomo $x$, $y$, $z$ ou $w$ e seu respectivo dual.
					
						Neste caso o átomo em questão não figura mais na teia de interação da estrutura e, portanto, claramente as subredes $\textsf{w}_{1}$ e $\textsf{w}_{2}$ não podem ser atingidas.
					
				\end{enumerate}
		
		\item[Casos indutivos:]
		
				\ \\
				Vamos argumentar baseados na última aplicação de regra $\rho$ sobre a estrutura \\
				$S\pars{P\aprs{X\{x\}, Y\{y\}},Q\aprs{Z\{z\}, W\{w\}}}$:
				
				$$
					\dernote{\rho}{}{S\pars{P\aprs{X\{x\}, Y\{y\}},Q\aprs{Z\{z\}, W\{w\}}}}
					{\leaf{S^{*}}}				
				$$
		
				Nos concentraremos nos casos em que $\rho$ é não-trivial.
		
				\begin{enumerate}				
				
					\item $\rho = \ruleaidown$
					
						\begin{enumerate}
							
							\item O redex está em $X\{ \ \}$
								
								$$
									\dernote{\ruleaidown}{}{S\pars{P\aprs{X\{x\}, Y\{y\}},Q\aprs{Z\{z\}, W\{w\}}}}
									{\leaf{S\pars{P\aprs{X^{*}\{x\}, Y\{y\}},Q\aprs{Z\{z\}, W\{w\}}}}}
								$$
								
								Na estrutura do contractum ainda temos a subrede de interação $\textsf{w}_{0}$, porém o número de possibilidades de aplicação de regras agora é menor. Podemos, então, aplicar a hipótese de indução.
								
							\item O redex está em $Y\{ \ \}$, $Z\{ \ \}$, $W\{ \ \}$, $P\{ \ \}$, $Q\{ \ \}$ ou $S\{ \ \}$
							
							Análogos ao caso anterior.
								
						\end{enumerate}
						
					\item $\rho = \rules$	
					
						\begin{enumerate}
							
							\item O redex está em $X\{ \ \}$
								
								$$
									\dernote{\rules}{}{S\pars{P\aprs{X\{x\}, Y\{y\}},Q\aprs{Z\{z\}, W\{w\}}}}
									{\leaf{S\pars{P\aprs{X^{*}\{x\}, Y\{y\}},Q\aprs{Z\{z\}, W\{w\}}}}}
								$$
								
								Na estrutura do contractum ainda temos a subrede de interação $\textsf{w}_{0}$, porém o número de possibilidades de aplicação de regras agora é menor. Podemos, então, aplicar a hipótese de indução.
								
							\item O redex está em $Y\{ \ \}$, $Z\{ \ \}$, $W\{ \ \}$, $P\{ \ \}$, $Q\{ \ \}$ ou $S\{ \ \}$
							
							Análogos ao caso anterior.
							
							\item O redex envolve $P\aprs{X\{x\}, Y\{y\}}$ e $Q\aprs{Z\{z\}, W\{w\}}$
							
								Sabemos que $P\{ \circ \} \neq \circ \neq Q\{ \circ \}$, pois senão teríamos o caso base.
								
								Vamos analisar as possibilidades para $P\aprs{X\{x\}, Y\{y\}}$:
								
								\begin{enumerate}
									
									\item $P\aprs{X\{x\}, Y\{y\}} = \pars{\aprs{X\{x\}, Y\{y\}}, P_{1}}$
									
										Mas então a estrutura seria da forma $S\pars{\aprs{X\{x\}, Y\{y\}}, P_{1}, Q\aprs{Z\{z\}, W\{w\}}}$, que é o equivalente a um dos casos base.
									
									\item $P\aprs{X\{x\}, Y\{y\}} = \aprs{X\{x\}, Y\{y\}, P_{1}}$
									
										Mas então podemos chamar $Y''\{y\} = \aprs{Y\{y\}, P_{1}}$ e temos que a estrutura é da forma $S\pars{\aprs{X\{x\}, Y''\{y\}},Q\aprs{Z\{z\}, W\{w\}}}$, que é o equivalente a um dos casos base.
									
									\item $P\aprs{X\{x\}, Y\{y\}} = \aprs{\pars{P''\aprs{X\{x\}, Y\{y\}}, P_{2}}, P_{1}}$
									
								\end{enumerate}
								
								As possibilidades para $Q\aprs{Z\{z\}, W\{w\}}$ são análogas. 
								
								Portanto, o único caso a ser analisado é o em que \\
								$P\aprs{X\{x\}, Y\{y\}} = \aprs{\pars{P''\aprs{X\{x\}, Y\{y\}}, P_{2}}, P_{1}}$ e\\
								$Q\aprs{Z\{z\}, W\{w\}} = \aprs{\pars{Q''\aprs{Z\{z\}, W\{w\}}, Q_{2}}, Q_{1}}$.\\
								Façamos $P''' = \pars{P''\aprs{X\{x\}, Y\{y\}}, P_{2}}$ e\\
								$Q''' = \pars{Q''\aprs{Z\{z\}, W\{w\}}, Q_{2}}$.
								
								As possibilidades de aplicação da regra switch sobre a estrutura copar $\aprs{P''', P_{1}}$:
								
								$$
									\dernote{\rules}{}{S\pars{\aprs{P''', P_{1}},\aprs{Q''', Q_{1}}}}
									{\leaf{S\aprs{\pars{P''', \aprs{Q''', Q_{1}}},P_{1}}}}
								$$
								
								Na estrutura do contractum ainda temos a subrede de interação $\textsf{w}_{0}$, porém o número de possibilidades de aplicação de regras agora é menor. Podemos, então, aplicar a hipótese de indução.
								
								$$
									\dernote{\rules}{}{S\pars{\aprs{P''', P_{1}},\aprs{Q''', Q_{1}}}}
									{\leaf{S\aprs{\pars{P_{1}, \aprs{Q''', Q_{1}}},P'''}}}
								$$
								
								Mas, com isso, $x \uparrow w$ e $y \uparrow z$ (o que vai contra a criação da rede $\textsf{w}_{1}$) e $x \uparrow z$ e $y \uparrow w$ (o que vai contra a criação da rede $\textsf{w}_{2}$). Dado que uma vez que criamos uma relação $\uparrow$ de baixo para cima em uma demonstração ela não pode mais ser desfeita, então não poderemos criar as desejadas $\textsf{w}_{1}$ ou $\textsf{w}_{2}$.
								
								As possibilidades de aplicação da regra switch sobre a estrutura copar $\aprs{Q''', Q_{1}}$ são análogas.								
								
						\end{enumerate}
					
				\end{enumerate}
		
	\end{description}
	
\end{proof}

\end{Lemma}

Agora podemos enunciar o teorema que nos fornece condições necessárias para uma estrutural ser demonstrável em \FBV.

\begin{Theorem}[Condições necessárias para demonstrabilidade]
\label{teorema:condicoes}

	Se uma estrutura $S$ é demonstrável em \FBV, então:
	
	\begin{enumerate}
	
		\item Condição $C_{1}$
		
			$\exists f : \occ{S} \rightarrow \occ{S}. \ f(a) = \dual{a} \ \  \wedge \ \ (a \downarrow \dual{a})$ em $S$
	
		\item Condição $C_{2}$
		
$$\forall a \in  \occ{S} . \ \not{\exists} \ q \in  \occ{S} \ \ . \ \ S[S'(A\{a\}, \ Q\{q\}), \ \ S''(A'\{\dual{a} \}, \ Q'\{\dual{q}\})]$$
onde $S\{ \ \}$, $S'\{ \ \}$, $S''\{ \ \}$, $A\{ \ \}$, $Q\{ \ \}$, $A'\{ \ \}$, $Q'\{ \ \}$ são contextos, possivelmente vazios.
			 
	\end{enumerate}
	
	\begin{Remark}
		De agora em diante, consideraremos $\dual{a} \equiv f(a)$ para todo $a \in \occ{S}$, considerando que a negação em \BV\ é involutiva ($\dual{\dual{a}} = a$) e que, se existe mais de uma ocorrência de um átomo $a$ em $S$, elas podem ser diferenciadas através do uso de índices.
	\end{Remark}
	
	\begin{Remark}
		O Teorema~\ref{teorema:condicoes} é equivalente a $\strpr{}{\FBV}{S} \implies C_{1} \wedge C_{2}$		
	\end{Remark}
	
	\begin{Remark}
	A condição $C_{2}$ é equivalente a:
	
	$\not \exists a, \dual{a}, q, \dual{q} \in \occ{S} . $
	
	$$
	\xymatrix{
		a \ar@{-}[r] \ar@{-}[d] \ar@{~}[dr]  & \dual{q} \ar@{-}[d] \\
		\dual{a} \ar@{-}[r] \ar@{~}[ur] &  q \\
	}	
	$$
	\end{Remark}

	\begin{proof}
	
		O contrapositivo do teorema diz que $\neg{C_{1}} \vee \neg{C_{2}} \implies \neg \left[ \vcenter{\strpr{}{\FBV}{S}} \right ]$.
		
		\begin{itemize}
		
			\item $\neg{C_{1}} \implies \neg \left[ \vcenter{\strpr{}{\FBV}{S}} \right]$
			
				A demonstração segue diretamente das regras do sistema \FBV. Numa visão de baixo para cima, nenhuma aplicação de regra de inferência pode criar uma relação $\downarrow$ entre dois átomos. Observe na figura~\ref{fig:condicao-c1} que as relações $\uparrow$ são sempre mantidas do contractum para o redex, e que as relações $\downarrow$ do contractum ou são mantidas ou convertidas em relações $\uparrow$. Em outras palavras, relações $\downarrow$ jamais são criadas de baixo para cima em uma demonstração.
				
				\begin{figure}[!htb]
					\center
					
					\begin{tabular}{|c||c|c|c|}
					
						\hline
						Regra & Antes da aplicação & $\downarrow$ mantidas & $\downarrow$ transformadas em $\uparrow$ \\
						\hline \hline
					
						\dernote{\circ \downarrow}{}{\circ}{\leaf{}} & - & - & - \\ \hline

						\dernote{\ruleaidown}{}{S\pars{a,\dual{a}}}{\leaf{S\{\circ\}}} & $a \downarrow a$ & - & $a - a$ \\ \hline
											
	
						\dernote{\rules}{}{S\pars{\aprs{R,R'},T}}{\leaf{S\aprs{\pars{R,T},R'}}} & $R \downarrow T$, $R' \downarrow T$, $R \uparrow R'$ & $R \downarrow T$ & $R' \uparrow T$\\ \hline
	
					\end{tabular}
					
					\label{fig:condicao-c1}				
					\caption{Regras de inferência não criam relações $\downarrow$ quando vistas de baixo para cima.}
				\end{figure}								
				
				Portanto, se existe um átomo $a \in \occ{S}$ para o qual não exista um $\dual{a} \in \occ{S}$ tal que $a \downarrow \dual{a}$, nunca será possível criar, via regras de inferência, uma relação $\downarrow$ entre estes átomos. Levando em conta que toda estrutura demonstrável deve cancelar seus átomos (relacionados por $\downarrow$) via instâncias da regra $\ruleaidown$, $a$ nunca poderá ser cancelado com $\dual{a}$ e $S$ não é, portanto, demonstrável. 
			
			\item $\neg{C_{2}} \implies \neg \left[ \vcenter{\strpr{}{\FBV}{S}} \right]$
			
				Se nós temos $\neg{C_{2}}$, então a estrutura $S$ é da forma:
				
				$$ S\pars{S'\aprs{A\{a\},Q\{q\}},S''\aprs{A'\{\dual{a}\},Q'\{\dual{q}\}}} $$			
				
				Note que a estrutura possui uma subrede de interação da forma:
	
				$$
					\xymatrix{
						a \ar@{-}[r] \ar@{-}[d] \ar@{~}[dr]  & \dual{q} \ar@{-}[d] \\
						\dual{a} \ar@{-}[r] \ar@{~}[ur] &  q \\
					}				
				$$
				
				Se a estrutura for demonstrável, então uma hora necessariamente deve haver uma aplicação de regra:
				
				$$
					\dernote{\ruleaidown}{}{S\pars{a,\dual{a}}}
					{\leaf{S\{\circ\}}}
				$$
				, onde podemos chamar $\pars{a,\dual{a}} = R$. (O caso em que desejássemos argumentar baseados em $q$ e $\dual{q}$ seria análogo.)
					
				Mas sabemos que, dada uma estrutura $S\{R\}$ e duas ocorrências de átomos $a$ em $S\{ \ \}$ e $b$ em $R$, se $a \triangleleft b$ (respectivamente, $a \triangleright b$, $a \downarrow b$, $a \uparrow b$) então $a \triangleleft c$ (respectivamente, $a \triangleright c$, $a \downarrow c$, $a \uparrow c$) para todas as ocorrências de átomos $c$ em $R$.
				
				Logo, se a estrutura for demonstrável, em um dado ponto da demonstração $a$ e $\dual{a}$ devem manter a mesma relação estrutural com \emph{todas} as ocorrências de átomos em $S$, em particular com $q$ e com $\dual{q}$.
				
				Vamos nos concentrar na subrede de interação:
				
				$$
					\xymatrix{
						a \ar@{-}[r] \ar@{-}[d] \ar@{~}[dr]  & \dual{q} \ar@{-}[d] \\
						\dual{a} \ar@{-}[r] \ar@{~}[ur] &  q \\
					}				
				$$
				
				Como não há maneira de transformar uma relação $\uparrow$ em $\downarrow$ na construção de baixo para cima de uma demonstração em \FBV, só nos resta tentar transformar as relações $\downarrow$ em $\uparrow$. Em particular, temos que criar pelo menos duas relações: $a \uparrow \dual{q}$ e $\dual{a} \uparrow q$.
				
				Vamos analisar os casos. Suponha que consigamos criar $a \uparrow \dual{q}$. Pela propriedade do quadrado, na teia de interação de uma estrutura, um quadrado não pode apresentar exatamente três lados representando a mesma relação estrutural: deve haver um quarto lado. Portanto, teremos um dos seguintes casos:
				
				$$
					\begin{array}{ccccc}
						
						& &

							\xymatrix{
								a \ar@{~}[r] \ar@{-}[d] \ar@{~}[dr]  & \dual{q} \ar@{-}[d] \\
								\dual{a} \ar@{-}[r] \ar@{~}[ur] &  q \\
							}

						& & \\
						
						& &
						
						\implies
						& & \\
						
						\framebox(95,85){
							I \ \			
							\xymatrix{
								a \ar@{~}[r] \ar@{~}[d] \ar@{~}[dr]  & \dual{q} \ar@{-}[d] \\
								\dual{a} \ar@{-}[r] \ar@{~}[ur] &  q \\
							}		
						}					
						
						&
						
						\vee
						
						& 
						
						\framebox(95,85){
							II \ \ 
							\xymatrix{
								a \ar@{~}[r] \ar@{-}[d] \ar@{~}[dr]  & \dual{q} \ar@{~}[d] \\
								\dual{a} \ar@{-}[r] \ar@{~}[ur] &  q \\
							}		
						}
							
						&
						
						\vee
						
						& 
						
						\framebox(95,85){
							III \ \ 
							\xymatrix{
								a \ar@{~}[r] \ar@{-}[d] \ar@{~}[dr]  & \dual{q} \ar@{-}[d] \\
								\dual{a} \ar@{~}[r] \ar@{~}[ur] &  q \\
							}			
						}
						
						\\
						
					\end{array}
				$$
				
				Mas no caso I, a estrutura não seria demonstrável, uma vez que $a \uparrow \dual{a}$. No caso II, $q \uparrow \dual{q}$ e também a estrutura não seria demonstrável. Já o caso III implicaria que, durante a construção de uma demonstração:
				
				$$
					\begin{tabular}{cccc}

						& $\textsf{w}_{0}$ &  & $\textsf{w}_{1}$ \\
						
						Partindo de
						
						&		
						
						$
							\xymatrix{
								a \ar@{-}[r] \ar@{-}[d] \ar@{~}[dr]  & \dual{q} \ar@{-}[d] \\
								\dual{a} \ar@{-}[r] \ar@{~}[ur] &  q \\
							}	
						$
						
						&
									
						chegamos a			
									
						&
						
						$
							\xymatrix{
								a \ar@{~}[r] \ar@{-}[d] \ar@{~}[dr]  & \dual{q} \ar@{-}[d] \\
								\dual{a} \ar@{~}[r] \ar@{~}[ur] &  q \\
							}	
						$
						\\				
					\end{tabular}
				$$
				, o que não é possível, pelo Lema \ref{lema:X}.
				
				Da mesma forma, se supusermos que conseguimos criar $\dual{a} \uparrow q$, teremos casos análogos.										
		\end{itemize}
	
	\end{proof}

\end{Theorem}

\subsubsection{Número de incoerência}
\label{section:numero-de-incoerencia}

As Definições~\ref{def:aincnumber-1} e~\ref{def:aincnumber-2} são a chave do nosso trabalho. Elas são equivalentes, e a diferença consiste na forma de apresentação: a Definição~\ref{def:aincnumber-2} é recursiva.

\begin{Definition}[Número de incoerência]

	\label{def:aincnumber-1}

	Seja $S$ uma estrutura em \FBV. Sejam $a$ e $b$ dois átomos $\in \occ{S}$. Sejam $\sigma_{1}, \sigma_{2} \in \{\downarrow, \uparrow\}$ possíveis relações entre átomos em uma estrutura. O \emph{número de incoerência} do par de átomos $\{a,b\}$ no contexto $S$ é definido da seguinte maneira:

\begin{enumerate}

	\item $\aincnumber{S}{a}{b} = |\{x \in \occ{S} \ | \ (a \ \sigma_{1} \ x) \ \wedge \ (b \ \sigma_{2} \ x) \ \wedge
 \ (\sigma_{1} \neq \sigma_{2}) \}|$, se $a \downarrow b$;
 
	\item $\aincnumber{S}{a}{b}= \infty$, se $a \uparrow b$;
	
\end{enumerate}
 
\end{Definition}

O número de incoerência entre dois átomos $a$ e $b$ dentro da estrutura $S$, por motivos que veremos adiante, só faz sentido se $a \downarrow b$. Nesse caso, ele é definido como número de ocorrências de átomos de $S$ que mantêm uma relação estrutural com $a$ e outra diferente com $b$ (daí o nome de \emph{número de incoerência}).

\begin{Example}

Seja a estrutura $S = \pars{a,b,\aprs{c,\pars{d,e}}}$. Assim temos:

$\aincnumber{S}{a}{b} = 0$;

$\aincnumber{S}{a}{c} = 2$, pois $a \downarrow d$ e $a \downarrow e$ mas $c \uparrow d$ e $c \uparrow e$;

$\aincnumber{S}{a}{d} = 1$, pois $a \downarrow c$ mas $d \uparrow c$;

$\aincnumber{S}{c}{d} = \infty$, pois $c \uparrow d$;

\end{Example}

\begin{Definition}[Número de incoerência]

	\label{def:aincnumber-2}

	Seja $S$ uma estrutura em \FBV. Sejam $a$ e $b$ dois átomos $\in \occ{S}$. O \emph{número de incoerência} do par de átomos $\{a,b\}$ no contexto $S$ é recursivamente definido da seguinte maneira:
	
	\begin{enumerate}
	
		\item $\aincnumber{S\pars{a,b}}{a}{b} = 0$
	
		\item $\aincnumber{S\pars{S'\{a\},S''\{b\}, P}}{a}{b} = \aincnumber{\pars{S'\{a\},S''\{b\}}}{a}{b}$
		
		\item $\aincnumber{S\pars{S'\{a\},\aprs{S''\{b\},P}}}{a}{b} = |\occ{P}| + \aincnumber{\pars{S'\{a\},S''\{b\}}}{a}{b}$ \ \ \ , onde $S''\{b\}$ não é um copar próprio.
		
		\item $\aincnumber{S\pars{\aprs{S'\{a\},P},S''\{b\}}}{a}{b} = |\occ{P}| + \aincnumber{\pars{S'\{a\},S''\{b\}}}{a}{b}$\ \ \ , onde $S'\{a\}$ não é um copar próprio.
		
		\item $\aincnumber{S\aprs{S'\{a\},S''\{b\}}}{a}{b} = \infty$
		
	\end{enumerate}
e, em todos os casos, $P$ é uma estrutura.

\end{Definition}

\begin{Example}

Seja a estrutura $S = \pars{\aprs{\pars{a,b},c},\aprs{d,\pars{e,f}}}$. Nesse caso temos:

	\begin{tabular}{lll}

		$\aincnumber{S}{a}{b} = $ & $\aincnumber{S'''\pars{a,b}}{a}{b} = 0$ & , pelo caso (1); \\ \\ 
		
		$\aincnumber{S}{a}{c} = $ & $\aincnumber{S'''\aprs{S'\{a\},S''\{c\}}}{a}{c} = \infty$ & , pelo caso (5); \\ \\
		
		$\aincnumber{S}{a}{d} = $ & $\aincnumber{\pars{\aprs{S'\{a\},P},S''\{d\}}}{a}{d} = $ & , pelo caso (4)\\
		                          & $\aincnumber{\pars{\aprs{S'\{a\},c},S''\{d\}}}{a}{d} = $ &                \\
		                          & $|\occ{P}| + \aincnumber{\pars{S'\{a\},S''\{d\}}}{a}{d}  = $ &                \\
		                          & $|\occ{c}| + \aincnumber{\pars{a,b,\aprs{d,\pars{e,f}}}}{a}{d}  = $ &         \\
		                          & $1 + \aincnumber{\pars{S'\{a\},S''\{d\},P'}}{a}{d} =$ & , pelo caso (2) \\
		                          & $1 + \aincnumber{\pars{a,\aprs{d,\pars{e,f}}}}{a}{d} =$ & \\
		                          & $1 + \aincnumber{\pars{S'\{a\},\aprs{S'''\{d\},P''}}}{a}{d} =$ & , pelo caso (3) \\
		                          & $1 + |\occ{P''}| + \aincnumber{\pars{a,d}}{a}{d} =$ &  \\
		                          & $1 + |\occ{\pars{e,f}}| + \aincnumber{\pars{a,d}}{a}{d} = $ &  \\
		                          & $1 + 2 + 0 = $ & , pelo caso (1)  \\
		                          & $3$ &  \\ \\
	
	\end{tabular}

\end{Example}

Considere a seguinte estrutura:

$$S = \pars{a,\aprs{b,\pars{c,d}}}$$

Pela definição de número de incoerência, $\aincnumber{S}{a}{b} = 2$. Entretanto, observe que se considerarmos a subestrutura $T = \pars{c,d}$, podemos reescrever $S$ como:

$$S = \pars{a,\aprs{b,T}}$$

Nesse caso, pode-se abstrair do tamanho de $T$, ou seja, quantas ocorrências de átomos $T$ possui, e focar no que realmente interessa: o número de incoerência entre os átomos $a$ e $b$ se deve às relações estruturais diferentes que estes átomos mantêm com toda a subestrutura $T$, não importando se $T$ é um átomo ou uma estrutura maior. Dessa forma, faria sentido generalizar a idéia de incoerência além de átomos, lidando com subestruturas. O número de incoerência entre $a$ e $b$ em $S$ seria $1$, pois $a$ e $b$ mantêm relações diferentes com a subestrutura $T$, independentemente de seu tamanho ou natureza. Essa é a idéia de número de incoerência módulo coerência.

\begin{Definition}[Número de incoerência módulo coerência]

	\label{def:incnumber}

	Seja $S$ uma estrutura em {\FBV}. Sejam $U$ e $V$ duas subestruturas disjuntas em $S$. O \emph{número de incoerência módulo coerência} do par de estruturas $\{U,V\}$ dentro da estrutura $S$ é recursivamente definido da seguinte maneira:
	
	\begin{enumerate}
	
		\item $\incnumber{S\pars{U,V}}{U}{V} = 0$
	
		\item $\incnumber{S\pars{S'\{U\},S''\{V\}, P}}{U}{V} = \incnumber{\pars{S'\{U\},S''\{V\}}}{U}{V}$
		
		\item $\incnumber{S\pars{S'\{U\},\aprs{S''\{V\},P}}}{U}{V} = 1 + \incnumber{\pars{S'\{U\},S''\{V\}}}{U}{V}$ \ \ \ , onde $S''\{V\}$ não é um copar próprio.
		
		\item $\incnumber{S\pars{\aprs{S'\{U\},P},S''\{V\}}}{U}{V} = 1 + \incnumber{\pars{S'\{U\},S''\{V\}}}{U}{V}$ \ \ \ , onde $S'\{U\}$ não é um copar próprio.
		
		\item $\incnumber{S\aprs{S'\{U\},S''\{V\}}}{U}{V} = \infty$
		
	\end{enumerate}
onde $P$ é uma estrutura.

\end{Definition}

\begin{Example}

Seja a estrutura $S = \pars{\aprs{\pars{a,b},c},\aprs{d,\pars{e,f}}}$. Nesse caso temos:

	\begin{tabular}{lll}

		$\incnumber{S}{a}{b} = $ & $\incnumber{S'''\pars{a,b}}{a}{b} = 0$ & , pelo caso (1); \\ \\ 
		
		$\incnumber{S}{a}{c} = $ & $\incnumber{S'''\aprs{S'\{a\},S''\{c\}}}{a}{c} = \infty$ & , pelo caso (5); \\ \\
		
		$\incnumber{S}{a}{d} = $ & $\incnumber{\pars{\aprs{S'\{a\},P},S''\{d\}}}{a}{d} = $ & , pelo caso (4)\\
		                          & $\incnumber{\pars{\aprs{S'\{a\},c},S''\{d\}}}{a}{d} = $ &                \\
		                          & $|\occ{P}| + \incnumber{\pars{S'\{a\},S''\{d\}}}{a}{d}  = $ &                \\
		                          & $|\occ{c}| + \incnumber{\pars{a,b,\aprs{d,\pars{e,f}}}}{a}{d}  = $ &         \\
		                          & $1 + \incnumber{\pars{S'\{a\},S''\{d\},P'}}{a}{d} =$ & , pelo caso (2) \\
		                          & $1 + \incnumber{\pars{a,\aprs{d,\pars{e,f}}}}{a}{d} =$ & \\
		                          & $1 + \incnumber{\pars{S'\{a\},\aprs{S'''\{d\},P''}}}{a}{d} =$ & , pelo caso (3) \\
		                          & $1 + 1 + \incnumber{\pars{a,d}}{a}{d} =$ &  \\
		                          & $2 + \incnumber{\pars{a,d}}{a}{d} = $ &  \\
		                          & $2 + 0 = $ & , pelo caso (1)  \\
		                          & $2$ &  \\ \\
	
	\end{tabular}

\end{Example}

Note que, para a estrutura $S = \pars{\aprs{\pars{a,b},c},\aprs{d,\pars{e,f}}}$, o número de incoerência entre os átomos $a$ e $d$ é $\aincnumber{S}{a}{b} = 3$, mas o número de incoerência módulo coerência é  $\incnumber{S}{a}{b} = 2$

\subsection{Resultados principais}
\label{section:resultados-principais}

Agora estamos prontos, para, usando o teorema de condições necessárias para demonstrabilidade e a idéia de número de incoerência módulo coerência, começar a propor a nossa tentativa de estratégia de demonstração para o sistema \FBV\ com pares de átomos distintos dois a dois. 

Antes de mais nada, entretando, vamos nos deter um momento para entender o porquê de no momento restringirmos a nossa estratégia ao sistema \FBV\ com pares de átomos dois a dois distintos. 

\subsubsection{Porque \FBV\ com pares de átomos dois a dois distintos}
\label{section:fbv-distintos}

Como vimos pelo Teorema \ref{teorema:condicoes}, a condição $C_{1}$ exige que, para uma estrutura demonstrável $S$ haja uma função de escolha 

$$f : \occ{S} \rightarrow \occ{S}. \ f(a) = \dual{a} \ \  \wedge \ \ (a \downarrow \dual{a}) \ \ em \ \ S$$

Entretanto, não é possível, a princípio, determinar trivialmente qual ocorrência de $a$ deve ser associada a qual ocorrência de $\dual{a}$ via a função $f$. Como mostram os exemplos a seguir, esta escolha não pode ser feita arbitrariamente.
 
\begin{Example}\label{exemplo:eliminacao_inicial}
A estrutura
$\pars{a,\dual{a},b,\dual{b},\aprs{a,b},\aprs{\dual{a},\dual{b}}}$
é demonstrável. 
Apenas com o objetivo de facilitar a argumentação, vamos diferenciar cada ocorrência de átomos repetidos indexando-os da seguinte forma: $\pars{a_{1},\dual{a}_{1},b_{1},\dual{b}_{1},\aprs{a_{2},b_{2}},\aprs{\dual{a}_{2},\dual{b}_{2}}}$.

Esta estrutura é demonstrável, como podemos ver a seguir:
$$\dernote{\rules}{}{\pars{a_{1},\dual{a}_{1},b_{1},\dual{b}_{1},\aprs{a_{2},b_{2}},\aprs{\dual{a}_{2},\dual{b}_{2}}}}
{\rootnote{\ruleaidown}{}{\pars{a_{1},b_{1},\dual{b}_{1},\aprs{\pars{a_{2},\dual{a}_{1}},b_{2}},\aprs{\dual{a}_{2},\dual{b}_{2}}}}
{\root{\ruleaidown}{\pars{a_{1},b_{1},\dual{b}_{1},b_{2},\aprs{\dual{a}_{2},\dual{b}_{2}}}}
{\root{\rules}{\pars{a_{1},b_{2},\aprs{\dual{a}_{2},\dual{b}_{2}}}}
{\root{\ruleaidown}{\pars{b_{2},\aprs{\pars{a_{1},\dual{a}_{2}},\dual{b}_{2}}}}
{\root{\ruleaidown}{\pars{b_{2},\dual{b}_{2}}} {\root{\ruleodown}{\circ}
{\leaf{}}}}}}}}
$$
o que equivale à função de escolha $f(a_{1}) = \dual{a}_{2}$, $f(a_{2}) = \dual{a}_{1}$, $f(b_{1}) = \dual{b}_{1}$ e $f(b_{2}) = \dual{b}_{2}$ (basta olhar os índices das ocorrências atômicas em cada instância de aplicação da regra $\ruleaidown$ na demonstração).

Se, por outro lado, optássemos por começar a demonstração pela aplicação da regra $\ruleaidown$ duas vezes, eliminando primeiro os átomos na estrutura par, cairíamos na seguinte situação:

$$\dernote{\ruleaidown}{}{\pars{a_{1},\dual{a}_{1},b_{1},\dual{b}_{1},\aprs{a_{2},b_{2}},\aprs{\dual{a}_{2},\dual{b}_{2}}}}
{\rootnote{\ruleaidown}{}{\pars{b_{1},\dual{b}_{1},\aprs{a_{2},b_{2}},\aprs{\dual{a}_{2},\dual{b}_{2}}}}
{\leaf{{\pars{\aprs{a_{2},b_{2}},\aprs{\dual{a}_{2},\dual{b}_{2}}}}}}}
$$

Mas a estrutura $\pars{\aprs{a,b},\aprs{\dual{a},\dual{b}}}$ não é demonstrável, pela condição $C_{2}$ do Teorema \ref{teorema:condicoes}. Portanto, não podemos começar eliminando todos os átomos duais da estrutura par mais externa, ou seja, a função de escolha $f(a_{1}) = \dual{a}_{1}$, $f(a_{2}) = \dual{a}_{2}$, $f(b_{1}) = \dual{b}_{1}$ e $f(b_{2}) = \dual{b}_{2}$ não é válida.

Apenas para ressaltar o quão complicada a escolha da função pode ficar, podemos mostrar que não é necessariamente preciso adiar a aplicação das regras $\ruleaidown$ para o final da construção da demonstração: é possível construir uma demonstração da mesma estrutura em que começamos com uma aplicação $\ruleaidown$ sobre apenas um dos pares de átomos da estrutura par mais externa, no caso, o par $a_{1}$ e $\dual{a}_{1}$:
$$\dernote{\ruleaidown}{}{\pars{a_{1},\dual{a}_{1},b_{1},\dual{b}_{1},\aprs{a_{2},b_{2}},\aprs{\dual{a}_{2},\dual{b}_{2}}}}
{\rootnote{\rules}{}{\pars{b_{1},\dual{b}_{1},\aprs{a_{2},b_{2}},\aprs{\dual{a}_{2},\dual{b}_{2}}}}
{\root{\ruleaidown}{\pars{b_{1},\aprs{a_{2},\pars{b_{2},\dual{b}_{1}}},\aprs{\dual{a}_{2},\dual{b}_{2}}}}
{\root{\rules}{\pars{b_{1},a_{2},\aprs{\dual{a}_{2},\dual{b}_{2}}}}
{\root{\ruleaidown}{\pars{b_{1},\aprs{\pars{a_{2},\dual{a}_{2}},\dual{b}_{2}}}}
{\root{\ruleaidown}{\pars{b_{1},\dual{b}_{2}}} {\root{\ruleodown}{\circ}
{\leaf{}}}}}}}}
$$
o que equivale à função de escolha $f(a_{1}) = \dual{a}_{1}$, $f(a_{2}) = \dual{a}_{2}$, $f(b_{1}) = \dual{b}_{2}$ e $f(b_{2}) = \dual{b}_{1}$.

O mesmo poderia ser feito se escolhêssemos eliminar $b_{1}$ e $\dual{b}_{1}$ primeiro. De fato, das quatro funções de escolha possíveis, três podem levar a uma demonstração, e uma necessariamente não leva.

Como vemos, escolher qual a função $f$ adequada para cada estrutura não parece ser uma tarefa trivial, e por isso, não foi focada neste trabalho. Eliminamos o problema de lidar com $f$ restringindo o sistema \FBV\ de modo a ter apenas pares de átomos dois a dois distintos. Desta forma, a cada ocorrência de $a \in \occ{S}$ só existe uma única possibilidade de mapeamento via função $f$, que consiste em $\dual{a} \in \occ{S}$.

Contornado o problema da função $f$, vamos agora justificar o porquê de trabalharmos com \FBV\ no momento, e não com o sistema \BV\ completo. Em primeiro lugar, porque queremos começar do subsistema mais simples para depois generalizar a estratégia para o supersistema mais complexo. Além disso, a presença da regra $\ruleqdown$ em \BV\ complica a escolha das regras de inferência a aplicar, uma vez que não é possível ordenar as instâncias de regra em uma demonstração de forma que se possa usar todas as instâncias de uma regra antes de todas as instâncias de outra regra. Mais especificamente, isso quer dizer que mesmo que os átomos da estrutura a ser demonstrada sejam dois a dois distintos, a presença da regra $\ruleqdown$ não permite que a aplicação da regra $\ruleaidown$ seja adiada de modo a ocorrer apenas na parte superior da derivação, depois que todas as instâncias $\ruleqdown$ já tenham sido aplicadas. De fato, a estrutura
$$\pars{\seqs{\aprs{\pars{d,\dual{d}},\seqs{a;b}} ; c } , \seqs{ \dual{a} ; \aprs{\seqs{\dual{b};\dual{c}},\pars{e,\dual{e}}}}}$$
é demonstrável mas, temos que aplicar a regra $\ruleaidown$ no início da demonstração senão a estrutura copar que envolve o seq não permitirá a aplicação da regra $\ruleqdown$.
\end{Example}

\subsubsection{A tentativa estratégia de demonstração}
\label{section:estrategia}

Vamos começar por um lema chave para a nossa estratégia.

\begin{Lemma}[``Esquecimento'']
\label{lema:forget}
Seja $S\{a,\dual{a}\}=S\{X\{a\},Y\{\dual{a}\}\}$ uma estrutura demonstrável em \BV\ com pares distintos de átomos, tal que $f(a) = \dual{a}$, pela condição $C_{1}$ do teorema de condições necessárias para demonstrabilidade (Teorema \ref{teorema:condicoes}). Então $S\{\circ\}=\{X\{\circ\},Y\{\circ\}\}$ é demonstrável.
\end{Lemma}

\begin{proof}
A demonstração segue facilmente por indução no tamanho da derivação $\Pi$
$$\strpr{\Pi}{\BV}{S\{a,\dual{a}\}}$$
Seja $\rho$ a última regra de $\Pi$. Se $\rho = \ruleaidown$:
$$\strpr{\Pi'}{\BV}{\inf{\ruleaidown}{S[a,\dual{a}]}{S\{\circ\}}}$$
o resultado segue imediatamente. Se não, $\Pi$ é da forma:
$$\strpr{\Pi'}{\BV}{\inf{\rho}{S\{a,\dual{a}\}}{S'\{a,\dual{a}\}}}$$
Por indução, $S'\{\circ\}$ é demonstrável. Portanto
$$\strpr{\Pi'}{\BV}{\inf{\rho}{S\{\circ\}}{S'\{\circ\}}}$$ é a demonstração desejada.
\end{proof}

Note que a demonstração do lema \ref{lema:forget} foi dada para uma estrutura em \BV, mas o resultado vale também para o sistema \FBV\ (a demonstração apresentada é igualmente válida para uma estrutura em \FBV).

Pelo justificado na seção \ref{section:fbv-distintos}, restringimos neste trabalho o nosso foco de interesse ao sistema \FBV\ com pares de átomos dois a dois distintos. Isto nos possibilita uma série de resultados interessantes. Vamos começar a expô-los agora.

\begin{Lemma}[\emph{Estratégia para $\incnumber{S}{a}{\dual{a}} = 0$}]
\label{lema:incnumber0}
	Se uma estrutura $S$ em \FBV\ só possui pares de átomos dois a dois distintos e $S=S'\pars{a,\dual{a}}$, então podemos aplicar a regra $\ruleaidown$ sobre $a$ e $\dual{a}$.
	
	\begin{proof}
		Como restringimos a estrutura ao sistema \FBV\ com pares de átomos dois a dois distintos, sabemos que necessariamente devemos eliminar a única ocorrência de $a$ com a única ocorrência de $\dual{a}$. Considere a seguinte derivação:
		
		$$
			\dernote{\ruleaidown}{}{S\pars{a,\dual{a}}}
			{\leaf{\strpr{}{}{S\{\circ\}}}}
		$$
		
		Observe que esta derivação equivale exatamente a, de baixo para cima, ``esquecer'' os átomos duais $a$ e $\dual{a}$. Pelo lema do ``esquecimento'' (Lema \ref{lema:forget}), podemos concluir que $S\pars{a,\dual{a}}$ é demonstrável se, e somente se, $S\{\circ\}$ o é. Logo, podemos aplicar a regra $\ruleaidown$.
			
	\end{proof}
\end{Lemma}

Este lema é particularmente útil por dizer que, sempre que encontrarmos um par de átomos duais que tenham número de incoerência igual a zero, podemos aplicar a regra $\ruleaidown$ sem comprometer a demonstrabilidade da estrutura.

\begin{Lemma}
\label{lema:incnumber1aux}
A estrutura $S\pars{a,\aprs{\dual{a},X}}$ é demonstrável em \FBV\ com apenas pares de átomos dois a dois distintos se, e somente se, $S\{X\}$ é demonstrável no mesmo sistema.

	\begin{proof}
		Como restringimos a estrutura ao sistema \FBV\ com pares de átomos dois a dois distintos, sabemos que necessariamente devemos eliminar a única ocorrência de $a$ com a única ocorrência de $\dual{a}$. Considere a seguinte derivação:
		
		$$
			\dernote{\rules}{}{S\pars{a,\aprs{\dual{a},X}}}
			{\rootnote{\ruleaidown}{}{S\aprs{\pars{a,\dual{a}},X}}
			{\leaf{S\{X\}}}}
		$$
		
		Observe que esta derivação equivale exatamente a, de baixo para cima, ``esquecer'' os átomos duais $a$ e $\dual{a}$. Pelo lema do ``esquecimento'' (Lema \ref{lema:forget}), podemos concluir que $S\pars{a,\aprs{\dual{a},X}}$ é demonstrável se, e somente se, $S\{X\}$ o é.
				
	\end{proof}

\end{Lemma}

\begin{Lemma}[\emph{Estratégia para $\incnumber{S}{a}{\dual{a}} = 1$}]
\label{lema:incnumber1}
	Em \FBV\ com apenas pares de átomos dois a dois distintos, $S\pars{a,\aprs{\dual{a},X}}$ é demonstrável se, e somente se, $S\aprs{\pars{a,\dual{a}},X}$ o é.
	
	\begin{proof}
	
	Pelo lema \ref{lema:incnumber1aux}, $S\pars{a,\aprs{\dual{a},X}}$ é demonstrável se, e somente se, $S\{X\}$ é demonstrável. Mas, pelo Lema \ref{lema:incnumber0}, $S\{X\}$ é demonstrável se, e somente se, $S\aprs{\pars{a,\dual{a}},X}$ é demonstrável. Logo, por transitividade, $S\pars{a,\aprs{\dual{a},X}}$ é demonstrável se, e somente se $S\aprs{\pars{a,\dual{a}},X}$ o é.
	
	\end{proof}

\end{Lemma}

Este lema é particularmente útil por dizer que, sempre que encontrarmos um par de átomos duais que tenham número de incoerência igual a um, podemos aplicar a regra $\rules$ sem comprometer a demonstrabilidade da estrutura:

$$\dernote{\rules}{}{S\pars{a,\aprs{\dual{a},X}}}
{\leaf{\strpr{}{}{S\aprs{\pars{a,\dual{a}},X}}}}
$$
onde chegamos à situação em que o número de incoerência entre o par de átomos duais diminui para zero. 

Pelo lema \ref{lema:incnumber0} podemos, se quisermos, aplicar a regra $\ruleaidown$:

$$\dernote{\rules}{}{S\pars{a,\aprs{\dual{a},X}}}
{\rootnote{\ruleaidown}{}{S\aprs{X,\pars{a,\dual{a}}}}
{\leaf{\strpr{}{}{S\{X\}}}}}
$$

Entretanto, se não quisermos, temos a liberdade de adiar a aplicação da regra $\ruleaidown$ para mais tarde na construção da demonstração.

Observe que os lemas \ref{lema:incnumber0} e \ref{lema:incnumber1} fornecem uma idéia da estratégia de demonstração para quando temos pares de átomos duais cujo número de incoerência seja zero ou um. Se a estrutura que se deseja demonstrar possuir apenas pares de átomos destes tipos, a estratégia de demonstração já está completa. Basta aplicar o lema \ref{lema:incnumber0} sempre que possível, eliminando os pares de átomos duais com número de incoerência igual a zero, e depois prosseguir com o lema \ref{lema:incnumber1} eliminando os demais átomos. 

Entretanto, o que fazer no caso de, para todo par de átomos duais na estrutura, o número de incoerência for maior que um? A situação em questão equivale a $\forall a \in \occ{S}, \incnumber{S}{a}{\dual{a}} > 1$, onde $S$ é uma estrutura demonstrável. Uma idéia seria escolher um par de átomos $a$ e $\dual{a}$ e tentar reduzir seu número de incoerência até que ele se torne $1$ ou $0$ para, assim, poder aplicar a estratégia já conhecida. 

Neste momento é importante notar que os resultados obtidos até agora, que consistem na estratégia para pares de átomos duais cujo número de incoerência seja igual a zero ou um, já estão demonstrados. A partir deste ponto utilizamos indícios obtidos através de observações e considerações teóricas para propor um caminho para avançar na busca por uma estratégia para átomos duais com número de incoerência maior que um. Vamos argumentar a favor do caminho proposto, embasando a conjectura de que ele seja um caminho válido, produzindo a nossa tentativa de estratégia final.

Iniciando nossa tentativa de solução, conjecturamos que para ``diminuir'' o número de incoerência de um par de átomos, podemos utilizar a idéia do lema do ``esquecimento'' (Lema \ref{lema:forget}). Além das aplicações já mostradas, este lema é particularmente interessante porque ele dá uma idéia de como seria possível reduzir o número de incoerência de um par de átomos: bastaria ``esquecer'' (retirar) subestruturas adequadas de forma que a estrutura resultante possuísse pelo menos um par de átomos duais com número de incoerência igual a zero ou 1. 

Voltando à idéia de estratégia de demonstração em \BV, notemos que se $S\{a,\dual{a}\}$ é demonstrável e os átomos em $S$ são pares distintos, então $\Pi$ é da forma

$$\strder{\Pi_1}{\BV}{S\{a,\dual{a}\}}
{\leaf{\strpr{\Pi_2}{\BV}{\inf{\ruleaidown}
{S\pars{a,\dual{a}}}{{S\{\circ\}}}}}}$$ 

Isto significa que as regras switch e seq devem ser aplicadas para aproximar átomos com polaridades opostas.

A idéia é então aplicar o switch quantas vezes forem necessárias para aproximar um átomo de sua forma negada, ``esquecendo'' do resto da estrutura. Como podemos ver no Lema \ref{lema:forget}, ``esquecer'' átomos não altera a demonstrabilidade da estrutura.

A questão é: demonstramos o Lema~\ref{lema:forget} mantendo a ``estrutura'' da demonstração. Podemos alterar a demonstração para retornar novamente com os átomos que retiramos? Se a resposta for sim, então claramente temos uma estratégia para demonstrar estruturas, que é o principal objetivo deste trabalho.

Antes, entretanto, precisamos de uma definição do que seria ``esquecer'' estruturas inteiras, uma vez que o lema \ref{lema:forget} se refere a ``esquecer'' apenas átomos.

\begin{Definition}
\label{definition:forget}
``Esquecer'' $R$ em $S\{R\}$  significa
$$\forall a \in \occ{R}. \ \ ``esquecer" \ \ a, \dual{a} \in \occ{S\{R\}}$$
Note que não necessariamente $\dual{a} \in \occ{R}$. Pode ser o caso em que $\dual{a}$ está em $\occ{S\{ \ \}}$.
\end{Definition} 

\begin{Example}
	Seja a estrutura $\pars{a,\aprs{\dual{b},c},\aprs{\dual{a},b},\dual{c}}$. ``Esquecer'' a subestrutura $\aprs{\dual{a},b}$ significa ``esquecer'' os pares de átomos $a$, $\dual{a}$ e $b, \dual{b}$, obtendo assim a estrutura $\pars{c,\dual{c}}$.
\end{Example}

Podemos agora voltar à busca da estratégia de demonstração, começando por um exemplo.

\begin{Example} Seja $S = \pars{\aprs{a,\dual{b}},\aprs{\dual{c},\pars{\dual{a},b}},\aprs{c,\pars{\dual{d},e}},\aprs{d,\dual{e}}}$. Há várias maneiras de se demonstrar $S$. Observe que todos os átomos possuem número de incoerência módulo coerência igual a $2$. Entretanto, os átomos $a, b, d$ e $e$
estão ``aninhados'', o que torna a escolha de por onde começar a demonstração mais complicada. A maneira mais clara de proceder seria aplicar um switch sobre $c$ e $\dual{c}$. Vamos fazer $R=\pars{\dual{a},b}$ e $T=\pars{\dual{d},e}$, de forma que $S = \pars{\aprs{a,\dual{b}},\aprs{\dual{c},R},\aprs{c,T},\aprs{d,\dual{e}}}$. 

Podemos então usar a idéia da Definição \ref{definition:forget} e ``esquecer'' a subestrutura $T$. Nesse caso, consideramos que a estrutura $S$ fica implicitamente transformada em $\pars{\aprs{a,\dual{b}},\aprs{\dual{c},R},c}$. Note que, com isso, o número de incoerência $\incnumber{S}{c}{\dual{c}}$ diminui para um. E é exatamente o que queríamos! Já sabemos como lidar com os casos em que há um par de átomos duais com número de incoerência módulo coerência igual a um (Lema \ref{lema:incnumber1}). Podemos então construir a seguinte demonstração da estrutura $S$:

$$
\dernote{\rules}{}
{\pars{\aprs{a,\dual{b}},\aprs{\dual{c},R},\aprs{c,T},\aprs{d,\dual{e}}}}
{
  \root{\rules}
  {\pars{\aprs{a,\dual{b}},\aprs{\pars{\dual{c},\aprs{c,T}},R},\aprs{d,\dual{e}}}}
  {\root{\ruleaidown}{\pars{\aprs{a,\dual{b}},\aprs{\aprs{\pars{\dual{c},c},T},R},\aprs{d,\dual{e}}}}
  {\leaf{\strpr{}{}{\pars{\aprs{a,\dual{b}},\aprs{T,R},\aprs{d,\dual{e}}}}}}}}
$$

Note que na demonstração não explicitamos o ``esquecimento'' de $T$.  Simplesmente usamos a idéia implícita de esquecimento para nos guiar por quem começar a demonstração, no caso, por uma instância switch sobre o par de átomos duais $c$ e $\dual{c}$, que possuem então um número de incoerência módulo coerência com que sabemos lidar (no caso, um).

É claro que poderíamos ``esquecer'' $R$ e o resultado seria equivalente. Neste caso, não importa a subestrutura que decidimos apagar, uma vez que as estruturas $\aprs{\dual{c},R}$ e $\aprs{c,T}$ são
similares.

\end{Example}

Após este exemplo, estamos preparados para formalizar a idéia de ``esquecimento'' como meio para diminuir números de incoerência módulo coerência maiores que um e, assim, obter uma estratégia.

\begin{Conjecture}[Estratégia para $\incnumber{S}{a}{\dual{a}} > 1$]
\label{conjecture:incnumber2}
Seja $S$ uma estrutura demonstrável em \FBV\ com átomos dois a dois distintos tal que $\incnumber{S}{a}{\dual{a}}>1$ para todo $a\in \occ{S}$. Seja $a\in \occ{S}$ o átomo com menor número de incoerência módulo coerência na estrutura.
				Então $S\{a,\dual{a}\}$ é da forma $S'\aprs{R\{a\},T\{\dual{a}\}}$ e temos um dos seguintes casos:
				
				\ \\				
				\begin{tabular}{lll}
					
					1 & $R\{\circ\} = T\{\circ\} = \circ$: & $S\{a,\dual{a}\} = S'\pars{\aprs{a,X},\aprs{\dual{a},Y}}$\\
								
					2 & $R\{\circ\} = \circ$ e $T\{\circ\} \neq \circ$: & $S\{a,\dual{a}\}= S'\pars{\aprs{a,X},\aprs{T\{\dual{a}\},Y}}$\\
								
					3 & $R\{\circ\} \neq \circ$ e $T\{\circ\} \neq \circ$: & $S\{a,\dual{a}\}= S'\pars{\aprs{R\{a\},X},\aprs{T\{\dual{a}\},Y}}$\\
				\end{tabular}
				
				\ \\
				, onde $X,Y \neq \circ$				
				
				\ \\				
				Além disso:
								
				\ \\
				\begin{tabular}{llll}
					1 & $S'\pars{\aprs{a,X},\aprs{\dual{a},Y}}$ é demonstrável & $\iff$ & $S'\aprs{X,Y}$ é demonstrável; \\
					2 & $S'\pars{\aprs{a,X},\aprs{T\{\dual{a}\},Y}}$ é demonstrável & $\iff$ & $S'\aprs{\pars{T\{\dual{a}\},\aprs{a,X}},Y}$ é demonstrável; \\
					3 & $S'\pars{\aprs{R\{a\},X},\aprs{T\{\dual{a}\},Y}}$ é demonstrável & $\iff$ &				$S'\aprs{\pars{T\{\dual{a}\},\aprs{R\{a\},X}},Y}$ ou \\
					  &                                                              &        &
					  $S'\aprs{\pars{R\{a\},\aprs{T\{\dual{a}\},Y}},X}$ é demonstrável. \\
				\end{tabular}
				
\end{Conjecture}
				
\emph{Argumentação a favor da conjectura:}

	A primeira parte da conjectura segue facilmente da definição de número de incoerência e da minimalidade de $a$.

Seja $S\{a,\dual{a}\}=\pars{X\{a\},Y\{\dual{a}\}}$. A demonstração da demonstrabilidade seria por indução sobre o número de incoerência de $\{a, \dual{a}\}$.

\begin{itemize}
\item Caso 1: \ \ $S'\pars{\aprs{a,X},\aprs{\dual{a},Y}}$

	$(\Rightarrow)$ Por indução no tamanho da derivação $\Pi$ de $S\pars{\aprs{a,X},\aprs{\dual{a},Y}}$. Se $\occ{X} \ \cap \ \occ{Y} \ = \ \emptyset$, então acreditamos que não seria criada nenhuma inconsistência. Se $\occ{X} \ \cap \ \occ{Y} \ \neq \ \emptyset$, a estrutura inicial conteria uma subestrutura equivalente a $\pars{\aprs{a,b},\aprs{\dual{a},\dual{b}}}$, que sabemos não ser demonstrável pela condição $C_{2}$ do Teorema~\ref{teorema:condicoes}.

$(\Leftarrow)$ Considere a demonstração
	$$
	\dernote {\rules }{}{S\pars{\aprs{a,X},\aprs{\dual{a},Y}}} {
	  \rootnote{\rules
	  }{}{S\aprs{\pars{a,{\aprs{\dual{a},Y}}},X}}
	  {\root{\ruleaidown}{S\aprs{\pars{a,\dual{a}},Y,X}}
	  {\leaf{S\aprs{Y,X}}}}}
	$$

\item Caso 2: \ \ $S'\pars{\aprs{a,X},\aprs{T\{\dual{a}\},Y}}$, \  \ $T\{\circ\} \neq \circ$

	Considere a derivação que ``esquece'' $X$:

	$$
		\dernote{\rules}{}{S'\pars{\aprs{a,X},\aprs{T\{\dual{a}\},Y}}}
		{\leaf{S'\aprs{\pars{T\{\dual{a}\},\aprs{a,X}},Y}}}
	$$			
	
	Isto se $\occ{X} \cap \occ{Y} = \emptyset$. Mas se $\occ{X} \cap \occ{Y} \neq \emptyset$ a estrutura já não seria demonstrável pela condição $C_{2}$ do Teorema \ref{teorema:condicoes}.	

	Acreditamos que o resultado seguiria por indução.

\item Case 3: \ \ \ $S'\pars{\aprs{R\{a\},X},\aprs{T\{\dual{a}\},Y}}$, \ \ $R\{\circ\} \neq \circ$ \ \ e \ \  $T\{\circ\} \neq \circ$

		Nesse caso há duas hipóteses, sendo que necessariamente pelo menos uma das duas ocorre:
		
		\begin{enumerate}
			\item $R\{a\} \cap Y = \emptyset$
				
				Nesse caso poderíamos usar o fato de que $R\{a\} \cap Y = \emptyset$, ``esquecer'' $X$ e fazer:
				
				$$
					\dernote{\rules}{}{S'\pars{\aprs{R\{a\},X},\aprs{T\{\dual{a}\},Y}}}
					{\leaf{S'\aprs{\pars{T\{\dual{a}\},\aprs{R\{a\},X}},Y}}}
				$$
				
				Isto se $\occ{X} \cap \occ{Y} = \emptyset$. Mas se $\occ{X} \cap \occ{Y} \neq \emptyset$ a estrutura já não seria demonstrável pela condição $C_{2}$ do Teorema \ref{teorema:condicoes}.
				
				Acreditamos que o resultado seguiria por indução.
				
			\item $T\{\dual{a}\} \cap X = \emptyset$
			
				Nesse caso poderíamos usar o fato de que $T\{\dual{a}\} \cap X = \emptyset$, ``esquecer'' $Y$ e fazer:
			
				$$
					\dernote{\rules}{}{S'\pars{\aprs{R\{a\},X},\aprs{T\{\dual{a}\},Y}}}
					{\leaf{S'\aprs{\pars{R\{a\},\aprs{T\{\dual{a}\},Y}},X}}}										
				$$
				
				Isto se $\occ{X} \cap \occ{Y} = \emptyset$. Mas se $\occ{X} \cap \occ{Y} \neq \emptyset$ a estrutura já não seria demonstrável pela condição $C_{2}$ do Teorema \ref{teorema:condicoes}.
				
				Acreditamos que o resultado seguiria por indução.
			
		\end{enumerate}
		
		Faltaria apenas mostrar que $R\{a\} \cap Y = \emptyset$ ou $T\{\dual{a}\} \cap X = \emptyset$ deve ocorrer. Imagine, por contradição, que $R\{a\} \cap Y \neq \emptyset$ e $T\{\dual{a}\} \cap X \neq \emptyset$. Então as seguintes asserções são verdadeiras:
		
		$\exists n \in \occ{Y}. \dual{n} \in \occ{R\{a\}} \implies 
			\left\{ 
				\begin{array}{ll}
					Y = Y'\{n\} \ \ \wedge \\
					R\{a\} = R'\{a,\dual{n}\} 
				\end{array}
			\right.		
		$
		
		$\exists m \in \occ{Y}. \dual{m} \in \occ{T\{ \dual{a} \}} \implies 
			\left\{ 
				\begin{array}{ll}
					X = X'\{m\} \ \ \wedge \\
					T\{\dual{a}\} = T'\{\dual{a},\dual{m}\} 
				\end{array}
			\right.		
		$
		
		Tomemos o caso de $R\{a\} = R'\{a,\dual{n}\}$. Há apenas duas possibilidades: 
		
		$R\{a\} = R'\{a,\dual{n}\} \implies 
			\left\{ 
				\begin{array}{ll}
					R'\{a,\dual{n}\} = R''\pars{P\{a\},P'\{\dual{n}\}} \ \ \vee \\
					R'\{a,\dual{n}\} = R''\aprs{P\{a\},P'\{\dual{n}\}}
				\end{array}
			\right.	
		$
		
		Similarmente, para $T\{\dual{a}\} = T'\{\dual{a},\dual{m}\}$ também há apenas duas possibilidades:
		
		$T\{\dual{a}\} = T'\{\dual{a},\dual{m}\} \implies 
			\left\{ 
				\begin{array}{ll}
					T'\{\dual{a},\dual{m}\} = T''\pars{Q\{\dual{a}\},Q'\{\dual{m}\}} \ \ \vee \\
					T'\{\dual{a},\dual{m}\} = T''\aprs{Q\{\dual{a}\},Q'\{\dual{m}\}}
				\end{array}
			\right.	
		$
		
		Combinando as possibilidades para $R\{a\}$ e $T\{\dual{a}\}$, há, então, quatro possibilidades a serem avaliadas:
		
		\begin{enumerate}
			\item $R'\{a,\dual{n}\} = R''\pars{P\{a\},P'\{\dual{n}\}}$ \ \ e \ \ $T'\{\dual{a},\dual{m}\} = T''\pars{Q\{\dual{a}\},Q'\{\dual{m}\}}$ \\
			
			Nesse caso $S = \pars{ \aprs{ R''\pars{P\{a\},P'\{\dual{n}\}} ,X'\{m\} }, \aprs{ T''\pars{Q\{\dual{a}\},Q'\{\dual{m}\}} , Y'\{n\} } } $.											
				
			\item $R'\{a,\dual{n}\} = R''\pars{P\{a\},P'\{\dual{n}\}}$ \ \ e \ \ $T'\{\dual{a},\dual{m}\} = T''\aprs{Q\{\dual{a}\},Q'\{\dual{m}\}}$ \\
			
			Nesse caso $S = \pars{ \aprs{ R''\pars{P\{a\},P'\{\dual{n}\}} , X'\{m\} }, \aprs{ T''\aprs{Q\{\dual{a}\},Q'\{\dual{m}\}} , Y'\{n\} } }$
			
			\item $R'\{a,\dual{n}\} = R''\aprs{P\{a\},P'\{\dual{n}\}}$ \ \ e \ \ $T'\{\dual{a},\dual{m}\} = T''\pars{Q\{\dual{a}\},Q'\{\dual{m}\}}$ \\
			
			Nesse caso $S = \pars{ \aprs{ R''\aprs{P\{a\},P'\{\dual{n}\}} , X'\{m\} }, \aprs{ T''\pars{Q\{\dual{a}\},Q'\{\dual{m}\}} , Y'\{n\} } }$
			
			\item $R'\{a,\dual{n}\} = R''\aprs{P\{a\},P'\{\dual{n}\}}$ \ \ e \ \ $T'\{\dual{a},\dual{m}\} = T''\aprs{Q\{\dual{a}\},Q'\{\dual{m}\}}$ \\
			
			Nesse caso $S = \pars{ \aprs{ R''\aprs{P\{a\},P'\{\dual{n}\}} , X'\{m\} }, \aprs{ T''\aprs{Q\{\dual{a}\},Q'\{\dual{m}\}} , Y'\{n\} } }$
				
		\end{enumerate}
		
		Em qualquer dos casos acima, temos que $S$ possui uma subestrutura da forma: 
			
			$$
				\xymatrix{
					m \ar@{-}[r] \ar@{-}[d] \ar@{~}[dr]  & n  \ar@{-}[d] \\
					\dual{m} \ar@{-}[r] \ar@{~}[ur] & \dual{n} \\
				}	
			$$
			o que, pela condição $C_{2}$ do Teorema \ref{teorema:condicoes}, mostra que a estrutura $S$ já não seria, desde o princípio, demonstrável, o que entra em contradição com a hipótese de que $S$ é demonstrável. Logo, não é o caso em que $R\{a\} \cap Y \neq \emptyset$ e $T\{\dual{a}\} \cap X \neq \emptyset$.

\end{itemize}

\emph{Fim da argumentação a favor da conjectura.}

Antes de enunciar o resultado principal, vamos apresentar um lema auxiliar que, apesar de não ser essencial, facilita a estratégia proposta.

\begin{Lemma}[Estratégia para $\incnumber{S}{a}{\dual{a}} = 0$ ou $\incnumber{S}{a}{\dual{a}} = 1$]
\label{lema:incnumber01}
Se $S$ é demonstrável em \FBV\ contendo apenas pares de átomos distintos dois a dois, podemos escolher eliminar \emph{qualquer} par de átomos duais $a,\dual{a} \in \occ{S}$ tal que $\incnumber{S}{a}{\dual{a}}=0$ ou $1$.
\end{Lemma}
\begin{proof}
Pelos Lemas \ref{lema:incnumber1} e \ref{lema:forget}.
\end{proof}

Agora sim estamos prontos para enunciar o resultado principal desta dissertação, através do teorema seguinte.

\begin{Conjecture}[Tentativa de estratégia de demonstração]
\label{conjectura:resultado-principal}
Existe um algoritmo que corresponde a uma estratégia de demonstração para toda estrutura demonstrável $S$ em \FBV\ contendo apenas pares de átomos distintos dois a dois. 
\end{Conjecture}

	\emph{Argumentação a favor da conjectura:}
	
	Considere a estratégia fornecida pelos Lemas \ref{lema:incnumber0}, \ref{lema:incnumber1} e \ref{lema:incnumber01} e pela Conjectura \ref{conjecture:incnumber2}. Comece a demonstração pela redução da subestrutura que contém ocorrências de átomos $a$ e $\dual{a}$ de número de incoerência módulo coerência igual a $0$ ou $1$ (pelo Lema \ref{lema:incnumber01} posso começar a por qual par de átomos duais cujo número de incoerência módulo coerência seja 0 ou 1. Pelos lemas \ref{lema:incnumber0} e \ref{lema:incnumber1} sei como agir em cada caso). Se não houver tal subestrutura, escolha o par de átomos com menor número de incoerência módulo coerência e ``reduza'' este número via ``esquecimento'' (pela Conjectura \ref{conjecture:incnumber2}). Se um estágio for atingido sobre o qual não se possa avançar, é porque a estrutura já não era demonstrável desde o início (pelos lemas \ref{lema:incnumber0}, \ref{lema:incnumber1} e pela conjectura \ref{conjecture:incnumber2}).
	
	\emph{Fim da argumentação a favor da conjectura.}
	
Vamos demonstrar a aplicação da tentativa de estratégia proposta na Conjectura \ref{conjectura:resultado-principal} através de um exemplo.

\begin{Example}

	Seja a seguinte estrutura demonstrável em \FBV\ com pares de átomos dois a dois distintos: $$\pars{\aprs{a,\dual{b}},\aprs{\dual{c},\pars{\dual{a},b}},\aprs{c,\pars{\dual{d},e}},\aprs{d,\dual{e}}}$$
	
	Vamos prosseguir com a estratégia proposta, passo a passo.
	
	\begin{enumerate}
		
		\item Estrutura a ser demonstrada:
		
		$$S = \pars{\aprs{a,\dual{b}},\aprs{\dual{c},\pars{\dual{a},b}},\aprs{c,\pars{\dual{d},e}},\aprs{d,\dual{e}}}$$
		
		Números de incoerência módulo coerência:
		
		\begin{center}
			\begin{tabular}{|c||c|c|c|c|c|}
				\hline
				Átomo $i$                    & $a$ & $b$ & $c$ & $d$ & $e$\\ \hline
				$\incnumber{S}{i}{\dual{i}}$ &  2  &  2  &  2  &  2  &  2 \\ \hline
			\end{tabular}	
		\end{center}
		
		Todos os valores são iguais a 2, portanto posso escolher qualquer um como sendo o menor. Escolhamos o par $a, \dual{a}$. 
		
		Como $\incnumber{S}{a}{\dual{a}} = 2$, recorremos então à conjectura \ref{conjecture:incnumber2}, que trata de número de incoerência módulo coerência maior que 1, e vemos que o nosso caso equivale a:
		
		$$
			\begin{array}{ll}
				
					S'\pars{\aprs{a,X},\aprs{T\{\dual{a}\},Y}}
								
				& 
				\left\}
				\begin{array}{c}
					X = \dual{b} \\
					Y = \dual{c} \\
					T\{\dual{a}\} = \pars{\dual{a},\dual{b}} \\	
				\end{array}
				\right .
				
			\end{array}
		$$	
		, de onde concluímos que devemos ``esquecer'' X e a regra de inferência a ser aplicada é um switch da forma:
		
		$$
			\dernote{\rules}{}{S'\pars{\aprs{a,X},\aprs{T\{\dual{a}\},Y}}}
			{\leaf{S'\aprs{\pars{T\{\dual{a}\},\aprs{a,X}},Y}}}
		$$	
		
		Assim, podemos começar a construção da nossa demonstração:

		$$
	\dernote{\rules}{}{\pars{\aprs{a,\dual{b}},\aprs{\dual{c},\pars{\dual{a},b}},\aprs{c,\pars{\dual{d},e}},\aprs{d,\dual{e}}}}
	{\leaf{\pars{\aprs{\pars{\aprs{a,\dual{b}},\dual{a},b},\dual{c}},\aprs{c,\pars{\dual{d},e}},\aprs{d,\dual{e}}}}
}
	$$				
		
		\item Agora temos que demonstrar a seguinte estrutura:
		
			$$S = \pars{\aprs{\pars{\aprs{a,\dual{b}},\dual{a},b},\dual{c}},\aprs{c,\pars{\dual{d},e}},\aprs{d,\dual{e}}}$$
		
		Números de incoerência módulo coerência:
		
		\begin{center}
			\begin{tabular}{|c||c|c|c|c|c|}
				\hline
				Átomo $i$                    & $a$ & $b$ & $c$ & $d$ & $e$\\ \hline
				$\incnumber{S}{i}{\dual{i}}$ &  1  &  1  &  2  &  2  &  2 \\ \hline
			\end{tabular}	
		\end{center}		
		
		Há dois valores iguais a 1, portanto, posso escolher $\incnumber{S}{a}{\dual{a}} = 1$.
			
		Recorremos, então, ao Lema \ref{lema:incnumber1}, que trata de número de incoerênica módulo coerência igual a 1, e vemos que o nosso caso equivale a:
		
		$$
			\begin{array}{ll}
				
				S'\pars{a,\aprs{\dual{a},X}}
								
				& 
				\left\}
				\begin{array}{c}
					X = \dual{b}\\		
				\end{array}
				\right .
				
			\end{array}
		$$			
		, de onde concluímos que devemos ``esquecer'' X e a regra de inferência a ser aplicada é um switch da forma:
		
		$$
			\dernote{\rules}{}{S'\pars{a,\aprs{\dual{a},X}}}
			{	\leaf{{S'\aprs{\pars{a,\dual{a}},X}}}}
		$$
		
		Assim, podemos continuar a construção da nossa demonstração:
		
		$$
	\dernote{\rules}{}{\pars{\aprs{a,\dual{b}},\aprs{\dual{c},\pars{\dual{a},b}},\aprs{c,\pars{\dual{d},e}},\aprs{d,\dual{e}}}}
	{\rootnote{\rules}{}{\pars{\aprs{\pars{\aprs{a,\dual{b}},\dual{a},b},\dual{c}},\aprs{c,\pars{\dual{d},e}},\aprs{d,\dual{e}}}}
	{\leaf{\pars{\aprs{\pars{b,\aprs{\pars{a,\dual{a}},\dual{b}}},\dual{c}},\aprs{c,\pars{\dual{d},e}},\aprs{d,\dual{e}}}}}}
	$$
	
	\item Agora a estrutura a ser demonstrada é a seguinte:
		
		$$S = \pars{\aprs{\pars{b,\aprs{\pars{a,\dual{a}},\dual{b}}},\dual{c}},\aprs{c,\pars{\dual{d},e}},\aprs{d,\dual{e}}}$$
		
		Números de incoerência módulo coerência:
		
		\begin{center}
			\begin{tabular}{|c||c|c|c|c|c|}
				\hline
				Átomo $i$                    & $a$ & $b$ & $c$ & $d$ & $e$\\ \hline
				$\incnumber{S}{i}{\dual{i}}$ &  0  &  1  &  2  &  2  &  2 \\ \hline
			\end{tabular}	
		\end{center}
		
		O menor valor é $\incnumber{S}{a}{\dual{a}} = 0$	
			
		Recorremos, então, ao Lema \ref{lema:incnumber0}, que trata de número de incoerênica módulo coerência igual a 0, e vemos que o nosso caso equivale a:
		
		$$S'\pars{a,\dual{a}}$$		
		, de onde concluímos que devemos aplicar uma instância de $\ruleaidown$ da forma:
		
		$$
			\dernote{\ruleaidown}{}{S'\pars{a,\dual{a}}}
			{\leaf{S'\{\circ\}}}
		$$
		
		Assim, podemos continuar a construção da nossa demonstração, já levando em conta a relação de equivalência sintática:
		
		$$
	\dernote{\rules}{}{\pars{\aprs{a,\dual{b}},\aprs{\dual{c},\pars{\dual{a},b}},\aprs{c,\pars{\dual{d},e}},\aprs{d,\dual{e}}}}
	{\rootnote{\rules}{}{\pars{\aprs{\pars{\aprs{a,\dual{b}},\dual{a},b},\dual{c}},\aprs{c,\pars{\dual{d},e}},\aprs{d,\dual{e}}}}
	{\root{\ruleaidown}{\pars{\aprs{\pars{b,\aprs{\pars{a,\dual{a}},\dual{b}}},\dual{c}},\aprs{c,\pars{\dual{d},e}},\aprs{d,\dual{e}}}}
	{\root{=}{\pars{\aprs{\pars{b,\aprs{\circ,\dual{b}}},\dual{c}},\aprs{c,\pars{\dual{d},e}},\aprs{d,\dual{e}}}}
	{\leaf{\pars{\aprs{\pars{b,\dual{b}},\dual{c}},\aprs{c,\pars{\dual{d},e}},\aprs{d,\dual{e}}}}}}}}
	$$
	
	\item Temos que demonstrar, então, a seguinte estrutura:
		
		$$S = \pars{\aprs{\pars{b,\dual{b}},\dual{c}},\aprs{c,\pars{\dual{d},e}},\aprs{d,\dual{e}}}$$
		
		Números de incoerência módulo coerência:
		
		\begin{center}
			\begin{tabular}{|c||c|c|c|c|}
				\hline
				Átomo $i$                    & $b$ & $c$ & $d$ & $e$\\ \hline
				$\incnumber{S}{i}{\dual{i}}$ &  0  &  2  &  2  &  2 \\ \hline
			\end{tabular}	
		\end{center}
		
		O menor valor é $\incnumber{S}{b}{\dual{b}} = 0$.
			
		Recorremos, então, ao Lema \ref{lema:incnumber0}, que trata de número de incoerênica módulo coerência igual a 0, e vemos que o nosso caso equivale a:
		
		$$S'\pars{b,\dual{b}}$$		
		, de onde concluímos que devemos aplicar uma instância de $\ruleaidown$ da forma:
		
		$$
			\dernote{\ruleaidown}{}{S'\pars{b,\dual{b}}}
			{\leaf{S'\{\circ\}}}
		$$
		
		Assim, podemos continuar a construção da nossa demonstração, já levando em conta a relação de equivalência sintática:
		
		$$
	\dernote{\rules}{}{\pars{\aprs{a,\dual{b}},\aprs{\dual{c},\pars{\dual{a},b}},\aprs{c,\pars{\dual{d},e}},\aprs{d,\dual{e}}}}
	{\rootnote{\rules}{}{\pars{\aprs{\pars{\aprs{a,\dual{b}},\dual{a},b},\dual{c}},\aprs{c,\pars{\dual{d},e}},\aprs{d,\dual{e}}}}
	{\root{\ruleaidown}{\pars{\aprs{\pars{b,\aprs{\pars{a,\dual{a}},\dual{b}}},\dual{c}},\aprs{c,\pars{\dual{d},e}},\aprs{d,\dual{e}}}}
	{\root{=}{\pars{\aprs{\pars{b,\aprs{\circ,\dual{b}}},\dual{c}},\aprs{c,\pars{\dual{d},e}},\aprs{d,\dual{e}}}}
	{\root{\ruleaidown}{\pars{\aprs{\pars{b,\dual{b}},\dual{c}},\aprs{c,\pars{\dual{d},e}},\aprs{d,\dual{e}}}}
	{\root{=}{\pars{\aprs{\circ,\dual{c}},\aprs{c,\pars{\dual{d},e}},\aprs{d,\dual{e}}}}
	{\leaf{\pars{\dual{c},\aprs{c,\pars{\dual{d},e}},\aprs{d,\dual{e}}}}
	}}}}}}
	$$
	
	\item Neste momento temos que a estrutura a ser demonstrada é:
		
		$$S = \pars{\dual{c},\aprs{c,\pars{\dual{d},e}},\aprs{d,\dual{e}}}$$
		
		Números de incoerência módulo coerência:
		
		\begin{center}
			\begin{tabular}{|c||c|c|c|}
				\hline
				Átomo $i$                    & $c$ & $d$ & $e$\\ \hline
				$\incnumber{S}{i}{\dual{i}}$ &  1  &  2  &  2 \\ \hline
			\end{tabular}	
		\end{center}		
		
		O menor valor é $\incnumber{S}{c}{\dual{c}} = 1$.
			
		Recorremos, então, ao Lema \ref{lema:incnumber1}, que trata de número de incoerênica módulo coerência igual a 1, e vemos que o nosso caso equivale a:
		
		$$
			\begin{array}{ll}
				
				S'\pars{c,\aprs{\dual{c},X}}
								
				& 
				\left\}
				\begin{array}{c}
					X = \pars{\dual{d},e}\\		
				\end{array}
				\right .
				
			\end{array}
		$$		
		, de onde concluímos que devemos ``esquecer'' X e a regra de inferência a ser aplicada é um switch da forma:
		
		$$
			\dernote{\rules}{}{S'\pars{c,\aprs{\dual{c},X}}}
			{	\leaf{{S'\aprs{\pars{c,\dual{c}},X}}}}
		$$
		
		Assim, podemos continuar a construção da nossa demonstração:
		
		$$
	\dernote{\rules}{}{\pars{\aprs{a,\dual{b}},\aprs{\dual{c},\pars{\dual{a},b}},\aprs{c,\pars{\dual{d},e}},\aprs{d,\dual{e}}}}
	{\rootnote{\rules}{}{\pars{\aprs{\pars{\aprs{a,\dual{b}},\dual{a},b},\dual{c}},\aprs{c,\pars{\dual{d},e}},\aprs{d,\dual{e}}}}
	{\root{\ruleaidown}{\pars{\aprs{\pars{b,\aprs{\pars{a,\dual{a}},\dual{b}}},\dual{c}},\aprs{c,\pars{\dual{d},e}},\aprs{d,\dual{e}}}}
	{\root{=}{\pars{\aprs{\pars{b,\aprs{\circ,\dual{b}}},\dual{c}},\aprs{c,\pars{\dual{d},e}},\aprs{d,\dual{e}}}}
	{\root{\ruleaidown}{\pars{\aprs{\pars{b,\dual{b}},\dual{c}},\aprs{c,\pars{\dual{d},e}},\aprs{d,\dual{e}}}}
	{\root{=}{\pars{\aprs{\circ,\dual{c}},\aprs{c,\pars{\dual{d},e}},\aprs{d,\dual{e}}}}
	{\root{\rules}{\pars{\dual{c},\aprs{c,\pars{\dual{d},e}},\aprs{d,\dual{e}}}}
	{\leaf{\pars{\aprs{\pars{c,\dual{c}},\pars{\dual{d},e}},\aprs{d,\dual{e}}}}
	}}}}}}}
	$$
	
	\item Agora a estrutura a ser demonstrada é a seguinte:
		
		$$S = \pars{\aprs{\pars{c,\dual{c}},\pars{\dual{d},e}},\aprs{d,\dual{e}}}$$
		
		Números de incoerência módulo coerência:
		
		\begin{center}
			\begin{tabular}{|c||c|c|c|}
				\hline
				Átomo $i$                    & $c$ & $d$ & $e$\\ \hline
				$\incnumber{S}{i}{\dual{i}}$ &  0  &  2  &  2 \\ \hline
			\end{tabular}	
		\end{center}
		
		O menor valor é $\incnumber{S}{c}{\dual{c}} = 0$	
			
		Recorremos, então, ao Lema \ref{lema:incnumber0}, que trata de número de incoerênica módulo coerência igual a 0, e vemos que o nosso caso equivale a:
		
		$$S'\pars{c,\dual{c}}$$		
		, de onde concluímos que devemos aplicar uma instância de $\ruleaidown$ da forma:
		
		$$
			\dernote{\ruleaidown}{}{S'\pars{c,\dual{c}}}
			{\leaf{S'\{\circ\}}}
		$$
		
		Assim, podemos continuar a construção da nossa demonstração, já levando em conta a relação de equivalência sintática:
		
		$$
	\dernote{\rules}{}{\pars{\aprs{a,\dual{b}},\aprs{\dual{c},\pars{\dual{a},b}},\aprs{c,\pars{\dual{d},e}},\aprs{d,\dual{e}}}}
	{\rootnote{\rules}{}{\pars{\aprs{\pars{\aprs{a,\dual{b}},\dual{a},b},\dual{c}},\aprs{c,\pars{\dual{d},e}},\aprs{d,\dual{e}}}}
	{\root{\ruleaidown}{\pars{\aprs{\pars{b,\aprs{\pars{a,\dual{a}},\dual{b}}},\dual{c}},\aprs{c,\pars{\dual{d},e}},\aprs{d,\dual{e}}}}
	{\root{=}{\pars{\aprs{\pars{b,\aprs{\circ,\dual{b}}},\dual{c}},\aprs{c,\pars{\dual{d},e}},\aprs{d,\dual{e}}}}
	{\root{\ruleaidown}{\pars{\aprs{\pars{b,\dual{b}},\dual{c}},\aprs{c,\pars{\dual{d},e}},\aprs{d,\dual{e}}}}
	{\root{=}{\pars{\aprs{\circ,\dual{c}},\aprs{c,\pars{\dual{d},e}},\aprs{d,\dual{e}}}}
	{\root{\rules}{\pars{\dual{c},\aprs{c,\pars{\dual{d},e}},\aprs{d,\dual{e}}}}
	{\root{\ruleaidown}{\pars{\aprs{\pars{c,\dual{c}},\pars{\dual{d},e}},\aprs{d,\dual{e}}}}
	{\root{=}{\pars{\aprs{\circ,\pars{\dual{d},e}},\aprs{d,\dual{e}}}}
	{\leaf{\pars{\dual{d},e,\aprs{d,\dual{e}}}}
	}}}}}}}}}
	$$
	
	\item Neste momento temos que a estrutura a ser demonstrada é:
		
		$$S = \pars{\dual{d},e,\aprs{d,\dual{e}}}$$
		
		Números de incoerência módulo coerência:
		
		\begin{center}
			\begin{tabular}{|c||c|c|}
				\hline
				Átomo $i$                    & $d$ & $e$\\ \hline
				$\incnumber{S}{i}{\dual{i}}$ &  1  &  1 \\ \hline
			\end{tabular}	
		\end{center}		
		
		Há dois valores iguais a 1, portanto, posso escolher $\incnumber{S}{d}{\dual{d}} = 1$.
			
		Recorremos, então, ao Lema \ref{lema:incnumber1}, que trata de número de incoerênica módulo coerência igual a 1, e vemos que o nosso caso equivale a:
		
		$$
			\begin{array}{ll}
				
				S'\pars{d,\aprs{\dual{d},X}}
								
				& 
				\left\}
				\begin{array}{c}
					X = \dual{e}\\		
				\end{array}
				\right .
				
			\end{array}
		$$			
		, de onde concluímos que devemos ``esquecer'' X e a regra de inferência a ser aplicada é um switch da forma:
		
		$$
			\dernote{\rules}{}{S'\pars{d,\aprs{\dual{d},X}}}
			{	\leaf{{S'\aprs{\pars{d,\dual{d}},X}}}}
		$$
		
		Assim, podemos continuar a construção da nossa demonstração:
		
		$$
	\dernote{\rules}{}{\pars{\aprs{a,\dual{b}},\aprs{\dual{c},\pars{\dual{a},b}},\aprs{c,\pars{\dual{d},e}},\aprs{d,\dual{e}}}}
	{\rootnote{\rules}{}{\pars{\aprs{\pars{\aprs{a,\dual{b}},\dual{a},b},\dual{c}},\aprs{c,\pars{\dual{d},e}},\aprs{d,\dual{e}}}}
	{\root{\ruleaidown}{\pars{\aprs{\pars{b,\aprs{\pars{a,\dual{a}},\dual{b}}},\dual{c}},\aprs{c,\pars{\dual{d},e}},\aprs{d,\dual{e}}}}
	{\root{=}{\pars{\aprs{\pars{b,\aprs{\circ,\dual{b}}},\dual{c}},\aprs{c,\pars{\dual{d},e}},\aprs{d,\dual{e}}}}
	{\root{\ruleaidown}{\pars{\aprs{\pars{b,\dual{b}},\dual{c}},\aprs{c,\pars{\dual{d},e}},\aprs{d,\dual{e}}}}
	{\root{=}{\pars{\aprs{\circ,\dual{c}},\aprs{c,\pars{\dual{d},e}},\aprs{d,\dual{e}}}}
	{\root{\rules}{\pars{\dual{c},\aprs{c,\pars{\dual{d},e}},\aprs{d,\dual{e}}}}
	{\root{\ruleaidown}{\pars{\aprs{\pars{c,\dual{c}},\pars{\dual{d},e}},\aprs{d,\dual{e}}}}
	{\root{=}{\pars{\aprs{\circ,\pars{\dual{d},e}},\aprs{d,\dual{e}}}}
	{\root{\rules}{\pars{\dual{d},e,\aprs{d,\dual{e}}}}
	{\leaf{\pars{\aprs{\pars{d,\dual{d}},\dual{e}},e}}
	}}}}}}}}}}
	$$
	
	\item Agora a estrutura a ser demonstrada é a seguinte:
	
		$$S = \pars{\aprs{\pars{d,\dual{d}},\dual{e}},e}$$
		
		Números de incoerência módulo coerência:
		
		\begin{center}
			\begin{tabular}{|c||c|c|}
				\hline
				Átomo $i$                    & $d$ & $e$\\ \hline
				$\incnumber{S}{i}{\dual{i}}$ &  0  &  1 \\ \hline
			\end{tabular}	
		\end{center}
		
		O menor valor é $\incnumber{S}{d}{\dual{d}} = 0$	
			
		Recorremos, então, ao Lema \ref{lema:incnumber0}, que trata de número de incoerênica módulo coerência igual a 0, e vemos que o nosso caso equivale a:
		
		$$S'\pars{d,\dual{d}}$$		
		, de onde concluímos que devemos aplicar uma instância de $\ruleaidown$ da forma:
		
		$$
			\dernote{\ruleaidown}{}{S'\pars{d,\dual{d}}}
			{\leaf{S'\{\circ\}}}
		$$
		
		Assim, podemos continuar a construção da nossa demonstração, já levando em conta a relação de equivalência sintática:
		
		$$
	\dernote{\rules}{}{\pars{\aprs{a,\dual{b}},\aprs{\dual{c},\pars{\dual{a},b}},\aprs{c,\pars{\dual{d},e}},\aprs{d,\dual{e}}}}
	{\rootnote{\rules}{}{\pars{\aprs{\pars{\aprs{a,\dual{b}},\dual{a},b},\dual{c}},\aprs{c,\pars{\dual{d},e}},\aprs{d,\dual{e}}}}
	{\root{\ruleaidown}{\pars{\aprs{\pars{b,\aprs{\pars{a,\dual{a}},\dual{b}}},\dual{c}},\aprs{c,\pars{\dual{d},e}},\aprs{d,\dual{e}}}}
	{\root{=}{\pars{\aprs{\pars{b,\aprs{\circ,\dual{b}}},\dual{c}},\aprs{c,\pars{\dual{d},e}},\aprs{d,\dual{e}}}}
	{\root{\ruleaidown}{\pars{\aprs{\pars{b,\dual{b}},\dual{c}},\aprs{c,\pars{\dual{d},e}},\aprs{d,\dual{e}}}}
	{\root{=}{\pars{\aprs{\circ,\dual{c}},\aprs{c,\pars{\dual{d},e}},\aprs{d,\dual{e}}}}
	{\root{\rules}{\pars{\dual{c},\aprs{c,\pars{\dual{d},e}},\aprs{d,\dual{e}}}}
	{\root{\ruleaidown}{\pars{\aprs{\pars{c,\dual{c}},\pars{\dual{d},e}},\aprs{d,\dual{e}}}}
	{\root{=}{\pars{\aprs{\circ,\pars{\dual{d},e}},\aprs{d,\dual{e}}}}
	{\root{\rules}{\pars{\dual{d},e,\aprs{d,\dual{e}}}}
	{\root{\ruleaidown}{\pars{\aprs{\pars{d,\dual{d}},\dual{e}},e}}
	{\root{=}{\pars{\aprs{\circ,\dual{e}},e}}
	{\leaf{\pars{\dual{e},e}}
	}}}}}}}}}}}}
	$$
	
	\item Chegamos a um ponto em que só falta demonstrar a estrutura:
	
		$$S = \pars{\dual{e},e}$$
		
		Números de incoerência módulo coerência:
		
		\begin{center}
			\begin{tabular}{|c||c|c|}
				\hline
				Átomo $i$                    & $e$\\ \hline
				$\incnumber{S}{i}{\dual{i}}$ &  0 \\ \hline
			\end{tabular}	
		\end{center}
		
		O menor valor é $\incnumber{S}{e}{\dual{e}} = 0$	
			
		Recorremos, então, ao Lema \ref{lema:incnumber0}, que trata de número de incoerênica módulo coerência igual a 0, e vemos que o nosso caso equivale a:
		
		$$S'\pars{e,\dual{e}}$$		
		, de onde concluímos que devemos aplicar uma instância de $\ruleaidown$ da forma:
		
		$$
			\dernote{\ruleaidown}{}{S'\pars{e,\dual{e}}}
			{\leaf{S'\{\circ\}}}
		$$
		
		Assim, podemos continuar a construção da nossa demonstração, já levando em conta a relação de equivalência sintática:
		
		$$
	\dernote{\rules}{}{\pars{\aprs{a,\dual{b}},\aprs{\dual{c},\pars{\dual{a},b}},\aprs{c,\pars{\dual{d},e}},\aprs{d,\dual{e}}}}
	{\rootnote{\rules}{}{\pars{\aprs{\pars{\aprs{a,\dual{b}},\dual{a},b},\dual{c}},\aprs{c,\pars{\dual{d},e}},\aprs{d,\dual{e}}}}
	{\root{\ruleaidown}{\pars{\aprs{\pars{b,\aprs{\pars{a,\dual{a}},\dual{b}}},\dual{c}},\aprs{c,\pars{\dual{d},e}},\aprs{d,\dual{e}}}}
	{\root{=}{\pars{\aprs{\pars{b,\aprs{\circ,\dual{b}}},\dual{c}},\aprs{c,\pars{\dual{d},e}},\aprs{d,\dual{e}}}}
	{\root{\ruleaidown}{\pars{\aprs{\pars{b,\dual{b}},\dual{c}},\aprs{c,\pars{\dual{d},e}},\aprs{d,\dual{e}}}}
	{\root{=}{\pars{\aprs{\circ,\dual{c}},\aprs{c,\pars{\dual{d},e}},\aprs{d,\dual{e}}}}
	{\root{\rules}{\pars{\dual{c},\aprs{c,\pars{\dual{d},e}},\aprs{d,\dual{e}}}}
	{\root{\ruleaidown}{\pars{\aprs{\pars{c,\dual{c}},\pars{\dual{d},e}},\aprs{d,\dual{e}}}}
	{\root{=}{\pars{\aprs{\circ,\pars{\dual{d},e}},\aprs{d,\dual{e}}}}
	{\root{\rules}{\pars{\dual{d},e,\aprs{d,\dual{e}}}}
	{\root{\ruleaidown}{\pars{\aprs{\pars{d,\dual{d}},\dual{e}},e}}
	{\root{=}{\pars{\aprs{\circ,\dual{e}},e}}
	{\root{\ruleaidown}{\pars{\dual{e},e}}
	{\leaf{\circ}
	}}}}}}}}}}}}}
	$$
	
	\item Para finalizar, temos que:
	
	$$S = \circ$$
	
	e podemos invocar o axioma lógico (unidade), concluindo, assim, a nossa demonstração:
	
	$$
	\dernote{\rules}{}{\pars{\aprs{a,\dual{b}},\aprs{\dual{c},\pars{\dual{a},b}},\aprs{c,\pars{\dual{d},e}},\aprs{d,\dual{e}}}}
	{\rootnote{\rules}{}{\pars{\aprs{\pars{\aprs{a,\dual{b}},\dual{a},b},\dual{c}},\aprs{c,\pars{\dual{d},e}},\aprs{d,\dual{e}}}}
	{\root{\ruleaidown}{\pars{\aprs{\pars{b,\aprs{\pars{a,\dual{a}},\dual{b}}},\dual{c}},\aprs{c,\pars{\dual{d},e}},\aprs{d,\dual{e}}}}
	{\root{=}{\pars{\aprs{\pars{b,\aprs{\circ,\dual{b}}},\dual{c}},\aprs{c,\pars{\dual{d},e}},\aprs{d,\dual{e}}}}
	{\root{\ruleaidown}{\pars{\aprs{\pars{b,\dual{b}},\dual{c}},\aprs{c,\pars{\dual{d},e}},\aprs{d,\dual{e}}}}
	{\root{=}{\pars{\aprs{\circ,\dual{c}},\aprs{c,\pars{\dual{d},e}},\aprs{d,\dual{e}}}}
	{\root{\rules}{\pars{\dual{c},\aprs{c,\pars{\dual{d},e}},\aprs{d,\dual{e}}}}
	{\root{\ruleaidown}{\pars{\aprs{\pars{c,\dual{c}},\pars{\dual{d},e}},\aprs{d,\dual{e}}}}
	{\root{=}{\pars{\aprs{\circ,\pars{\dual{d},e}},\aprs{d,\dual{e}}}}
	{\root{\rules}{\pars{\dual{d},e,\aprs{d,\dual{e}}}}
	{\root{\ruleaidown}{\pars{\aprs{\pars{d,\dual{d}},\dual{e}},e}}
	{\root{=}{\pars{\aprs{\circ,\dual{e}},e}}
	{\root{\ruleaidown}{\pars{\dual{e},e}}
	{\root{\ruleodown}{\circ}
	{
	}}}}}}}}}}}}}}
	$$

	\end{enumerate}

\end{Example}

\chapter{Implementação}
\label{cap:implementacao}

A tentativa de estratégia de demonstração proposta para o sitema \FBV\ com pares de átomos distintos dois a dois foi implementada em uma ferramenta chamada de \cosprover\ (demonstrador de cálculo das estruturas). A implementação foi feita em Java e a documentação se encontra disponível em \cite{cosprover-documentacao}.

O formato de entrada e saída de dados para o \cosprover\ é apresentado a seguir.

\section{Entrada de dados}
\label{section:entrada}

		A entrada para o \cosprover\ é uma $<$estrutura$>$ em \FBV \ gerada pela seguinte gramática:\\

			\begin{tabular}{lll}
				$<$unidade$>$    & ::= & \unidade \\
				$<$letra$>$      & ::= & a $\hdots$ z \ $\mid$ \ A $\hdots$ Z \\
		 		$<$algarismo$>$  & ::= & 0 $\hdots$ 9 \\
		 		$<$atm$>$        & ::= & $<$letra$>$ \ $\mid$ \ $<$algarismo$>$ \ $\mid$ \ $<$atm$>$ $<$letra$>$ \ $\mid$ \\ 
		 		             &     & $<$atm$>$ $<$algarismo$>$  \\		 		
		 		$<$atomo$>$      & ::= & $<$atm$>$ \ $\mid$ \ \negsymbol{$<$atm$>$} \\
		 		$<$estruturas$>$ & ::= & $<$estrutura$>$ \ $\mid$ \ $<$estruturas$>$ , $<$estrutura$>$ \\
		 		$<$estrutura$>$  & ::= & $<$unidade$>$ \ $\mid$ $<$atomo$>$ \ $\mid$ \ [$<$estruturas$>$] \ $\mid$ \ ($<$estruturas$>$) \\
		 	\end{tabular}\\ \\
		
		A tabela~\ref{tab:ex-gramatica} apresenta exemplos de construções válidas da gramática acima.
		
		\begin{table}[!htb]
			\center
			\begin{tabular}{|l|l|}
		
				\hline
				$<$unidade$>$ & \unidade \\ \hline
				$<$atomo$>$   & a, \negsymbol{b1}, aToMo, \negsymbol{123} \\ \hline
				$<$estrutura$>$ & \unidade, \negsymbol{a}, \pr{a,\negsymbol{b1}}, \cpr{c,\pr{\negsymbol{a},b}} \\ \hline		
			\end{tabular}
			
			\caption{Exemplos das principais construções da gramática de estruturas para o \cosprover.}
			\label{tab:ex-gramatica}
			
		\end{table}
				
		São exemplos de entradas válidas:
		
		\begin{enumerate}
		
			\item \pr{\negsymbol{a}, \cpr{a,\negsymbol{b}}, \cpr{b,\negsymbol{c}}, \cpr{c,\negsymbol{d}}, \cpr{d,\negsymbol{e}}, \cpr{e,\negsymbol{f}}, f }
			
			\item \pr{\cpr{a,b,c}, \negsymbol{a}, \negsymbol{b}, \negsymbol{c}}
			
			\item \cpr{a, \pr{\negsymbol{a}}, \unidade}
			
			\item \unidade
			
			\item \pr{\pr{a,b}, \unidade}
		
		\end{enumerate}
		
		Note que a estrutura de entrada não precisa estar na forma normal. As estruturas (3) e (5) possuem unidades, a estrutura (3) apresenta um \emph{singleton} e a estrutura (5) apresenta associatividade explícita. 
		
		Além disso, é importante observar que as únicas estruturas que podem estar negadas são átomos. Por exemplo, a estrutura \negsymbol{\cpr{a,b}}, apesar de ser bem formada para o sistema \FBV, não é uma entrada permitida. Uma estrutura equivalente permitida seria \pr{\negsymbol{a}, \negsymbol{b}}
		
\section{Saída de dados}
\label{section:saida}
	
		A saída é uma mensagem dizendo que a estrutura não é provável (se for o caso), ou então uma demonstração em \FBV\ para a estrutura dada.
		
		Caso haja uma demonstração, cada instância de regra:
		
		\begin{center}
			\fbox{
				$$
				\dernote{\rho}{}{contractum}{\leaf{redex}}
				$$
			}
		\end{center}
		
		é representada por
		
		\begin{center}
			\fbox{
				\begin{tabular}{ll}
					\textsf{regra} & \textsf{redex} \\
					           & \textsf{contractum} \\
				\end{tabular}
			}
		\end{center}
		
		Toda estrutura que aparece na saída está na forma normal. As instâncias de regras triviais (equivalência =) são omitidas.
		
		A Tabela~\ref{tab:regras} apresenta a representação do nome das regras.
		
		\begin{table}[!htb]
			\center
			\begin{tabular}{|c|c|}
				\hline
				Regra & Representação \\
				\hline \hline
				$\ruleodown$ & oi \\ \hline
				$\ruleaidown$ & ai \\ \hline
				$\rules$ & s \\ \hline				
			\end{tabular}
			\caption{Representação de regras na saída.}
			\label{tab:regras}
		\end{table}
		
		São exemplos de saídas:
		
		\begin{enumerate}
		
			\item
			
				\begin{tabular}{ll}
					oi & \\
					ai &  \unidade \\
					ai &  \pr{\negsymbol{c},c} \\
					 s &  \pr{\negsymbol{c},\cpr{\pr{b,\negsymbol{b}},c}} \\
					ai &  \pr{\negsymbol{b},\negsymbol{c},\cpr{b,c}} \\
					 s &  \pr{\negsymbol{b},\negsymbol{c},\cpr{\pr{a,\negsymbol{a}},b,c}} \\
					   & \pr{\cpr{a,b,c},\negsymbol{a},\negsymbol{b},\negsymbol{c}} \\
				\end{tabular}
			
			\item 
			
				\begin{tabular}{ll}
					oi &  \\
					ai &  \unidade \\
					ai &  \pr{f,\negsymbol{f}} \\ 
					 s &  \pr{f,\cpr{\pr{e,\negsymbol{e}},\negsymbol{f}}} \\
					ai &  \pr{\cpr{e,\negsymbol{f}},f,\negsymbol{e}} \\
					 s &  \pr{\cpr{e,\negsymbol{f}},f,\cpr{\pr{d,\negsymbol{d}},\negsymbol{e}}} \\
					ai &  \pr{\cpr{d,\negsymbol{e}},\cpr{e,\negsymbol{f}},f,\negsymbol{d}} \\
					 s &  \pr{\cpr{d,\negsymbol{e}},\cpr{e,\negsymbol{f}},f,\cpr{\pr{c,\negsymbol{c}},\negsymbol{d}}} \\
					ai &  \pr{\cpr{c,\negsymbol{d}},\cpr{d,\negsymbol{e}},\cpr{e,\negsymbol{f}},f,\negsymbol{c}} \\
					 s &  \pr{\cpr{c,\negsymbol{d}},\cpr{d,\negsymbol{e}},\cpr{e,\negsymbol{f}},f,\cpr{\pr{b,\negsymbol{b}},\negsymbol{c}}} \\
					ai &  \pr{\cpr{b,\negsymbol{c}},\cpr{c,\negsymbol{d}},\cpr{d,\negsymbol{e}},\cpr{e,\negsymbol{f}},f,\negsymbol{b}} \\
					 s &  \pr{\cpr{b,\negsymbol{c}},\cpr{c,\negsymbol{d}},\cpr{d,\negsymbol{e}},\cpr{e,\negsymbol{f}},f,\cpr{\pr{a,\negsymbol{a}},\negsymbol{b}}} \\
					   & \pr{\negsymbol{a},\cpr{a,\negsymbol{b}},\cpr{b,\negsymbol{c}},\cpr{c,\negsymbol{d}},\cpr{d,\negsymbol{e}},\cpr{e,\negsymbol{f}},f} \\
				\end{tabular}

				\item
				
					\begin{tabular}{ll}
					
						oi &  \\
						ai &  \unidade \\
						ai &  \pr{e,\negsymbol{e}} \\
						 s &  \pr{e,\cpr{\pr{d,\negsymbol{d}},\negsymbol{e}}} \\
						ai &  \pr{\cpr{d,\negsymbol{e}},\negsymbol{d},e} \\
						 s &  \pr{\cpr{d,\negsymbol{e}},\cpr{\pr{c,\negsymbol{c}},\pr{\negsymbol{d},e}}} \\
						ai &  \pr{\cpr{c,\pr{\negsymbol{d},e}},\cpr{d,\negsymbol{e}},\negsymbol{c}} \\
						ai &  \pr{\cpr{c,\pr{\negsymbol{d},e}},\cpr{d,\negsymbol{e}},\cpr{\pr{b,\negsymbol{b}},\negsymbol{c}}} \\
						 s &  \pr{\cpr{c,\pr{\negsymbol{d},e}},\cpr{d,\negsymbol{e}},\cpr{\pr{b,\cpr{\pr{a,\negsymbol{a}},\negsymbol{b}}},\negsymbol{c}}} \\
						 s &  \pr{\cpr{c,\pr{\negsymbol{d},e}},\cpr{d,\negsymbol{e}},\cpr{\pr{\negsymbol{a},b,\cpr{a,\negsymbol{b}}},\negsymbol{c}}} \\
						   & \pr{\cpr{a,\negsymbol{b}},\cpr{\negsymbol{c},\pr{\negsymbol{a},b}},\cpr{c,\pr{\negsymbol{d},e}},\cpr{d,\negsymbol{e}}} \\							
					\end{tabular}
					
				\item
				
					\begin{tabular}{ll}
					
						 & \pr{\cpr{a,b},\cpr{\negsymbol{a},\negsymbol{b}}} \\
				 		 & The structure is not provable. \\
					
					\end{tabular}
					
				\item
				
					\begin{tabular}{ll}
					
						  &\pr{\cpr{a,\negsymbol{b}},\cpr{b,\negsymbol{c}},\cpr{c,\negsymbol{d}},\cpr{d,\negsymbol{e}},\cpr{e,\negsymbol{f}},\cpr{f,\negsymbol{a}}} \\
				 		 & The structure is not provable. \\
					
					\end{tabular}								
		
		\end{enumerate}
		
\section{Complexidade computacional}
\label{section:complexidade}

Uma questão importante é analisar a complexidade computacional em termos de tempo do algoritmo proposto. Duas questões centrais para a implementação da estratégia de demonstração são:

\begin{enumerate}
	\item Como calcular o número de incoerência módulo coerência.	
	\item Como resolver a questão da igualdade sob a definição de relação de equivalência sintática.	
\end{enumerate}

Vamos discutir em mais detalhes a complexidade de uma dessas questões.

\subsubsection{Cálculo do número de incoerência módulo coerência}
\label{section:incnumber}

		Seja $n = |\occ{S}|$, ou seja, o tamanho da estrutura a ser demonstrada $S$. A complexidade de se calcular o número de incoerência módulo coerência entre dois átomos $a$ e $b$ no contexto $S$ é $O(n^{4})$. 
		A estratégia precisa que se calcule o menor número de incoerência módulo coerência, portanto é preciso calcular $\incnumber{S}{i}{\dual{i}}$ para cada um dos $n/2$ pares de átomos duais $(i,\dual{i})$. Dessa, forma, a complexidade de se calcular o menor $\incnumber{S}{i}{\dual{i}}$ é $O(n/2 * n^{4}) = O(n^{5})$, o que é, claramente, um custo polinomial.
		
\subsubsection{Igualdade sob a definição de relação de equivalência sintática}
\label{section:igualdade}

		Uma possível fonte de não-determinismo é a que se oculta sob a relação de equivalência sintática (ver Figura \ref{fig:equivalencia-sintatica}). Uma mesma estrutura pode se apresentar em diversas formas equivalentes. Por exemplo:
		
		$$\pars{a,\aprs{b,c}} = \pars{\aprs{a,\circ},\aprs{b,\pars{c,\circ}}} = \pars{\pars{a},\aprs{b,\aprs{c}}}$$

		A rigor, esse não determinismo não causa nenhum problema do ponto de vista da aplicação da estratégia proposta. O número de incoerência módulo coerência é calculado sobre a teia de interação da estrutura, e o Teorema \ref{theorem:equivalencia} garante que duas estruturas equivalentes possuem a mesma teia de interação. Logo, como só se lida com a teia de interação, pouco importa a maneira como a estrutura está apresentada.
		
		Entretanto, a fim de uniformizar a representação utilizada, toda estrutura dentro do \cosprover\ deve estar na forma normal (ver Seção \ref{section:estruturas}). Ou seja, em toda estrutura é preciso que:
		
		\begin{itemize}
			\item as únicas estruturas negadas sejam átomos;
			\item não haja unidades ($\circ$);
			\item nenhum delimitador possa ser eliminado mantendo a equivalência.
		\end{itemize}
		
		A opção por colocar, sempre que possível, as estruturas em sua forma normal tem duas vantagens principais. Em primeiro lugar, é esteticamente mais agradável a apresentação de estruturas na forma mais compacta que é a forma normal. Em segundo lugar, a representação em forma normal evita subestruturas desnecessárias, o que representa uma economia de memória na sua representação interna ao \cosprover.
		
		A estrutura de entrada não precisa estar na forma normal, entretanto, as únicas estruturas negadas devem ser átomos (ver Seção \ref{section:entrada}). Além disso, a cada aplicação de regra de inferência, o \emph{redex} é colocado na forma normal antes de dar prosseguimento ao algoritmo. 
		
		Portanto, é preciso uma função que coloque uma estrutura na forma normal. Como nenhuma regra de inferência lida com negação, e a estrutura de entrada só pode possuir átomos negados, o algoritmo que coloca uma estrutura na forma normal dentro do \cosprover\ precisa lidar com os seguintes casos:
		
		\begin{enumerate}
			\item Eliminar unidades.\\
				Exemplo: $\pars{a,b,\circ,c} \implies \pars{a,b,c}$
			\item Eliminar delimitadores supérfluos. Divide-se em dois casos:
				\begin{enumerate}
					\item Eliminar \emph{singletons}.\\
						Exemplo: $\pars{a,\aprs{b},c} \implies \pars{a,b,c}$
					\item Eliminar associatividade explícita.\\
						Exemplo: $\pars{a,\pars{b,c}} \implies \pars{a,b,c}$
				\end{enumerate}
		\end{enumerate}
		
		O algoritmo que coloca estruturas na forma normal realiza os três passos seguintes:
		
		\begin{enumerate}
			\item elimina unidades a um custo $O(n^{2})$;
			\item elimina \emph{singletons} a um custo $O(n^{2})$;
			\item elimina associatividade explícita a um custo $O(n^{2})$;
		\end{enumerate}
		
		Como os três passsos são seqüenciais, o custo final de colocar uma estrutura na forma normal é de $O(n^{2})$, o que é, claramente, um custo polinomial.
		
%
		
\section{Qualidade dos resultados}
\label{section:qualidade}

Como visto no capítulo ~\ref{cap:aspectos-computacionais-cos}, o algoritmo proposto consiste em uma \emph{tentativa de estratégia}, no sentido de que a demonstração matemática de seu funcionamento ainda não está completa neste estágio da pesquisa. Entretanto, para todos os testes realizados, o \cosprover \ obteve o resultado correto, fornecendo uma demonstração no caso de a estrutura ser demonstrável, ou indicando que a mesma não o é. 

Sabemos que a ausência de contra-exemplos não equivale a uma demonstração de correção, mas em se tratando de um campo de pesquisa novo e de um formalismo cujo comportamento operacional ainda não é completamente  compreendido, este fato é um indício de o caminho escolhido possa ser promissor ou, no mínimo, que ainda é cedo para descartá-lo sem mais investimento em sua pesquisa.

\chapter{Conclusão}
\label{cap:conclusao}
O impacto que o cálculo das estruturas provocou na comunidade de Teoria da Demonstração ainda não foi completamente absorvido. Como um formalismo em muitos casos mais poderoso que o cálculo de seqüentes, aquele mais utilizado até então, o cálculo das estruturas deixou a comunidade muito interessada em entender como ele poderia ser melhor explorado. Após a formalização de diversas lógicas em cálculo das estruturas, como a lógica clássica, a intuicionista, a modal, etc., a atenção tem se voltado para o problema de \emph{como implementar} um sistema que utilize o cálculo das estruturas. A questão central reside no alto grau de não-determinismo que o formalismo apresenta. 

Uma idéia é encontrar uma estratégia de demonstrações para o cálculo das estruturas, que torne possível sua implementação, inspirada no conceito de demonstrações uniformes para cálculo de seqüentes. Neste trabalho propusemos uma tentativa de estratégia que atende esse objetivo para um subconjunto do sistema \FBV, um subsistema do sistema \BV, que corresponde ao fragmento multiplicativo da lógica linear mais a regra \mix \ (\MLL \ + \mix + seq). Dizemos \emph{tentativa de estratégia} porque, apesar de se basear em resultados teóricos apresentados formalmente e de ter funcionado perfeitamente em todos os testes que executamos (através da implementação computacional da tentativa de estratégia), a demonstração de sua correção ainda está em curso (apesar de a maior parte do trabalho neste sentido já estar pronta).

Se comprovada, a nossa tentativa de estratégia apresentará como principais contribuições relevantes:

\begin{enumerate}

	\item A estratégia teria uma fundamentação teórica, e não apenas operacional. O melhor trabalho no sentido de se melhorar a implementação do cálculo das estruturas até então se baseia na redução do não-determinismo do formalismo pela restrição da maneira como se aplicam regras de inferência. Esta é uma abordagem predominantemente operacional, que não captura a essência do processo de demonstração.
	
	\item A estratégia não exigiria \emph{bactracking}: uma vez que se decidisse aplicar uma regra de inferência, ela não precisaria ser reconsiderada mais em nenhum momento. A estratégia manteria a demonstrabilidade da estrutura em cada passo, de forma que seria sempre possível avançar na busca pela demonstração. Se não for possível avançar a partir de um certo ponto, é porque a estrutura já não era demonstrável desde o princípio.
	
\end{enumerate}	

Como pontos a serem melhorados, podemos levantar as seguintes questões:

\begin{enumerate}

	\item Ela funcionaria somente para um subsistema do sistema \FBV\ (lógica linear multiplicativa \MLL\ mais a regra \mix) em que não houvesse pares de átomos repetidos. É preciso ainda mais trabalho para suportar átomos repetidos e incluir na estratégia a noção de como lidar com a estrutura seq, no sentido de expandí-la para ser usada no sistema \BV\ completo.
	
	\item As estruturas que poderiam ser demonstradas deveriam ter pares de átomos distintos, ou seja, não seriam permitidos dois átomos iguais. Se a estrutura possuísse átomos iguais, seria possível diferenciá-los atribuindo índices diferentes aos átomos. Entretanto, o trabalho de atribuir estes índices parece não ser trivial, pois é preciso ter uma função de escolha $f : \occ{S} \rightarrow \occ{S}. \ f(a) = \neg{a} \ \  \wedge \ \ a \downarrow  \dual{a}$, que é uma condição necessária para uma estrutura ser provável. 

\end{enumerate}	

Para finalizar, podemos dizer que o trabalho representa um passo, apesar de restrito a um subsistema do sistema \BV, importante para o entendimento dos aspectos computacionais do cálculo das estruturas e, sem dúvida, terá certa visibilidade na comunidade de Teoria da Demonstração.




\bibliographystyle{amsalpha}
\bibliography{msalvim_dissertacao}


\end{document}